\documentclass[a4paper,11pt]{article}
\pdfoutput=1
\usepackage{jheppub}
\usepackage{graphicx}
\usepackage{amssymb}
\usepackage{amsmath}
\usepackage{amsfonts}
\usepackage{latexsym}
\usepackage{hyperref}
\usepackage{physics}
\usepackage{bbm}
\usepackage{empheq}
\usepackage{tikz}
\usepackage{cancel}
\usepackage{hhline}
\usepackage{soul}
\usepackage{float}
\usepackage{multirow}
\usepackage{hhline}
\usepackage{algorithm}
\usepackage{algpseudocode}
\usepackage{appendix}
\usepackage{subcaption}
\usepackage{makecell}
\usepackage[normalem]{ulem}

\usepackage[T1]{fontenc}

\usepackage{mathtools}
\usepackage{comment}
\usepackage[ansinew]{inputenc}
\usepackage{anysize}
\usepackage{titlesec}
\usepackage{enumerate}
\usepackage{enumitem}

\usepackage{empheq}

\usepackage{scalerel}[2016-12-29]

\usetikzlibrary{arrows.meta}
\usetikzlibrary{decorations.markings}
\tikzset{->-/.style={decoration={
			markings,
			mark=at position #1 with {\arrow{>}}},postaction={decorate}}}

\newcommand{\R}{\mathbb{R}}

\newcommand\AdS{\mathrm{AdS}}

\newcommand\rO{\mathrm{O}}
\newcommand\SO{\mathrm{SO}}

\newcommand\UU{\mathrm{U}}
\newcommand\SU{\mathrm{SU}}

\newcommand\cG{\mathcal{G}}

\newcommand\cL{\mathcal{L}}
\newcommand\cM{\mathcal{M}}
\newcommand\cN{\mathcal{N}}
\newcommand\cO{\mathcal{O}}
\newcommand\cP{\mathcal{P}}
\newcommand\cQ{\mathcal{Q}}

\newcommand\cS{\mathcal{S}}

\newcommand\cZ{\mathcal{Z}}

\newcommand{\be}{\begin{equation}}
\newcommand{\ee}{\end{equation}}

\newcommand\CO{\mathcal{O}}
\newcommand\CZ{\mathcal{Z}}
\newcommand{\rme}{\mathrm e}
\newcommand{\rmd}{\mathrm d}
\newcommand{\rmi}{\mathrm i}
\usepackage{array}   
\newcolumntype{C}{>{$}c<{$}}
\definecolor{purple2}{rgb}{0.4,.2,0.7}

\makeatletter
\newcommand*{\rom}[1]{\expandafter\@slowromancap\romannumeral #1@}
\makeatother

\marginsize{5.2cm}{0.4cm}{3.91cm}{-0.76cm}
\begin{document}

\thispagestyle{empty}

~\\[-1.5cm]

\begin{center}
\vskip 1.2truecm 
{\Large\bf
%
{\LARGE CFTs on Squashed Spheres \\ and the Thermal Effective Action}}\\
\vskip 1.95truecm
	{\bf Klaas Parmentier$^{1,2}$ and Nikolay Bobev$^{3}$
	}
		\vskip 0.44truecm
 
	{\it	$^1$Department of Physics, Columbia University,\\ New York, NY 10027, USA\\\vskip .2truecm}
		\vskip .2truecm   
	{\it	$^2$Department of Applied Mathematics and Theoretical Physics,\\
University of Cambridge, Cambridge CB3 0WA, UK\\\vskip .2truecm}
		\vskip .2truecm	
	{\it	$^3$Instituut voor Theoretische Fysica and Leuven Gravity Institute, KU Leuven, \\
	Celestijnenlaan 200D, B-3001 Leuven, Belgium\\\vskip .2truecm}
		\vskip .2truecm   

\vskip 0.25truecm		

\href{mailto:kp659@cam.ac.uk}{{\tt kp659@cam.ac.uk}}\,,\,  \href{nikolay.bobev@kuleuven.be}{{\tt nikolay.bobev@kuleuven.be}}

\end{center}

\vskip 0.75truecm

\centerline{\bf Abstract}
\vskip .5truecm
 \noindent We study three-dimensional CFTs on compact Euclidean manifolds in two complementary limits: small deformations of the round $S^3$ and the small-fiber, large-squashing limit of Seifert manifolds. Near the round sphere, conformal perturbation theory expresses the free-energy response through integrated stress-tensor correlators. We derive a harmonic-space formula for the universal quadratic response to arbitrary metric squashing and show that it is proportional to the $c_T$ coefficient of the stress-tensor two-point function. For unitary CFTs this establishes that the sphere free energy is a local maximum in the space of metric deformations. This result extends to conserved spin-$s$ currents, whose quadratic response alternates in sign. For the specific case of squashing the Hopf fiber, we find an explicit form for the cubic response, and in addition obtain the leading correction to the two-point function of scalar operators. In the small-fiber limit, corresponding to the large temperature regime of the CFT, the partition function is governed by a two-dimensional thermal effective action constructed out of the Weyl-rescaled base metric and a Kaluza--Klein field strength. The thermal effective action relates CFT free energies on different backgrounds, including squashed Lens spaces, which we discuss in detail. We also explicitly determine the Wilson coefficients of this effective action, to various orders in the high-temperature expansion, for free fields, the large-$N$ critical O($N$) model, and holographic CFTs.

\newpage
\setcounter{page}{0}\setcounter{footnote}{0}
\setcounter{tocdepth}{2}
\tableofcontents
\thispagestyle{empty}
\newpage

\section{Introduction}\label{sec:intro}
\begin{figure}
    \centering
    \href{https://youtu.be/kj4OzViCMIQ?si=yr6uOlCma3zQukMe&t=124}{\includegraphics[width=0.41\linewidth]{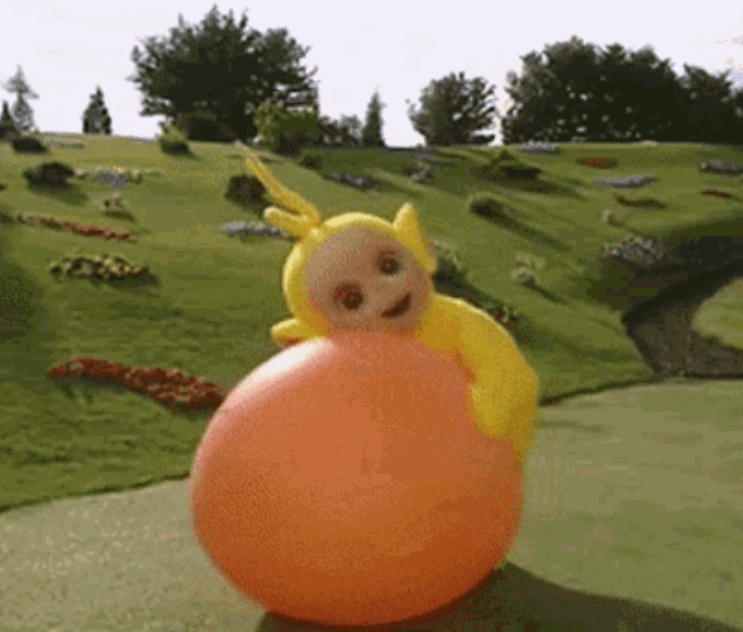}}
    \caption{The squashing of spheres has brought great joy to many generations of physicists.}
    \label{fig:squash}
\end{figure}

Coupling a quantum field theory to background fields gives a direct way to probe its dynamics.  For a conformal field theory (CFT), the metric provides a universal background field: every CFT has a conserved stress tensor, and placing the theory on a curved manifold amounts to sourcing this operator. In odd dimensions this is especially clean due to the absence of Weyl anomalies. Curved backgrounds are also very natural from the holographic point of view, as they simply change the boundary conditions for the dual gravitational system which in turn results in deformations of the AdS$_4$ geometry.

Moreover, in three dimensions, the finite part of the free energy on the round sphere plays a distinguished role as a measure of degrees of freedom. The $F$-theorem states that it should decrease along RG flows \cite{Jafferis:2010un,Klebanov:2011gs,Casini:2012ei}. It is therefore natural to ask how the free energy changes under deformations of the background geometry itself.

There are two controlled regimes in which the dependence of a three-dimensional CFT partition function on the background geometry becomes especially tractable. Near a conformally flat background, conformal perturbation theory expresses variations of the free energy in terms of integrated CFT correlators. Since the stress-tensor two-point function $\langle TT \rangle$ is fixed up to the positive coefficient $c_T$, the leading metric response is universal \cite{Osborn:1993cr,Bobev:2017asb,Fischetti:2017sut}. The next correction probes the finite set of CFT data entering $\langle TTT\rangle$ \cite{Osborn:1993cr,Erdmenger:1996yc,Hofman:2008ar,Buchel:2009sk}. At the opposite extreme, when a three-manifold has a small circle fiber, its free energy is determined by a local thermal effective action on the two-dimensional base \cite{Bhattacharyya:2007vs,Banerjee:2012iz,Benjamin:2023qsc}. The same curved-space observable therefore organizes CFT data in two rather different ways: through integrated zero-temperature correlators near the round sphere, and through Wilson coefficients of a lower-dimensional effective action in the small-fiber limit.

The goal of this paper is to study both descriptions for CFTs on curved three-manifolds. Near the round $S^3$, we use conformal perturbation theory and a decomposition in spherical harmonics to study general small metric deformations. In the large-squashing regime, we use thermal effective field theory (EFT) to study Seifert manifolds $S^1\times_{\mathrm f}\mathcal{M}_2$, where the shrinking fiber is interpreted as a thermal circle. We illustrate the resulting structures in free CFTs, in the large-$N$ critical $\rO(N)$ model, and in holographic CFTs dual to Einstein gravity.

\paragraph{Small squashing} A useful reference point is the Hopf squashing of $S^3$ --- a homogeneous but anisotropic deformation along the Hopf fiber of the round three-sphere which preserves four out of the six Killing vectors ---  for which the quadratic free-energy response is controlled by $c_T$ \cite{Bobev:2017asb,Zhang:2022nim}. We extend this analysis to arbitrary scalar, vector, and metric sources. In section~\ref{sec:quad}, we derive harmonic-space expressions for the quadratic response, using the decompositions of two-point kernels collected in appendix~\ref{app:spherharm}. This reproduces the Hopf-squashing result, yields the double-squashing formula previously inferred numerically, and extends directly to conserved spin-$s$ currents, whose quadratic response alternates in sign with $s$. Appendix~\ref{app:C} provides a complementary Ward-identity derivation of the relevant harmonic recurrence relation. We also discuss the supersymmetric squashing backgrounds of \cite{Hama:2011ea, Imamura:2011wg, Martelli:2013aqa} in section~\ref{sec:susy-squashings}.

At cubic order, the free energy depends on integrated three-point functions. For scalar deformations, we revisit these integrals on $S^d$ and give their harmonic decomposition on $S^3$ in appendix~\ref{app:3pt}. For metric deformations, we focus on Hopf squashing. Section~\ref{sec:TTT} explains how contact terms in different definitions of $\langle TTT\rangle$ enter the third metric variation of $\log \CZ$. This leads to a derivation of the cubic Hopf-squashing response proportional to $c_Tt_4$, previously conjectured based on free field and holographic calculations \cite{Bueno:2018yzo}. Finally, in section~\ref{sec:squashed-TOO} we compute the leading Hopf-squashing correction to the scalar two-point function. The regulated $\langle T\mathcal{O}\mathcal{O}\rangle$ distribution used there is detailed in appendix~\ref{app:TOOreg}. Our result agrees with that obtained via the ambient space formalism \cite{Parisini:2023nbd} and illustrates the role of contact terms in integrated correlators.

\paragraph{Large squashing} The opposite regime is the small-fiber limit of Seifert manifolds $S^1_\beta\times_{\mathrm f}\mathcal{M}_2$, which connects the problem to thermal physics. As we review in section~\ref{sec:ThEFT}, at small $\beta$, or large temperature, the equilibrium properties of the theory are captured by a local effective action on $\mathcal{M}_2$, organized in a derivative expansion. This goes back to the equilibrium partition function approach to hydrodynamics \cite{Bhattacharyya:2007vs,Banerjee:2012iz} and has recently been applied to extract asymptotic CFT data, including densities of states and heavy-state averaged OPE data \cite{Benjamin:2023qsc,Allameh:2024qqp}. In supersymmetric settings, such small-circle effective actions have been used to study the superconformal index of 4d $\mathcal{N}=1$ SCFTs, see~\cite{Cassani:2021fyv,DiPietro:2014bca}. 

Weyl invariance organizes the small-$\beta$ partition function into a thermal effective action built from the Weyl-rescaled metric on $\mathcal{M}_2$, its Ricci scalar $R$, and the Kaluza--Klein field strength $F$. In two dimensions, the non-derivative terms take the form $c_{n,m}R^nF^{2m}$, while higher orders also contain derivative invariants such as $(\nabla F)^2$. The theory-dependent Wilson coefficients $c_{n,m}$ provide a coarse-grained encoding of high-energy CFT data. We determine these coefficients in several complementary settings and apply the effective action formalism to relate the free energy on different backgrounds, including squashed Lens spaces. This illustrates an important point of the analysis. One can use one Seifert manifold for a given CFT to derive the thermal EFT coefficients and then apply the results to other, potentially more complicated, Seifert manifolds. 

For conformally coupled scalars and massless fermions in sections~\ref{sec:free-scalar} and \ref{sec:free-fermion}, we extract the curvature coefficients $c_{n,0}$ from the heat-kernel expansion on $S^1 \times S^2$ \cite{Bytsenko:1994bc,McKean:1967xf,DeWitt:1964mxt}. Turning on an angular fugacity $\Omega$ on $S^2$ gives access to coefficients involving $F^2$. Through the thermal effective action, these reproduce the large-squashing expansion on Lens spaces $S^3/\mathbb{Z}_p$ \cite{DeFrancia:2000xm}. 

In section~\ref{sec:criticalON}, we turn to the large-$N$ critical $\rO(N)$ model. Extending the analysis of~\cite{Hartnoll:2005yc} to Lens spaces and to higher order in the squashing parameter, we can extract the local EFT data through order $\beta^4$. The first few Wilson coefficients agree with recent results of \cite{David:2024pir,Mauro:2026zus} obtained from the theory on $S^1_\beta\times_\Omega S^2$. The leading $1/N$ correction to the free energy density can be found in \cite{Diatlyk:2023msc}. The Lens space approach allows us to further determine the coefficients $c_{1,1}$ and $c_{3,0}$ corresponding to the $RF^2$ and $R^3$ terms in the thermal EFT. 

Section~\ref{sec:holo} concerns holographic CFTs with Einstein gravity dual. We find in section~\ref{sec:taubbolt} that the AdS-Taub-Bolt geometry determines all of the non-derivative thermal EFT data. AdS-Kerr provides a consistency check presented in section~\ref{sec:kerr}, and further fixes the leading derivative coefficient while constraining subleading combinations. Higher-derivative bulk corrections are discussed in section~\ref{sec:higherderiv}; they simply rescale the thermal EFT and add an Euler-characteristic term, see also \cite{Bobev:2023ggk}. In appendix~\ref{app:AdSkerrbolt} we also discuss the more general AdS-Kerr-Bolt geometry.

\paragraph{Structure of the paper}

In section~\ref{sec:SquashS3} we develop conformal perturbation theory around the round sphere and compute quadratic corrections for scalar, vector, and metric sources. We analyze the cubic correction under Hopf squashing and the leading order correction to the scalar two-point function. In section~\ref{sec:ThEFT} we introduce the thermal EFT for $S^1\times_{\mathrm f}\mathcal{M}_2$, determining its coefficients for free fields, the large-$N$ critical $\rO(N)$ model, and holographic CFTs. Appendix~\ref{app:spherharm} collects harmonic decompositions of two-point functions on $S^d$, while appendix~\ref{app:C} derives a Ward-identity recurrence relation for the eigenvalues. Appendix~\ref{app:3pt} discusses integrated scalar three-point functions and appendix~\ref{app:TOOreg} has details on the regulated $\langle T\CO\CO\rangle$ correlator. Appendix~\ref{app:RxS2} presents related results for $\mathbb{R}\times S^2$, and finally, appendix~\ref{app:AdSkerrbolt} concerns the more general AdS-Kerr-Bolt geometry. Unless otherwise specified, we assume that the CFTs discussed in this work are parity invariant and unitary.

\section{CFTs on the squashed \texorpdfstring{$S^3$}{}}
\label{sec:SquashS3}
In this section we study conformal perturbation theory for Euclidean CFTs on the round $S^3$ coupled to small deformations of its background sources. For a deformation of order $\epsilon$, the order-$\epsilon^n$ contribution to the free energy is governed by an integrated $n$-point function of the operator to which the source couples. We will treat scalar, vector, and metric sources in a common notation, while keeping in mind an important distinction between them. A scalar source can represent a genuine coupling and can therefore initiate an RG flow, whereas a source for a conserved current, or the metric source for the stress tensor, is a background field probing the CFT at the same fixed point.\footnote{This distinction is particularly transparent in the momentum-space analysis of possible three-point-function singularities \cite{Bzowski:2015pba,Bzowski:2018fql}. Generic scalar three-point functions can contain semilocal singularities; removing them by source renormalisation induces a beta function \cite{Bzowski:2015pba}. For correlators built solely from conserved currents and/or stress tensors, the Ward identities forbid the corresponding gauge- or diffeomorphism-covariant source renormalisation, and these beta functions vanish.}

For the sources considered below, the linear term vanishes by conformal invariance. The quadratic correction is reviewed in section~\ref{sec:quad}, where we decompose both the sources and the two-point kernels into spherical harmonics. For metric deformations, this gives the response to an arbitrary smooth squashing of $S^3$. It is controlled entirely by the stress-tensor two-point function and hence by $c_T$. The required technical results on the harmonic decomposition of the two-point functions are collected in appendix~\ref{app:spherharm}, while appendix~\ref{app:C} gives a complementary derivation of the relevant recurrence relation from the stress-tensor Ward identity.

The metric result, derived in section~\ref{sec:quad2}, generalizes the Hopf-squashing calculation of~\cite{Bobev:2017asb} and gives a direct CFT derivation of the perturbative holographic result of~\cite{Fischetti:2017sut}. We illustrate the general formula with several examples, including the double-squashing background studied numerically in~\cite{Bobev:2016sap,Bobev:2017asb}. Since the quadratic variation is negative, the round sphere is a local maximum of the free energy under metric deformations. This may be interpreted as an energetic preference towards crumpling of the spatial geometry~\cite{Fischetti:2018shp}; in the dS/CFT setting this corresponds to having a peak of the Hartle--Hawking wavefunction near the round sphere \cite{Bobev:2016sap,Anninos:2012ft}. We also extend the analysis to $S^{2n+1}$ with $n>1$, recovering the higher-dimensional Hopf-squashing results of~\cite{Bobev:2017asb, Zhang:2022nim}. 

Analogous calculations are performed for deformations by operators of spin $s$. Section~\ref{sec:quad0} treats scalar deformations as a warm-up, while section~\ref{sec:quad1} studies conserved spin-one currents. For conserved currents, the general result exhibits an alternating sign with spin: in three dimensions the spin-one contribution is positive, whereas the stress-tensor contribution is negative, as also observed in~\cite{matijn}. Such opposite signs allow cancellations between background-field contributions, as is familiar from supersymmetric backgrounds, where the metric and gauge field deformations are related. We discuss these separately in section~\ref{sec:susy-squashings}.

The cubic correction, considered in section~\ref{sec:TTT}, is more subtle because it involves the integrated stress-tensor three-point function. In the parity-even sector relevant here, the independent three-dimensional $\langle TTT\rangle$ data are conventionally parametrized by $c_T$ and $c_Tt_4$ \cite{Osborn:1993cr,Erdmenger:1996yc,Hofman:2008ar,Buchel:2009sk}. We explain how contact terms enter the third metric variation of the free energy and, for the $\UU(1)\times\SU(2)$-invariant Hopf squashing, fix the resulting integrated correlator by holographic matching. This establishes the conjecture of~\cite{Bueno:2018yzo} that the cubic Hopf-squashing response is proportional to $c_Tt_4$. Appendix~\ref{app:3pt} contains some technical observations on integrated scalar three-point functions.

In section~\ref{sec:squashed-TOO} we then apply the same framework to the linear response of a scalar two-point function under Hopf squashing. This again requires careful treatment of contact terms; the regulated $\langle T\CO\CO\rangle$ distribution is presented in appendix~\ref{app:TOOreg}. 

Finally, in appendix~\ref{app:RxS2}, we apply the harmonic-decomposition method to vacuum-energy shifts on $\mathbb{R}\times S^2$ under scalar and metric deformations. The latter agrees with \cite{Fischetti:2017sut}, while the scalar result appears to be new.

\subsection{Quadratic correction to the free energy}\label{sec:quad}
We deform a CFT on $S^d$ with action $S_0$, by turning on a source $j$ for a given operator $\CO$:
\begin{equation}
    S = S_0 - \epsilon \int_{S^d} j \cdot \CO \,.
\end{equation}
Conformal invariance implies that the one-point function $\langle \CO\rangle$ vanishes on the conformally flat $S^d$ --- we are assuming $d$ is odd, so there is no Weyl anomaly. The leading change in the free energy $F=-\log \CZ$ is then quadratic in $\epsilon$:
\begin{equation}
   F(\epsilon) = F_0 + \frac{\epsilon^2}{2}F''(0) + O(\epsilon^3)\,,
\end{equation}
where 
\begin{equation}
    F''(0) = -\int_{S^d} \int_{S^d} j(x)\cdot \langle \CO(x) \CO(y)\rangle \cdot j(y)\,.
\end{equation}
To calculate this integral in general, we decompose both $j$ and $\langle \CO(x) \CO(y)\rangle$ into spherical harmonics. The two-point function on the sphere is related to the one in flat space by a conformal transformation. In stereographic coordinates, the metric on the unit sphere is
\begin{equation}\label{eq:stereo}
    \rmd s^2_{S^d} = \Omega^{-2}(x) \rmd s^2_{\mathbb{R}^d}\, , \qquad \Omega = \frac{1+x^2}{2}\,,
\end{equation}
and in our normalizations, the scalar two-point function is then given by
\begin{equation}
    \langle \CO(x) \CO(y)\rangle_{S^d} = \Omega^\Delta(x) \Omega^\Delta(y)\langle \CO(x) \CO(y)\rangle_{\mathbb{R}^d} = \frac{\Omega^\Delta(x) \Omega^\Delta(y)}{|x-y|^{2\Delta}}\,.
\end{equation}
The decomposition of this two-point function in spherical harmonics is reviewed in appendices~\ref{app:spherharm} and \ref{app:C}. Next, we study its implications.

\subsubsection{Spin-0 deformation}\label{sec:quad0}
Let us focus on $S^3$ first. If $j(x)=\sum a_{nlm}Y^{nlm}$, then $F''(0) = -\sum \lambda^\Delta_n |a_{nlm}|^2$, with $\lambda^\Delta_n$ the eigenvalues of the two-point function $\langle \CO \CO\rangle$ given in \eqref{eq:S32pt0}:
\begin{equation}\label{eq:F2scal}
   F''(0) = -\sum_{nlm} 4 \pi \, \Gamma(2-2\Delta)\sin(\pi \Delta) \frac{\Gamma(n+\Delta)}{\Gamma(n+3-\Delta)}|a_{nlm}|^2 \, ,
\end{equation}
With a constant source, only the $n=0$ mode contributes. Take for instance $j=1$, then $|a_{0}|^2=\cS_3 = 2\pi^2$, where $\cS_d=\text{Vol}(S^d)$. For a slightly relevant $\Delta=3-\varepsilon$, we get: 
\begin{equation}
    F''(0) = -\frac{8\pi^4 \, \Gamma(2\varepsilon-4)}{\Gamma(\varepsilon)\Gamma(\varepsilon-2)}\,,
\end{equation} 
having used the Euler reflection formula to simplify the result. This agrees with (17) in \cite{Klebanov:2011gs}.

The sign of the correction \eqref{eq:F2scal} is negative when $\Delta \in [\frac12,\frac32)$ (close to free field) as well as when $\Delta \in (\frac52,3)$ (slightly relevant). When $\Delta=3$ (marginal), the constant mode has vanishing contribution. For $\Delta \in (3,\frac72)$ (slightly irrelevant), the constant mode has instead positive sign, while those of higher $n$ do not. The above expressions allow an $F$-theorem-like conclusion for slightly relevant and almost free fields. At the special values $\Delta\in 3/2+\mathbb{N}$, we hit a pole and the scalar two-point function must be understood as a distribution. This pole is local and can be removed by a counterterm for the source, while the finite renormalized answer contains a nonlocal logarithmic piece.\footnote{In flat space this is the standard singularity of the distribution $|x-y|^{-2\Delta}$ at $\Delta=d/2+k$, proportional to $\Box^k\delta^{(d)}(x-y)$, which can be cancelled by a source counterterm $S_{\rm ct}\propto
\,\int  j \,\Box^k j$. Equivalently, in momentum space, the renormalized two-point function contains the nonlocal term $p^{2k}\log(p^2)$, together with scheme-dependent local terms proportional to $p^{2k}$ \cite{Osborn:1993cr, Bzowski:2013sza, Bzowski:2015pba}. On the round sphere the local operator $\Box^k$ is replaced by its conformally covariant completion, the GJMS operator $P_{2k}$ \cite{Graham:1992}. The result is visible directly from the harmonic eigenvalues \eqref{eq:Sd2pt0} when expanding $\Delta=\frac d2+k+\varepsilon$. One finds schematically:
\begin{equation}
\lambda_n^\Delta= \frac{A_k(n)}{\varepsilon}+ A_k(n) \Big[\psi\left(n+\frac d2+k\right)
+\psi\left(n+\frac d2-k\right)+\texttt{ct}\Big]+O(\varepsilon),\quad A_k(n)\propto
\frac{\Gamma(n+\frac d2+k)}{\Gamma(n+\frac d2-k)} .
\end{equation}
The residue $A_k(n)$ is the eigenvalue of $P_{2k}$, so that both the pole and $\texttt{ct}$ can be removed by counterterms, while the digamma terms are the compact-space analogue of $p^{2k}\log p^2$, with large-$n$ behavior $n^{2k}\log n^2$.}

In general dimensions, using \eqref{eq:Sd2pt0}, one finds instead
\begin{equation}
    F''(0)= - \sum_{nlm}2^{d-2\Delta}\pi^{\frac{d}{2}}\frac{\Gamma(\frac{d}{2}-\Delta)}{\Gamma(\Delta)}\frac{\Gamma(n+\Delta)}{\Gamma(n+d-\Delta)}|a_{nlm}|^2\,.
\end{equation}
For a slightly relevant $\Delta=d-\varepsilon$ this becomes  
\begin{equation}
    F''(0) = -\sum_{nlm}2^{-d+2\varepsilon}\pi^{\frac{d}{2}}\,\frac{\Gamma(-\frac{d}{2}+\varepsilon)}{\Gamma(d-\varepsilon)}\frac{\Gamma(n+d-\varepsilon)}{\Gamma(n+\varepsilon)}|a_{nlm}|^2\, ,
\end{equation}
reducing to (17) in \cite{Klebanov:2011gs} at $n=0$. Note that the sign alternates like $(-1)^{\tfrac{d-1}{2}}$ for odd $d$, as expected from the $F$-theorem \cite{Jafferis:2010un}. The same alternating sign is present for the $\Delta= d-2$ mass term of a free field. In the next subsection, we will similarly consider spin-$s$ deformations corresponding to conserved currents, with $\Delta=d+s-2$.

\subsubsection{Spin-1 deformation}\label{sec:quad1}
For a vector source $A_\mu$, the quadratic response is
\begin{equation}
    F''(0) = -\int_{S^d} \int_{S^d} A^\mu(x)\cdot \langle J_\mu(x) J_\nu(y)\rangle \cdot A^\nu(y)\,.
\end{equation}

\paragraph{Conserved current} Spin-one conserved currents have scaling dimension $\Delta = d-1$. In stereographic coordinates \eqref{eq:stereo}, their normalized two-point function on the unit sphere is
\begin{equation}\label{eq:JJ2pt}
    \langle J_\mu(x) J_\nu(y)\rangle_{S^d} = c_J\, \Omega^{d-2}(x) \Omega^{d-2}(y)\,\frac{\mathcal{I}_{\mu\nu}(x-y)}{|x-y|^{2d-2}}\, ,
\end{equation}
where we have defined
\begin{equation}\label{eq:OmIdef}
    \Omega(x) = \frac{1+x^2}{2}\, , \qquad \mathcal{I}_{\mu\nu}(x) = \delta_{\mu \nu} - 2\frac{x_\mu x_\nu}{x^2}\,.
\end{equation}

Let us first focus on the case $d=3$. Expanding $A^\mu$ in transverse vector spherical harmonics on $S^3$, using \eqref{eq:S3cons}:
\begin{equation}
    F''(0) = \sum_{nlm} c_J\frac{\,\pi^2 }{2}(n+1)|a_{nlm}|^2\, .
\end{equation}
In the above, $c_J$ is the normalization of the conserved current two-point function, which satisfies $c_J>0$ in unitary CFTs. In contrast to the scalar case, the quadratic result is therefore always positive, in agreement with the observations of \cite{matijn}. More generally, from \eqref{eq:S3cons} one can see that the sign for a conserved spin-$s$ deformation alternates with $s$. 

In general (odd) dimensions, using \eqref{eq:Sdspin1}, a conserved current deformation results in 
\begin{equation}
    F''(0) = \sum_{nlm} c_J \pi^{\frac{d}{2}}\,\frac{\Gamma(2-\frac{d}{2})\Gamma(n+d-1)}{2^{d-3}\Gamma(d)\Gamma(n+1)}|a_{nlm}|^2\, .
\end{equation}
Note again the alternation of the sign with dimensions, making it natural to consider $(-1)^{\frac{d+1}{2}}F$.

As an example, consider the case where $A^\mu$ is a Killing vector of $S^d$. This means taking $n=1$. We find that a unit-normalized KV deformation in (odd) dimension $d$ results in
\begin{equation}\label{eq:JJF2}
    F''_{\text{\tiny KV}}(0) = c_J \pi^{\frac{d}{2}}\,2^{3-d}\,\Gamma(2-\tfrac{d}{2})\,.
\end{equation}

In $d=3,5$ we can compare to \cite{matijn}. On $S^3$, with normalization $\int A^2 = 8\pi^2$ as in their (2.90), we get  $F''(0)= 8\pi^4 c_J$ , which is their (2.91).  On $S^5$, with normalization $\int A^2= \pi^3$ as in their (2.92), we read off $F''(0)=-\frac{\pi^6 c_J}{2}$, which is their (2.93). 

\paragraph{Contact terms}
In $d=3$, conserved currents can have a parity-odd contact term \cite{Closset:2012vp}:
\begin{equation}\label{eq:JJdelta}
    \langle J_\mu(x) J_\nu (0)\rangle_{\mathbb{R}^3} = \frac{\rmi \kappa}{2\pi}\epsilon_{\mu\nu\rho}\partial^\rho\, \delta^{(3)}(x)\,.
\end{equation}
It gives an imaginary contribution and can be understood as coming from a Chern-Simons term in the EFT of the source $A^\mu$. Counterterms compatible with large gauge transformations can only shift the integer part of $\kappa$, making it a meaningful observable mod 1. 

Using \eqref{eq:JJ2pt}, we find the following contribution from the contact term:
\begin{equation}
    F''(0) = \frac{8\rmi\kappa}{\pi}\int \frac{\rmd^3 x}{(1+x^2)^{4}}\epsilon_{\mu\nu\rho}A^\nu(x)\partial^\rho A^\mu(x)\,.
\end{equation}
For the specific Killing vector deformation $\partial_\psi$ on \eqref{eq:S3metric} considered in \cite{matijn}
\begin{equation}\label{eq:Adpsi}
    A^\mu = \frac12 \epsilon^{\mu\sigma 3}x^\sigma + \frac12 x^\mu x^3 + \frac14\delta^\mu_3(1-x^2)\,,
\end{equation}
the contact term contribution becomes
\begin{equation}
    F''(0) = -\,\frac{8\rmi \kappa }{\pi}\int \frac{\rmd^3 x}{(1+x^2)^{3}} =  -2\pi\rmi\kappa\,.
\end{equation}

\subsubsection{Spin-2 deformation}\label{sec:quad2}
Considering a deformation $h_{\mu\nu}$ of the background metric, we find
\begin{equation}\label{eq:Fspin2}
    F''(0) = - \frac14 \int_{S^d}\int_{S^d} h^{\mu\nu}(x)\cdot \langle T_{\mu\nu}(x) T_{\rho\sigma}(y)\rangle \cdot h^{\rho\sigma}(y)\,.
\end{equation}
The numerical prefactor is due to the standard definition of the stress tensor $T^{\mu\nu} = -\frac{2}{\sqrt{g}}\frac{\delta S}{\delta{g^{\mu\nu}}}$. Since $T^{\mu\nu}$ is traceless and conserved, the quadratic response is only sensitive to the transverse traceless part of $h^{\mu\nu}$. We now study this in more detail, using the results of appendix~\ref{app:spherharm}. In stereographic coordinates, the stress tensor two-point function in our normalization is
\begin{equation}
    \langle T_{\mu\nu}(x) T_{\rho\sigma}(y)\rangle_{S^d} = c_T\, \Omega^{d-2}(x)\Omega^{d-2}(y)\,\frac{\mathcal{I}_{\mu\nu,\rho\sigma}(x-y)}{|x-y|^{2d}}\,,
\end{equation}
where we have defined
\begin{equation}
    \mathcal{I}_{\mu\nu,\rho\sigma}(x-y) = \frac12\left(\mathcal{I}_{\mu\rho}(x) \mathcal{I}_{\nu\sigma}(x)+\mathcal{I}_{\mu\sigma}(x) \mathcal{I}_{\nu\rho}(x)\right) - \frac{1}{d} \delta_{\mu\nu}\delta_{\rho\sigma}\,.
\end{equation}
with the conformal factor $\Omega$ and tensor structure $\mathcal{I}_{\mu\nu}$ as in \eqref{eq:OmIdef}.

\paragraph{Squashing \texorpdfstring{$S^3$}{}}
Using \eqref{eq:S3cons}, we find that a general squashing results in\footnote{We are using the same conventions for $c_T$ as \cite{matijn, Fischetti:2017sut}, differing by a factor $\cS^2_{d-1}$ from \cite{Bobev:2017asb}.} 
\begin{equation}\label{eq:FS3spin2}
    F''(0) = - \sum_{nlm} \frac{\pi^2 c_T}{96} n(n+1)(n+2)|a_{nlm}|^2\,.
\end{equation}
Here, the stress-tensor two-point coefficient $c_T$ satisfies $c_T>0$ for unitary CFTs.
The result above agrees with the holographic calculation (B.13) of \cite{Fischetti:2017sut}. 

Let us discuss three deformations of particular interest.
\begin{enumerate}
    \item We can squash $S^3$ along its Hopf fiber, as in (3.1) of \cite{Bobev:2017asb}, keeping an $\SU(2)\times \UU(1)$ subgroup of the isometries. In terms of the left-invariant one-forms $\sigma_i$
    \begin{equation}
        \sigma_{1}+\rmi \sigma_2 = \rme^{\rmi \psi}(\rmi\rmd \theta+\sin\theta \rmd \phi ) \, , \qquad\sigma_3 = \rmd \psi + \cos\theta \rmd \phi\,,
    \end{equation}
    the metric on the Hopf-squashed $S^3$ is given by:
    \begin{equation}\label{eq:squashmet0}
        \rmd s^2 = \frac{1}{4}\left(\sigma^2_1 + \sigma^2_2 + \frac{1}{1+\alpha}\, \sigma^3_3\right)\,,
    \end{equation}
    More explicitly, with the metric on the unit $S^3$ written out as
\begin{equation}\label{eq:S3metric}
    \rmd s^2 = \frac14(\rmd \theta^2 + \rmd \phi^2 + \rmd \psi^2 +2\cos\theta \rmd \phi\rmd\psi)\,,
\end{equation}
the deformation of interest is $h^{\psi\psi} = -4$, which satisfies a Killing property:
\begin{equation}
    \nabla_{(\mu}h_{\nu\sigma)}=0\,.
\end{equation}
Only the tracefree part $\tilde{h}_{\mu\nu}= h_{\mu\nu}+\frac13g_{\mu\nu}$ contributes in \eqref{eq:Fspin2}. One verifies that
\begin{equation}\label{eq:KTprops}
    \nabla^2 \tilde{h}^{\mu\nu} =-6 \tilde{h}^{\mu\nu} \, ,\qquad \int_{S^{3}}\tilde{h}^{\mu\nu} \tilde{h}_{\mu\nu} = \frac{2}{3} \cS_3= \frac{4\pi^2}{3}\,.
\end{equation}
From this, we use \eqref{eq:S3harmonics} to read off the mode number $n=2$ --- the lowest one possible for $s=2$ --- and the normalization $|a_2|^2= \frac{4\pi^2}{3}$. Then \eqref{eq:FS3spin2} becomes $F''(0)=-\frac{\pi^4c_T}{3}$. Given our $c_T$ convention, this reproduces (1.8) of \cite{Bobev:2017asb}. 
    \item The following double squashing 
    \begin{equation}\label{eq:squashmetdouble}
        \rmd s^2 = \frac{1}{4}\left(\sigma^2_1 + \frac{1}{1+\beta}\,\sigma^2_2 + \frac{1}{1+\alpha}\, \sigma^3_3\right)\,,
    \end{equation}
    keeps an $\SU(2)$ subgroup of the isometries and was analyzed numerically in \cite{Bobev:2016sap, Bobev:2017asb}. It corresponds to applying the same squashing as in example 1, but in two different directions, $\alpha h_{\mu\nu}+ \beta d_{\mu\nu}$. Therefore, the normalizations of the tracefree parts $\tilde{h}^{\mu\nu}\tilde{h}_{\mu\nu}= \tilde{d}^{\mu\nu}\tilde{d}_{\mu\nu} = 2/3$ and mode numbers $n=2$ remain the same. The only new information is the overlap $\tilde{h}^{\mu\nu}\tilde{d}_{\mu\nu}=-1/3$, which leads to
    \begin{equation}
        F(\alpha,\beta)-F(0) \approx  -\frac{\pi^4 c_T}{6}(\alpha^2+ \beta^2-\alpha\beta)\,.
    \end{equation}
    This is precisely the relation conjectured in (5.7) of \cite{Bobev:2017asb} based on numerics.
    \item The $\UU(1)\times \UU(1)$ symmetric (ellipsoid) squashing of \cite{Hama:2011ea}. Here, the metric becomes\footnote{In the notation of \cite{Hama:2011ea}, we have $\ell^2/\tilde{\ell}^2 = 1+\epsilon$ with $\epsilon \ll 1$.}
    \begin{equation}\label{eq:HHL}
        \rmd s^2= ((1+\epsilon)\sin^2\theta + \cos^2\theta)\,\rmd \theta^2 + \sin^2\theta\,\rmd \varphi^2 + (1-\epsilon)\cos^2\theta \,\rmd \chi^2\,.
    \end{equation}
   The traceless part of the deformation is purely longitudinal:
    \begin{equation}
        \tilde h_{\mu\nu} = \nabla_{\mu}\nabla_{\nu} (\tfrac14\cos2\theta)+\tfrac23 \cos2\theta g_{\mu\nu}\,.
    \end{equation}
    Based on its eigenvalue $6-n(n+2)=-2$, we find it corresponds to $n=2$, see \eqref{eq:S3harmonics}. We conclude there is no $\epsilon^2$ correction to the partition function coming from this metric deformation. Note however that in the susy context of \cite{Hama:2011ea}, there is an accompanying background gauge field and therefore the full partition function will still receive a quadratic correction. We discuss this further in section~\ref{sec:susy-squashings}. 
\end{enumerate}

\paragraph{Squashing \texorpdfstring{$S^d$}{}}
In general (odd) dimensions, we find from \eqref{eq:Fspin2} and \eqref{eq:Sd2ptspin2} that
\begin{equation}
    F''(0)= \frac{(-1)^{\frac{d-1}{2}}\cS_{d+1}c_T}{(d+1)2^{d+3}\Gamma(d-1)}\sum_{nlm} (n)_d|a_{nlm}|^2\,.
\end{equation}

For a Killing deformation, $n=2$. Normalizing $\int \tilde{h}^2 = \frac{d-1}{d}\cS_d$ in analogy with \eqref{eq:KTprops} gives
\begin{equation}\label{eq:KVspin2gend}
    F''(0) = \frac{(-1)^{\frac{d-1}{2}}\cS_{d+1}\cS_d (d-1)^2c_T}{2^{d+3}} = \frac{(-1)^{\frac{d-1}{2}}\pi^{d+1}(d-1)^2 c_T}{2 \cdot d!}\,.
\end{equation}
For the special $\UU(1)$ Hopf-fiber squashing, this result was derived in \cite{Zhang:2022nim}, and before that for $S^3$ and $S^5$ in \cite{Bobev:2017asb}.\footnote{The normalization in \cite{Zhang:2022nim} follows from $h^{\mu\nu}h_{\mu\nu}=1$ and $h^{\mu\nu}g_{\mu\nu}=-1$, leading to $\tilde{h}^{\mu\nu}\tilde{h}_{\mu\nu}=\frac{d-1}{d}$.} It was conjectured for general $d$  based on holographic calculations~\cite{Bueno:2018yzo}. The derivation above is slightly more general: by decomposing an arbitrary metric source into tensor harmonics, it shows that the same result applies to any $n=2$ transverse-traceless deformation with the same normalization, not only to the Hopf-squashing representative.

\subsubsection{Supersymmetric two-parameter squashing backgrounds} \label{sec:susy-squashings}
Supersymmetry relates the metric and vector sources that were considered separately above.  A useful illustrative example is the two-parameter family of rigid $\mathcal{N}=2$ backgrounds constructed in~\cite{Martelli:2013aqa}.  Up to an overall conformal factor, the metric is
\begin{equation}\label{eq:MPmetric}\begin{split}
\rmd s^2&= \frac{\rmd\theta^2}{f(\theta)} +f(\theta)\sin^2\theta\,\rmd\hat\phi^2
+\left(\rmd\hat\psi+
\big(\cos\theta+a\sin^2\theta\big)\rmd\hat\phi\right)^2\,,\\
f(\theta)&=v^2-a^2\sin^2\theta-2a\cos\theta\,,
\end{split}
\end{equation}
while the gauge field dual to the $R$-symmetry current in the dual CFT takes the form
\begin{equation}\label{eq:MPgauge}
A^{(3)}=Q\left(\rmd\hat\psi+\cos\theta\,\rmd\hat\phi\right)\,.
\end{equation}
The hatted angles are local coordinates, whose global identifications are discussed in \cite{Martelli:2013aqa}.  At the round point $a=0$, $v=1$, they reduce to the standard Euler coordinates used in~\eqref{eq:S3metric}. Note that the gauge field is proportional to the one-form dual to the Hopf-fiber Killing vector. It corresponds to the $n=1$ example of section~\ref{sec:quad1}. To first order, the $v$-deformation turns on precisely the transverse-traceless Hopf tensor harmonic of the first $s=2$ example, normalized in \eqref{eq:KTprops}.  The $a$-deformation, on the other hand, is equivalent to the longitudinal $n=2$ deformation encountered in the ellipsoid example \eqref{eq:HHL} and has no projection onto the $\langle TT\rangle$ kernel. Since there is no mixed $\langle TJ_R\rangle$ contribution on the round sphere, it then follows directly from \eqref{eq:FS3spin2}, \eqref{eq:KTprops}, and \eqref{eq:JJF2} that
\begin{equation}\label{eq:MPquadraticcombined}
\Delta F^{(2)}
=-\frac{\pi^4c_T}{6}(v^2-1)^2
+4\pi^4c_{J_R}Q^2\,.
\end{equation}
For complex Euclidean backgrounds, this expression is understood by analytic continuation.\footnote{In particular, the current term is proportional to $Q^2$, rather than $|Q|^2$.} For an $\mathcal N=2$ SCFT, the stress tensor and superconformal $R$-current belong to the same multiplet, implying \cite{Closset:2012ru}:
\begin{equation}\label{eq:N2cTcJ}
c_T=6c_{J_R}\,.
\end{equation}

Supersymmetric completions of a fixed metric are
distinguished by discrete choices of $Q$.  For Type~I, the two terms in \eqref{eq:MPquadraticcombined} cancel:
\begin{equation}
Q_{\rm I}=\pm\frac{v^2-1}{2}\,, \qquad \Delta F_{\rm I}^{(2)}
=\pi^4\left(c_{J_R}-\frac16c_T\right)(v^2-1)^2=0\,.
\end{equation}
This includes as a special case at $a=0$ the familiar $\frac14$-BPS $\SU(2)\times\UU(1)$ background of \cite{Hama:2011ea}. 

For Type~II, a smooth parametrization near the round sphere is provided by $(a,Q)$, with
\begin{equation}\label{eq:MPtypeIIrelation}
v^2-1=\pm 4 aQ-4Q^2\,.
\end{equation}
Hence the transverse-traceless metric source begins only at second order in $a,Q$, while the current source is linear in $Q$.  Therefore
\begin{equation}\label{eq:MPtypeIIquadratic}
\Delta F_{\rm II}^{(2)} =4\pi^4c_{J_R}Q^2 =\frac{2\pi^4c_T}{3}Q^2\,.
\end{equation}
For the two-derivative holographic theories of \cite{Martelli:2013aqa}, where $F_{\rm round}=\pi^4c_{J_R}=\pi^4c_T/6$, this agrees with the leading order correction to $F_{\rm II} =F_{\rm round}/(1-4Q^2)$. The special case $a=0$ is the $\frac12$-BPS $\SU(2)\times \UU(1)$ background of \cite{Imamura:2011wg}.\footnote{This background takes in particular purely imaginary $Q= iu/2$. Note that $\Delta F \sim Q^2 \sim -u^2$ is still real.} The case $v=1, Q=a$ corresponds to the $\UU(1)\times\UU(1)$ ellipsoid of \cite{Hama:2011ea}. 

More generally, three-dimensional $\mathcal N=2$ theories may have parity-odd contact terms, described by supersymmetric background Chern--Simons terms~\cite{Closset:2012vp,Closset:2012ru}. These give imaginary contributions to $\log \CZ$, and are not captured by the separated-point $\langle TT\rangle$ and $\langle J_RJ_R\rangle$ calculation above, see also section~\ref{sec:quad1}.

\subsection{Cubic correction to the free energy}\label{sec:cub}
We now consider cubic corrections to the free energy. These involve integrated three-point functions. A metric deformation thus probes $\langle TTT\rangle$ data which, in $d=3$, besides $c_T$ contains an additional  coefficient $t_4$. We did not find a general decomposition in spherical harmonics, although this is possible for the scalar case, see appendix~\ref{app:3pt}. We therefore restrict to the lowest harmonic compatible with the spin: the constant mode for scalars, Killing-vector modes for currents, and the lowest transverse-traceless tensor for metric deformations. Concretely, in section~\ref{sec:TTT}, we are interested in proving the conjecture of \cite{Bueno:2018yzo} regarding the integrated $\langle TTT\rangle$ correlator on the Hopf-squashed $S^3$. Note that at this order, mixed correlators can also make an appearance. In particular, in the next section~\ref{sec:squashed-TOO}, we find the leading squashing correction to the scalar two-point function from the integrated $\langle T\CO\CO\rangle$ correlator.

\subsubsection{Spin-0 deformation}
For identical scalar primaries, the conformal three-point function admits a factorization into three two-point kernels of dimension $\Delta/2$. When $d=3$, we can then combine the harmonic decomposition of the latter with the triple overlap formula of \cite{Cutkosky:1983jd} to arrive at a decomposition of the 3-point function in spherical harmonics. This analysis is motivated by the ultimate goal of studying metric deformations, but it quickly becomes cumbersome due to the presence of $3j$- and $6j$-symbols --- we leave the details for appendix~\ref{app:3pt}. Let us just mention that for a constant $\Delta=d-\varepsilon$ deformation, as in \eqref{eq:3sum}, we retrieve the known result \cite{Cardy:1988cwa}:
\begin{equation}\label{eq:3ptint}
  F^{(3)}(0)=  -\int_{S^d}\int_{S^d}\int_{S^d} \langle \CO(x)\CO(y)\CO(z)\rangle
  = -\,C_{\scriptscriptstyle{\CO\CO\CO}}\,\frac{8\pi^{\tfrac{3}{2}(d+1)}\Gamma(\frac{3\varepsilon-d}{2})}{\Gamma(d)\Gamma(\frac{1+\varepsilon}{2})^3}\,.
\end{equation}

\subsubsection{Spin-1 deformation}
Consider now the conserved current three-point function $\langle J^\mu(x) J^\nu(y) J^\sigma(z)\rangle$ integrated against an $n=1$ mode, meaning a Killing vector $K^\mu$. Since one can always find a symmetry of the metric that flips the sign of $K\to -K$, we conclude that not only the integrated three-point function, but indeed all odd derivatives of the free energy under the Killing vector deformation vanish. In \cite{matijn}, a specific choice $K = \partial_\psi$ was made, on $S^3$ in coordinates \eqref{eq:S3metric}. The symmetry which leaves the metric untouched but flips the sign of $A$ is simply $(\theta, \phi,\psi)\to (\theta, -\phi,-\psi)$. It was also shown by direct computation in \cite{matijn} that the odd derivatives of the free energy of the conformally coupled scalar on $S^3$ and $S^5$ vanish under this deformation. 


\subsubsection{Spin-2 deformation}\label{sec:TTT}
We are interested in the cubic correction to the free energy
\begin{equation}\begin{split}\label{eq:cft3g}
    F^{(3)}(0) &= -\int \hspace{-0.1cm} \int\hspace{-0.1cm}  \int_{S^3} h^{\mu\nu}(x)h^{\rho \sigma}(y)h^{\alpha \beta}(z)\frac{\delta}{\delta g^{\mu\nu}}\frac{\delta}{\delta g^{\rho\sigma}}\frac{\delta}{\delta g^{\alpha \beta}}\log \cZ\\ 
    &= \frac18 \int\hspace{-0.1cm}  \int\hspace{-0.1cm}  \int_{S^3} h^{\mu\nu}(x)h^{\rho \sigma}(y)h^{\alpha \beta}(z)\, \langle T_{\mu\nu}(x)T_{\rho\sigma}(y)T_{\alpha\beta}(z) \rangle_{\scriptscriptstyle{\text{OP}}} \,.
    \end{split}
\end{equation}
As reflected by the subscript, the three-point correlator defined above via the third functional derivative of the generating function $\log \CZ$ follows the definition of Osborn and Petkou \cite{Osborn:1993cr}, see in particular their (6.6). This correlator has known tensor structures \cite{Osborn:1993cr, Erdmenger:1996yc}. There are three parity-even tensor structures, typically written with coefficients $c_T, t_2, t_4$ \cite{Hofman:2008ar, Buchel:2009sk}. In $d=3$ there is no $t_2$ structure. The cubic correction \eqref{eq:cft3g} is then found by integrating this three-point function --- with the correct regularization so that it satisfies the Ward identities ---  which will therefore give a result of the form:
\begin{equation}\label{eq:formF3}
    F^{(3)}(0)=c_T\,\#_1+c_Tt_4\,\#_2\,.
\end{equation}

The remaining task is to evaluate these two integrated tensor structures. This appears to be quite complicated --- for a slightly simpler calculation of an integrated regulated three-point function, see section~\ref{sec:squashed-TOO}. For now, we will deduce the result of the integrals in \eqref{eq:formF3} from holography, focusing in particular on the specific Hopf squashing \eqref{eq:squashmet0}. Since pure Einstein gravity has $t_4=0$, we import the Einsteinian cubic gravity result \cite{Bueno:2018xqc}:
\begin{equation}\label{eq:F3eps}
    F^{(3)}(\epsilon=0)=\frac{\pi^4}{630} \,c_T t_4\,.
\end{equation}
At cubic order it is important to distinguish $\epsilon = -\alpha/(1+\alpha)$, the parameter such that $g_{\mu\nu}=\bar{g}_{\mu\nu}+\epsilon h_{\mu\nu}$, from the squashing parameter $\alpha$ which appears linearly in $g^{\mu\nu}= \bar{g}^{\mu\nu}+\alpha h^{\mu\nu}$. In particular the third derivative with respect to $\alpha$ is:
\begin{equation}\label{eq:F3alpha}
    F^{(3)}(\alpha=0) = \pi^4c_T\left(2 - \frac{t_4}{630}\right) \implies \#_1 = 2\pi^4 \, , \quad \#_2 = -\frac{\pi^4}{630}\,,
\end{equation}
having figured out the result of the integrated tensor structures by comparing against \eqref{eq:formF3}.

The main point of this section is that we argued from a purely CFT point of view that the result takes the form \eqref{eq:formF3}, thereby proving the conjecture of \cite{Bueno:2018yzo} that \eqref{eq:F3eps}, or equivalently \eqref{eq:F3alpha} holds for any CFT (with a Lagrangian formulation) on the Hopf-squashed $S^3$. 

It was crucial that $\langle TTT\rangle_{\rm OP}$ denotes the correlator defined by three functional derivatives of $\log \CZ$. It differs from the correlator defined by three separate stress-tensor insertions by semilocal contact terms involving $\langle T\,\delta T/\delta g\rangle$ and $\langle TT\rangle$. These terms are already included in the third metric derivative above and must not be added separately. Being schematic about index contractions, they are related as follows, see also (5.48) of \cite{Bzowski:2013sza}:
    \begin{equation}\begin{split}\label{eq:TTT2}
        \langle TTT \rangle_{\scriptscriptstyle{\text{OP}}} = \left(\frac{-2}{\sqrt{g}}\right)^3 \frac{\delta^3}{\delta g^3} \log\CZ &=\left(\frac{-2}{\sqrt{g}}\frac{\delta}{\delta g}\right)^3 \log\CZ+3\langle TT\rangle\\
        &= \langle TTT\rangle - 6 \langle \delta T T\rangle+3\langle TT\rangle\, .
        \end{split}
    \end{equation}
The terms on the right hand side are precisely those written down in expansion (2.10) of \cite{Bobev:2017asb}, where the stress tensor correlators were indeed defined by separate insertions of $T^{\mu\nu}$. 


\subsection{Squashing correction to the two-point function}\label{sec:squashed-TOO}
A useful application of conformal perturbation theory is to derive the linear response to squashing of a scalar two-point function on $S^3$. To leading order, the metric deformation inserts an extra stress tensor in the scalar correlator, so that the correction follows from the integrated three-point function:
\begin{equation}
\delta \langle\CO\CO\rangle = \frac{1}{2}  \int_{S^3} \delta g_{\mu\nu}\, \langle T^{\mu\nu}\CO \CO\rangle\,.
\label{eq:squashed-TOO-response}
\end{equation}
The three-point function above must be understood as a distribution --- this is crucial in order to satisfy the Ward identities \cite{Osborn:1993cr}. The explicit calculations below therefore also serve to illustrate the importance of contact terms. As an example, the correlator at separated points is traceless, incorrectly suggesting a Weyl deformation gives no response. In fact, $\int \langle T \CO\CO\rangle $ is non-zero precisely because of the contact term needed to satisfy the Ward identity.\footnote{Note that $\int \langle T \CO\CO\rangle $ is also the vacuum energy induced at second order in a scalar deformation. RG flows can indeed induce a cosmological constant, as illustrated by the free field examples in appendix~A of \cite{Klebanov:2011gs}.}

In what follows we again focus on the Hopf squashing of $S^3$, restricting moreover to two $\CO$ insertions lying on a common Hopf fiber. The same response admits an independent geometric prediction, derived via the ambient-space construction of \cite{Parisini:2023nbd}. We verify their prediction against a direct integration of $\langle T^{\mu\nu}\CO\CO\rangle$. The required distributional extension of this correlator, together with the check of the Ward identities, is given in appendix~\ref{app:TOOreg}.

\paragraph{Setup}
Consider $S^3$ squashed along its Hopf fiber:
\begin{equation}
    \rmd s^2=\rmd\theta^2+\sin^2\theta\,\rmd\phi^2+\frac{1}{1+\alpha}(\rmd\psi+\cos\theta\,\rmd\phi)^2\, .
    \label{eq:too-berger}
\end{equation}
At $\alpha=0$ this reduces to the round $S^3$ with radius $2$. We are after the response of the scalar two-point function to first order in $\alpha$. To keep things tractable, we place the scalar insertions at equal $\theta=0$, where the non-degenerate fiber coordinate is $\chi$:
\begin{equation}
    \theta_1=\theta_2=0\,,\qquad \chi=\frac{\psi_2+\phi_2}{2}\,,\qquad 0<\chi<\pi\, .
    \label{eq:too-locus}
\end{equation}
In stereographic coordinates on $\R^3$ the round sphere metric is given by
\begin{equation}
    \bar g_{ij}=\Omega^2\delta_{ij},\qquad \Omega(z)=\frac{4}{1+z^2}\,,
    \label{eq:too-stereo}
\end{equation}
and we are putting the first insertion at the origin and the second at
\begin{equation}
   z=(0,0,q)\,,\qquad q=\tan\frac{\chi}{2}\,.
    \label{eq:too-stereo2}
\end{equation}
The normalized flat frame propagator is
\begin{equation}
    G_0(z)\equiv\langle\CO(0)\CO(z)\rangle_{\R^3} = |z|^{-2\Delta}=q^{-2\Delta}\, .
    \label{eq:too-G0}
\end{equation}
Expanding \eqref{eq:too-berger} as $g_{\mu\nu}=\bar g_{\mu\nu}+\alpha h_{\mu\nu}+O(\alpha^2)$ gives $h=-\zeta^2$, where $\zeta=\rmd\psi+\cos\theta\,\rmd\phi$.  Correspondingly, the flat-frame source takes the following simple form:
\begin{equation}
    k_{ij}=\Omega^{-2}h_{ij}=-\Omega^2K_iK_j , 
    \label{eq:too-kij}
\end{equation}
where the Killing vector $K=\partial_\psi$, see also \eqref{eq:Adpsi}, becomes
\begin{equation}
    K=\frac{z_2+z_1z_3}{2}\,\partial_{z_1} +\frac{-z_1+z_2z_3}{2}\,\partial_{z_2} +\frac{1+z_3^2-z_1^2-z_2^2}{4}\,\partial_{z_3}.
    \label{eq:too-K}
\end{equation}
It satisfies the following properties:
\begin{equation}
    K^2=\frac{(1+z^2)^2}{16},\qquad k^i{}_i=-1 .
    \label{eq:too-trace}
\end{equation}
The relative change in the propagator at order $\alpha$ is then the following integrated correlator:
\begin{equation}\label{eq:too-change}
   \frac{\delta G}{ G}= \frac{\delta\langle\CO(0)\CO(z)\rangle_{S^3}}{\langle\CO(0)\CO(z)\rangle_{S^3}} = \frac{\alpha}{2} \,q^{2\Delta}\int \rmd^3 w\, \, k^{ij}\,\langle T_{ij}(w) \CO(0)\CO(z) \rangle_{\mathbb{R}^3}\,.
\end{equation}
Note that the $w$-independent Weyl factors have canceled out in the relative correction. 

\paragraph{The regulated $\langle T\CO\CO\rangle$ distribution}
Following \cite{Osborn:1993cr}, we must be careful in regulating the correlator at coincident points in a way that respects the Ward identities. Let
\begin{equation}
    r=|w|\,,\qquad s=|w-z|\,,\qquad q=|z|\, ,\qquad
    D_{ij}=\partial_i\partial_j-\frac13\delta_{ij}\partial^2 .
    \label{eq:too-Dij}
\end{equation}
The operator $D_{ij}$ is the flat-space traceless second-derivative operator; below it will always act on the coordinate $w$.  The Osborn-Petkou differential-regularization formula \cite{Osborn:1993cr} yields 
\begin{align}
   q^{2\Delta}\,\langle T_{ij}(w) \CO(0)\CO(z) \rangle_{\mathbb{R}^3} ={}&\frac{3\Delta q}{8\pi}D_{ij}\!\left(\frac1{rs}\right) -\frac{\Delta q}{2\pi} \left(    \frac1sD_{ij}\!\left(\frac1r\right)     +\frac1rD_{ij}\!\left(\frac1s\right)
    \right)\nonumber\\
    &-\frac{\Delta}{3}\delta_{ij}  \left(\delta^{(3)}(w)+\delta^{(3)}(w-z)\right)\,.
    \label{eq:too-kernel}
\end{align}
The last line is the trace contact term.\footnote{In the functional-derivative convention of \cite{Osborn:1993cr}, the trace contact coefficient is $(3-\Delta)/3$.  Converting to the present convention for scalar insertions shifts this coefficient by $-1$, giving $-\Delta/3$.} The differential expression above agrees with the familiar conformal expression at separated points and is its unique distributional extension satisfying the Ward identities \cite{Osborn:1993cr}. See appendix~\ref{app:TOOreg} for more details.

The distributional prescription is
\begin{align}
    \left\langle D_{ij}\!\left(\frac1r\right)\!,\,f\right\rangle   &=  \lim_{\varepsilon \to 0}
   \; \int_{r>\varepsilon}\rmd^3w\;    \frac{3n_in_j-\delta_{ij}}{r^3}\,f(w)\,,  \qquad n_i=\frac{w_i}{r}\,,
    \label{eq:too-fp-one}\\
    \left\langle D_{ij}\!\left(\frac1{rs}\right)\!,\,f\right\rangle  &= \int \rmd^3w\,\frac1{rs}D_{ij}f(w)\,.
    \label{eq:too-fp-prod}
\end{align}
Since the angular average of $3n_in_j-\delta_{ij}$ vanishes, the first finite part is indeed well defined with a spherical excision. 

\paragraph{Reduction to scalar integrals}
The non-contact terms in \eqref{eq:too-kernel} reduce to three pairings,
\begin{align}
    \mathcal I_0&=   \left\langle D_{ij}\!\left(\frac1{rs}\right)\!,\,k_{ij}\right\rangle,
    &
    \mathcal I_1&=  \left\langle D_{ij}\!\left(\frac1r\right)\!,\,\frac{k_{ij}}s\right\rangle,
    &
    \mathcal I_2&=  \left\langle D_{ij}\!\left(\frac1s\right)\!,\,\frac{k_{ij}}r\right\rangle.
    \label{eq:too-Idef}
\end{align}
The first one reduces to an ordinary integral by \eqref{eq:too-fp-prod}. Parametrizing $w=(\varrho\cos\varphi,\varrho\sin\varphi,t)$
\begin{equation}
    D_{ij}k_{ij}   = \frac{8(\varrho^2-2t^2)}{(1+\varrho^2+t^2)^2},     \qquad     \mathcal I_0=
    \int\frac{D_{ij}k_{ij}}{rs}\,\rmd^3 w .
    \label{eq:too-Dk}
\end{equation}
The non-contact and contact pieces of the response \eqref{eq:too-change} are therefore
\begin{equation}
    \left.\frac{\delta G}{ G}\right|_{\rm non-ct}   =   \frac{\alpha\Delta q}{16\pi}
    \left(3\mathcal I_0-4\mathcal I_1-4\mathcal I_2\right)\,,  \qquad  \left.\frac{\delta G}{G}\right|_{\rm ct}   =   \frac{\alpha\Delta}{3}\,.
    \label{eq:too-split}
\end{equation}
The second equality follows directly from $k^i{}_i=-1$.

\paragraph{Evaluation}
For $\mathcal I_0$ and $\mathcal I_1$ use polar coordinates around the insertion at
the origin,
\begin{equation}
    t=Ru\,,\qquad  \varrho=R\sqrt{1-u^2}\,,\qquad  s=\sqrt{R^2+q^2-2qRu}\,.
    \label{eq:too-polar-zero}
\end{equation}
The angular integral needed in both terms is
\begin{equation}
    \int_{-1}^{1} \frac{1-3u^2}{\sqrt{R^2+q^2-2qRu}}\,\rmd u   =
    \begin{cases}
    -\dfrac{4R^2}{5q^3}\,,&0<R<q\,,\\[2mm] 
    -\dfrac{4q^2}{5R^3}\,,&R>q\,.
    \end{cases}
    \label{eq:too-angular}
\end{equation}
This leads to the following intermediate results
\begin{equation}\label{eq:too-I0}
    \mathcal I_0   =  \frac{16\pi}{5q^3} \left (2q^5\arctan q-\pi q^5+2q^4-4q^2+4\log(1+q^2)
    \right)\,,\qquad\mathcal I_1  =-\frac{4\pi}{3q}\,.
\end{equation}
For $\mathcal I_2$ it is simplest to use polar coordinates around the second insertion,
\begin{equation}
    t=q+Ru\,,\qquad \varrho=R\sqrt{1-u^2}\,,\qquad     r=\sqrt{R^2+q^2+2qRu}\,.
    \label{eq:too-polar-q}
\end{equation}
The finite-part prescription gives
\begin{equation}
    \mathcal I_2   =  -2\pi\lim_{\varepsilon\to 0}\int_\varepsilon^\infty\frac{\rmd R}{R}     \int_{-1}^{1}
    \frac{\mathcal P(R,u;q)}{r(1+r^2)^2}\,\rmd u,
    \label{eq:too-I2-integral}
\end{equation}
where
\begin{equation}
    \mathcal P(R,u;q)=3\Big(u\big(1+R^2+q^2\big)+2qR\Big)^2 - \Big(1+R^2+q^2+2qRu\Big)^2.
    \label{eq:too-P}
\end{equation}
The angular average removes the apparent singularity at $R=0$. Eventually one finds
\begin{equation}
    \mathcal I_2= \frac{4\pi}{15q^3} \left(48q^5\arctan q-24\pi q^5+48q^4-41q^2  +36\log(1+q^2)
    \right)\,.
    \label{eq:too-I2}
\end{equation}
Finally, substituting \eqref{eq:too-I0} and \eqref{eq:too-I2} into \eqref{eq:too-split} the logarithms cancel to give
\begin{equation}
    \frac{\delta G}{G}  = -\alpha\Delta\left( 2q^2-1+(2\arctan q-\pi)q^3  \right).
    \label{eq:too-answer-q}
\end{equation}
This agrees with the prediction of the ambient space formalism, see (6.40) in \cite{Parisini:2023nbd}.\footnote{In their notation $2\arctan q=\arccos v$.} To leading order the response of the two-point function is only sensitive to the change in geodesic length.

As a short distance check, consider $q\to 0$. Then \eqref{eq:too-answer-q} becomes $\delta G/G \to\alpha\Delta$.  This is the expected result for two nearly coincident insertions separated along the Hopf fiber, since the fiber line element changes by $1-\alpha/2$ at first order as can be seen from \eqref{eq:too-berger}.  

\section{Thermal EFT and CFTs on \texorpdfstring{$S^1 \times_{\text f} \mathcal{M}_2$}{}}
\label{sec:ThEFT}

We now consider CFTs on three-manifolds admitting a $\UU(1)$ isometry with compact orbits of length $\beta$ --- in other words, a circle bundle of the form $S^1 \times_{\text f} \mathcal{M}_2$. We focus on the regime where the circle becomes small. In this limit, the partition function is organized by a thermal effective theory on the two-dimensional base $\mathcal{M}_2$   \cite{Bhattacharyya:2007vs, Banerjee:2012iz, Benjamin:2023qsc}. As we will review, its action contains Weyl-invariant local geometric data on $\mathcal{M}_2$, built from the spatial metric, Kaluza--Klein (KK) gauge field, and local temperature determined by $\beta^{-1}$. Variations of this action determine the equilibrium stress tensor and currents. The existence of such a local hydrodynamic expansion is natural from the point of view of dimensional reduction, provided there are no protected gapless sectors.

We will discuss several concrete CFTs: the conformally coupled scalar in section~\ref{sec:free-scalar}, the massless fermion in section~\ref{sec:free-fermion}, the critical $\rO(N)$ model at strong coupling in section~\ref{sec:criticalON}, and finally, holographic CFTs in section~\ref{sec:holo}. For each of these theories, we use two complementary classes of backgrounds to determine the Wilson coefficients of the effective action. 

The first is $S^1_\beta\times_\Omega S^2$, with angular fugacity $\Omega$ on $S^2$. For holographic CFTs this background is associated with the AdS-Kerr geometry, while for free conformally coupled scalars and fermions its small-$\beta$ free-energy expansion can be obtained analytically.

The second class consists of squashed Lens spaces $S^3/\mathbb{Z}_p$ in the limit where the Hopf fiber becomes small. For holographic CFTs, this regime lies above the Hawking--Page transition and is governed by the AdS-Taub-Bolt saddle rather than AdS-Taub-NUT \cite{Chamblin:1998pz}. For free fields, the corresponding large-squashing expansions can be compared to the results of \cite{DeFrancia:2000xm}.

The agreement between the Wilson coefficients extracted from these two classes of backgrounds provides a non-trivial check of the thermal EFT description. Once these coefficients are extracted for a given CFT, they yield the small-$\beta$ expansion of the free energy on general backgrounds of the form $S^1_\beta \times_{\rm f} \mathcal{M}_2$. This will be the main thread through the rest of this section. We begin by briefly reviewing the thermal EFT formalism --- which also fixes our notation --- and by determining the form of the thermal effective action on $S^1_\beta\times_\Omega S^2$ and $S^3/\mathbb{Z}_p$. 

\paragraph{Thermal EFT expansion}
Consider $S^1\times_{\text f} \mathcal{M}_2$ with metric written in KK form:
\begin{equation}\label{eq:KKmet}
    \rmd s^2 = \rme^{2\sigma}(\rmd \tau+ A)^2 + \rmd s^2_2\, .
\end{equation}
The fiber is parametrized by $\tau\in(0,1)$, the dilaton $\sigma$ sets the local temperature, the KK gauge field $A$ is the connection on the circle bundle, and $\rmd s^2_2$ is the metric on the base $\cM_2$.\footnote{We focus on regular Seifert manifolds: principal circle bundles over a smooth base $\mathcal{M}_2$. In more general Seifert fibrations, $\mathcal{M}_2$ is an orbifold, see appendix~\ref{sec:spindle}, and the EFT may require additional defect data \cite{Benjamin:2024kdg}.} The thermal partition function depends on this background through Weyl-invariant geometric data built from the Weyl-rescaled metric $\rmd \hat s^2=\rme^{-2\sigma}\rmd s^2_2$ and the KK field strength $F=\rmd A$. Since the base is two-dimensional, the EFT expansion is relatively simple:
\begin{equation}\label{eq:genEFT}
    \log \cZ_{\text{EFT}} = -\int \rmd^2x \sqrt{\hat{g}}\left(\sum_{n,m} c_{n,m}R^nF^{2m}+ \cL_\partial \right)\,.
\end{equation}
The Wilson coefficients $c_{n,m}$ capture all terms involving non-negative powers of $R, F^2$ without additional derivatives acting on them. We will therefore refer to it as the non-derivative sector. Instead, $\cL_\partial$ collects terms with additional derivatives acting on $R, F$:
\begin{equation}\label{eq:Lder}\begin{split}
    \cL_\partial = \,&d_1 (\nabla F)^2 + d_2 (\nabla R)^2 + d_3F^2(\nabla F)^2+ d_4 R(\nabla F)^2 \\
    &+ d_5 (\nabla\nabla F)^2 + d_6 \nabla R \nabla F^2+\dots\,.
    \end{split}
\end{equation}
Since it is less clear how to enumerate a complete basis for these derivative terms, we just chose convenient labels $d_i$ for the leading terms. Here and below, contractions and covariant derivatives are always constructed from the Weyl-rescaled metric $\hat g$. In particular, the shorthand notation above stands for
\begin{equation}
(\nabla F)^2\equiv
\hat\nabla_\rho F_{\mu\nu}\hat\nabla^\rho F^{\mu\nu}\,,
\quad
(\nabla\nabla F)^2\equiv
\hat\nabla_\rho\hat\nabla_\sigma F_{\mu\nu}
\hat\nabla^\rho\hat\nabla^\sigma F^{\mu\nu}\,\quad \nabla R \nabla F^2 \equiv \hat\nabla_\mu R \hat \nabla^\mu F^2 \,.
\end{equation}


Most previous discussions in the literature focused on the first three coefficients $c_{0,0}$, $c_{1,0}$, and $c_{0,1}$, more conventionally denoted by $-f$, $c_1$, and $c_2$ respectively. The leading coefficient $f$ has a clear interpretation as the thermal free-energy density on $\mathbb{R}^{d-1}$.  Equivalently, exchanging thermal and spatial circles, it is the Casimir-energy density on $S^1_\beta\times\mathbb{R}^{d-2}$ \cite{Benjamin:2023qsc,Allameh:2024qqp}.

The Wilson coefficients in \eqref{eq:genEFT} encode flat-space CFT data in a coarse-grained sense --- for instance, after inverse Laplace transform $f$ controls the asymptotic density of states in the Hilbert space on $S^{d-1}$, while more general thermal one-point functions determine OPE data averaged over heavy states \cite{Benjamin:2023qsc,Allameh:2024qqp}. 

\paragraph{Thermal EFT for \texorpdfstring{$S^1_\beta\times_\Omega S^2$}{}}
Introducing an angular fugacity $\Omega$, the relevant metric is:
\begin{equation}\label{eq:kerr1}
    \rmd s^2 = \beta^2 \rmd \tau^2 + \rmd \theta^2 + \sin^2\theta (\rmd \phi-\beta \Omega \rmd \tau)^2\,.
\end{equation}
Comparing to the KK form \eqref{eq:KKmet}, we read off the Weyl-rescaled metric on $S^2$ \cite{Benjamin:2023qsc}:
\begin{equation}
    \rmd \hat{s}^2 = \frac{\rmd \theta^2}{\beta^2\left(1+\Omega^2 \sin^2\theta\right)} + \frac{\sin^ 2\theta\,\rmd\phi^2}{\beta^2\left(1+\Omega^2 \sin^2\theta\right)^2} \, , \quad \sqrt{\hat g}= \frac{\sin \theta }{\beta ^2 \left(1+\Omega ^2 \sin ^2\theta \right)^{3/2}}\,\,  ,
\end{equation}
with Ricci scalar
\begin{equation}\label{eq:kerrR}
    R = \frac{2+10\Omega^2\cos^2\theta-2\Omega^4\sin^2\theta}{1+\Omega^2\sin^2\theta}\beta^2\, .
\end{equation}
The KK gauge field and associated field strength are
\begin{equation}\label{eq:kerrF}
    A= -\frac{\Omega\sin^2\theta \,\rmd \phi}{\beta\left(1+\Omega^2\sin^2\theta\right)}\, , \quad F^2 =  \frac{8 \beta^2 \Omega^2\cos^2\theta}{1+\Omega^2\sin^2\theta}\, .
\end{equation}
Integrating the first few terms in the thermal EFT over $S^2$ leads to
\begin{equation}\label{eq:EFTexpK}
    \log \CZ = \frac{-4\pi }{\beta^2(1+\Omega^2)}\left(c_{0,0}+\beta^2\left(2\left(1+\Omega^2\right)c_{1,0}+\tfrac{8}{3}\Omega^2c_{0,1}\right)+\dots\right) \, ,
\end{equation}
which can also be found in \cite{Benjamin:2023qsc}. We now want to extend this to the next few orders in $\beta$. For this purpose, as well as for comparing to the Kerr metric in section~\ref{sec:kerr}, it will be convenient to perform the following coordinate redefinition:
\begin{equation}\label{eq:thetachange}
    \sin\theta = \frac{\sin\vartheta}{\sqrt{1+\Omega^2 \cos^2\vartheta}}\, , \quad \sin\vartheta = \frac{\sqrt{1+\Omega^2}\sin\theta}{\sqrt{1+\Omega^2\sin^2\theta}} \,, \qquad \theta,\vartheta \in (0,\pi)\, .
\end{equation}
This leads to a slightly simpler form of the Weyl rescaled metric
\begin{equation}\label{eq:weyl2}
    \rmd \hat{s}^2 = \frac{\rmd \vartheta^2}{\beta ^2 \left(1+\Omega ^2 \cos^2 \vartheta\right)}+ \frac{\sin ^2\vartheta  \left(1+\Omega ^2 \cos ^2\vartheta \right)\rmd \phi^2}{\beta ^2 \left(1+\Omega ^2\right)^2} \, , \quad \sqrt{\hat g} = \frac{\sin\vartheta}{\beta^2(1+\Omega^2)}\, , 
\end{equation}
with following Ricci scalar and KK gauge field and field strength:
\begin{equation}\label{eq:kerr_alt}
    R = 2\beta^2(1+(5-6\sin^2\vartheta)\Omega^2)\, , \quad A= -\frac{\Omega \sin^2\vartheta \,\rmd \phi}{\beta(1+\Omega^2)}  \, , \quad F^2 = 8\beta^2 \Omega^2\cos^2\vartheta\, .
\end{equation}
Moreover, the first few derivative terms are now found to be:
\begin{align}
(\nabla F)^2 &=8 \beta ^4 \Omega ^2 \sin ^2 \vartheta  \left(1+\Omega ^2 \cos^2\vartheta\right)\,,\\
   (\nabla R)^2 &= 144 \beta ^6 \Omega ^4 \sin ^2(2 \vartheta ) \left(1+\Omega ^2 \cos^2\vartheta\right)\,,\\
     (\nabla\nabla F)^2&= 16 \beta ^6 \Omega ^2 \cos ^2\vartheta  \left(1+\Omega ^2 \cos 2 \vartheta \right)^2\,,\\
     \nabla R \nabla F^2 &= 96 \beta ^6 \Omega ^4 \sin ^2(2 \vartheta ) \left(1+\Omega ^2 \cos^2\vartheta\right)\,. 
\end{align}
The coordinate change does not change $\log \cZ$, so we still find the leading contributions \eqref{eq:EFTexpK}. However, the integrals become somewhat easier to perform. In particular, going to order $\beta^2$,
\begin{equation}\begin{split}\label{eq:EFTK2}
    \log \cZ^{(2)} = &-\int \sqrt{\hat g} \left (c_{2,0} R^2 + c_{1,1}R F^2 + c_{0,2} F^4+d_1 (\nabla F)^2\right ) \, \\
    = &-\frac{16 \pi  \beta ^2}{15 \left(1+\Omega ^2\right)} \Big(4 \Omega ^2 \left(12 \Omega ^2 c_{0,2}+\left(13 \Omega ^2+5\right) c_{1,1}+d_1 \left(\Omega ^2+5\right)\right)\\
    &\qquad\qquad\qquad\;\;+3 \left(21 \Omega ^4+10 \Omega ^2+5\right) c_{2,0}\Big)\,.
    \end{split}
\end{equation}
The next contribution, at order $\beta^4$, is given by 
\begin{align}\label{eq:EFTK4}
    \log \cZ^{(4)} &= -\int \sqrt{\hat g} \Big(c_{3,0} R^3 + c_{2,1}R^2 F^2 + c_{1,2} R F^4+c_{0,3}F^6\nonumber\\
    &\qquad\qquad\quad+ d_2 (\nabla R)^2 + d_3F^2(\nabla F)^2 + d_4 R(\nabla F)^2 + d_5 (\nabla\nabla F)^2 +d_6 \nabla R \nabla F^2\Big) \,\nonumber \\
    &=  -\frac{32 \pi  \beta ^4}{105 \left(1+\Omega ^2\right)}\bigg ( 960 \Omega ^6 c_{0,3}+3 \left(499 \Omega ^6+441 \Omega ^4+105 \Omega ^2+35\right) c_{3,0}\\
    &\qquad\qquad\quad\;+48 \left(23 \Omega ^2+7\right) \Omega ^4 c_{1,2}+4 \left(323 \Omega ^4+182 \Omega ^2+35\right) \Omega ^2 c_{2,1}\nonumber\\
    &\qquad\;+2\Omega ^2 \left( (9d_2+d_3+6d_6) \left(24 \Omega ^2+56\right) \Omega ^2+\left(2 d_4+d_5\right) \left(11 \Omega ^4+14 \Omega ^2+35\right)\right)\bigg)\,.\nonumber
\end{align}

\paragraph{Thermal EFT for squashed Lens spaces}
The metric on squashed $S^3/\mathbb{Z}_p$ takes the form
\begin{equation}\label{eq:squashmetbdy}
    \rmd s^2 = \left(\rmd \theta^2 +\sin^2\theta \rmd \phi^2\right) + \frac{1}{1+\alpha}\left(\rmd \psi + \cos \theta \rmd \phi\right)^2,
\end{equation}
where $\theta\in(0,\pi)$, $\phi\in (0,2\pi)$ and $\psi \in (0,4\pi/p)$. The partition function at large squashing $\alpha$ is constrained by the terms allowed in a thermal EFT expansion. Note that the above metric is already in KK form. To bring it in the required form \eqref{eq:KKmet} we just have to rescale $\psi = 4\pi \tau/p$ so that $\tau\in(0,1)$. This yields the following thermal EFT quantities:
\begin{equation}\label{eq:squashmet}
    \rmd \hat{s}^2 = \beta^{-2}(\rmd \theta^2 +\sin^2\theta \,\rmd \phi^2)\,,\quad A = \frac{p}{4\pi}\cos\theta \,\rmd \phi\,,\quad \rme^{2\sigma} = \beta^2 = \frac{16\pi^2}{p^2(1+\alpha)}\,.
\end{equation}
 The non-derivative terms in the EFT expansion are of the form $c_{n,m}R^n F^{2m}$, with
\begin{equation}\label{eq:squashRF}
    R = 2 \beta^2\, , \quad F^2 = \,\frac{p^2\beta^4}{8\pi^2}\,. 
\end{equation}
Since these are constant, all of the derivative terms vanish, in contrast to the $S^1\times_\Omega S^2$ case. Yet another difference is that now $R$ and $F^2$ appear at different orders in the $\beta^2$ expansion. The thermal EFT formalism then predicts
\begin{equation}\label{eq:squashEFT}
    \log \CZ = -4\pi\beta^{-2}\,\sum_{n,m} c_{n,m} \,2^{n-3m}\left(\frac{p}{\pi}\right)^{2m}\beta^{2n+4m} \,.
\end{equation}

Finally, it is useful to distinguish a second way of making the Hopf fiber small, by taking $p\to\infty$ at fixed $\alpha$.\footnote{We thank David Tong for suggesting this limit to us.} In this limit, it is more natural to eliminate $p$ in favor of $\beta$ and $\alpha$ using~\eqref{eq:squashmet}. Although the total KK flux grows with $p$, its local field strength in the Weyl-rescaled geometry remains small.\footnote{The large-background regime provided by CFTs in strong external magnetic fields \cite{Boyack:2023uml,Herzog:2025ddq} is instead more closely related to the opposite limit $\alpha\to-1$ where the Hopf fiber decompactifies.} The local invariants on the Weyl-rescaled base become
\begin{equation}
R=2\beta^2\, \qquad F^2=\frac{2\beta^2}{1+\alpha}\, \qquad \frac{F^2}{R}=\frac{1}{1+\alpha}\,.
\end{equation}
Note that in this limit $R$ and $F^2$ do vanish at the same rate $\sim \beta^2\sim p^{-2}$; their ratio fixed in terms of the squashing $\alpha$. The thermal expansion \eqref{eq:squashEFT} reorganizes accordingly:
\begin{equation}
\log \CZ =-4\pi\sum_{N=0}^{\infty}2^N\beta^{2N-2} \sum_{m=0}^{N} \frac{c_{N-m,m}}{(1+\alpha)^m} \,.
\end{equation}

The leading order extensive result depends only on the free energy density $f=-c_{0,0}$, and is insensitive to the fibration, which lies encoded in the subleading $m>0$ terms.\footnote{Similarly, for $d=4$ CFTs, to leading order at large $p$ one has
$\log \CZ_{S^1_{2\pi/p}\times S^3} \approx
\log \CZ_{S^1_{2\pi}\times S^3/\mathbb{Z}_p}$ \cite{Shaghoulian:2016gol, Allameh:2024qqp}.}



\subsection{Conformally coupled scalar}\label{sec:free-scalar}
As a first explicit example,  we consider the partition function of a conformally coupled scalar. The heat-kernel expansion determines the pure-curvature thermal EFT coefficients at zero fugacity, while the exact spectrum on $S^1\times_\Omega S^2$ as well as on squashed Lens spaces provides independent checks of the high-temperature expansion, including its fugacity dependence.
\subsubsection{Turning off fugacity}
Consider first the partition function $\CZ$ of a conformally coupled scalar on the ultrastatic background $S^1_\beta \times \mathcal{M}_2$. Its high-temperature expansion captures the asymptotic spectrum of the Laplacian $\nabla^2$. In heat-kernel representation, see (2.60) of the review \cite{Bytsenko:1994bc}\footnote{Here we only compute the $\beta$-dependent part of the free energy, as in \cite{Benjamin:2023qsc}. Thus, we do not include the vacuum energy, which is a contribution to $\log \CZ$ linear in $\beta$, namely the $\zeta_R$ term in the LHS of (2.60) in \cite{Bytsenko:1994bc}.}, we have: 
\begin{equation}\label{eq:KtoZ}
    \log \CZ = \frac{ \beta}{2\sqrt{\pi}}\sum^\infty_{k=1}\int_0^\infty \rmd t\,t^{-\frac32}\rme^{-\frac{k^2\beta^2}{4t}}K(t|\nabla^2-R/8)\,, \quad K(t|D)= \tr \rme^{D t}\,.
\end{equation}
The asymptotic spectrum lies encoded in the small-$t$ expansion of the heat kernel. Assuming that $\mathcal{M}_2$ has constant curvature $R$, one finds \cite{McKean:1967xf}:
\begin{equation}\label{eq:mckean}
    K(t\to 0|\nabla^2) = \frac{A}{4\pi t}+ \frac{1}{24\pi}\int R+ \frac{t}{240\pi}\int R^2 +\dots\, .
\end{equation}
The leading growth is known as the Weyl law. The next term is topological and equals $\chi/6$. 
Introducing the conformal coupling results in\footnote{For the first few unintegrated heat-kernel coefficients $k_n(x)$ of $\nabla^2 - V(x)$, see (B.1) and (B.2) of \cite{Bytsenko:1994bc}. This goes back to earlier work of \cite{DeWitt:1964mxt, Gilkey:1975iq}.}
\begin{equation}\label{eq:mckean2}
    K(t\to 0|\nabla^2- R/8) = \frac{A}{4\pi t}+ \frac{1}{96\pi}\int R+\frac{7 t}{7680\pi}\int R^2 +\dots\, .
\end{equation}
We then have two expansions, one for the heat kernel and one for the thermal EFT:
\begin{equation}\label{eq:hkexp}
    K =  t^{-1}\sum k_n t^n \int R^n\, , \quad \log \CZ = -\beta^{-2}\sum c_{n,0}\beta^{2n}\int R^n\,.
\end{equation}
Their coefficients $k_n$ and $c_{n,0}$ are related via \eqref{eq:KtoZ}:
\begin{equation}\label{eq:ktoc}
    4^{1-n} \zeta (3-2 n) \Gamma \left(\tfrac{3}{2}-n\right)\,k_n= -\sqrt{\pi }\,c_{n,0}\,.
\end{equation}
The standard integrated heat-kernel expansion \eqref{eq:mckean2} thus yields:
\begin{equation}\label{eq:scalfreec}
    c_{0,0} =-\frac{\zeta(3)}{2\pi} \, , \quad c_{1,0}=-\frac{\zeta(1)}{96\pi} \, , \quad c_{2,0}= -\frac{7}{184320\pi}\,.
\end{equation}
These are the values also obtained in \cite{Benjamin:2023qsc}. One should note that $\zeta(1)$ is infinite. This is due to the presence of a gapless mode for the scalar, which, when treated properly, will introduce a $\log \beta$ term in the EFT expansion, as we will see in \eqref{eq:zeta1_reg}. For the free fermion in section\,\ref{sec:free-fermion} there will be no such zero-mode.
Our goal is now to also determine the other $c_{n,0}$.

\paragraph{Thermal EFT coefficients from \texorpdfstring{$S^1\times \mathbb{H}^2$}{}}
The Selberg trace formula, see (3.50) in \cite{Bytsenko:1994bc}, gives a useful expression for the heat kernel on compact hyperbolic manifolds $\mathbb{H}^2/\Gamma$. It has a universal `identity' contribution in terms of the Plancherel measure on $\mathbb{H}^2$:
\begin{equation}\label{eq:hkI}
    K_I(t) = \frac{A}{2\pi}\int^\infty_0 \rmd r\, \rme^{-tr^2}r\tanh(\pi r)\, .
\end{equation}
The only thing this term knows about the Fuchsian subgroup $\Gamma$ is the area of the Riemann surface $\mathbb{H}^2/\Gamma$. It determines all perturbative terms in the small-$t$ expansion: \footnote{ The other terms in the Selberg trace formula are a sum over the generators of $\Gamma$, and can be interpreted as a semiclassical sum over orbits. They are non-perturbatively small as $t\to 0$.}
\begin{equation}
   K_I(t\to 0) = A\cdot\left(\frac{1}{4\pi t} +\sum_{n=1}^\infty t^{n-1}(-1)^n 2^{-2 n-1} \left(4^n-2\right) \pi ^{-2 n-\frac{3}{2}} \zeta (2 n) \Gamma \left(n+\tfrac{1}{2}\right)\right)\,.
\end{equation}
Comparing to \eqref{eq:hkexp}, and inserting $R=-2$, we read off the heat-kernel coefficients
\begin{equation}
    k_n = \frac{4^n-2}{2^{3 n+1} \pi ^{2 n+\frac{3}{2}}}  \zeta (2 n) \Gamma \left(n+\tfrac{1}{2}\right)\,.
\end{equation}
Applying \eqref{eq:ktoc} we then find the thermal EFT coefficients for the conformally coupled scalar:
\begin{equation}\label{eq:cn0}
    c_{n,0} = \left(\frac{-1}{32}\right)^n \left(4^n-2\right) (2 n-1) \pi ^{-2 n-1} \zeta (3-2 n) \zeta (2 n) \,.
\end{equation}
 
\paragraph{Thermal EFT coefficients from \texorpdfstring{$S^1\times S^2$}{}}
We can compare the values \eqref{eq:cn0} to those found by expanding the partition function on $S^1\times S^2$. For a free bosonic theory with one-particle frequencies $\omega_l$ and degeneracies $d_l$ one has:
 \begin{equation}\begin{split}\label{eq:logZsgen}
     \log \CZ_s(\beta\to 0)= \sum_{l,k}\frac{d_l}{k}\, \rme^{-\beta k \omega_l}&= \sum_{l,k}\frac{d_l}{2\pi\rmi k} \int_{c+\rmi \mathbb{R}} \rmd s\, \Gamma(s)(\beta k \omega_l)^{-s}\\&= \sum \text{Res}\left( \beta^{-s}\Gamma(s) \zeta(s+1) \sum_n d_l \omega^{-s}_l\right) \,.
     \end{split}
 \end{equation}
 The conformally coupled scalar on $S^2$ has $d_l=2l+1$ and $\omega_l= l+1/2$. Its spectral $\zeta$-function equals $\sum_l d_l \omega^{-s}_l =\left(2^s-2\right) \zeta (s-1)$. Non-zero residues then come from $\Gamma(s)$ evaluated at negative even integers, yielding the same $c_{n,0}$ as before in \eqref{eq:cn0} for $n>1$, as well as from $\zeta(s-1)$ evaluated at $s=2$, which gives the correct leading $\beta^{-2}$ term corresponding to $c_{0,0}$. Finally, the double pole at $s=0$ contributes
 \begin{equation}\label{eq:zeta1_reg}
     c_{1,0}\,\beta^0 = \frac{\log (2 \beta )+12\zeta'(-1)}{96 \pi }\, ,
 \end{equation}
 providing a regularization --- the same one as in \cite{Benjamin:2023qsc} --- of $c_{1,0}= -\zeta(1)/96\pi$ in \eqref{eq:cn0}.

\subsubsection{Turning on fugacity}
Once we turn on a fugacity $\Omega$, the field strength $F$ in the thermal EFT is no longer zero. 
\paragraph{Thermal EFT coefficients from \texorpdfstring{$S^1\times_\Omega S^2$}{}}  The first few $c_{n,m}$ coefficients can be found by comparing the thermal EFT expansion \eqref{eq:EFTexpK} to the expansion of $\log \CZ_s(\beta,\Omega)$, as given for instance in tables 6 and 8 of \cite{Bobev:2023ggk} as well as \eqref{eq:logZscal} below. This determines the first few coefficients $c_{n,m}$ as follows: 
\begin{equation}\label{eq:cnmCS}
    \begin{tabular}{|C|C|C|}
c_{n,m} & 0 & 1  \\\hline
0 & -\frac{\zeta(3)}{2\pi}  & -\frac{\zeta(1)}{128\pi}    \\
1 & -\frac{\zeta(1)}{96\pi} &    \\
2 & -\frac{7}{184320\pi}  &     \\
\end{tabular}
\end{equation}
Additional $c_{n,0}$ are given by \eqref{eq:cn0}. It is understood that the formal $\zeta(1)$ contribution should be interpreted as in \eqref{eq:zeta1_reg}, which then also tells us what to make of $c_{0,1}$. Even though \eqref{eq:logZscal} allows an expansion to all orders in powers of $\beta^2$ and $\Omega^2$, we can still only extract the leading few $c_{n,m}$. This is because they end up contributing at the same order as the derivative coefficients $d_i$ in \eqref{eq:genEFT}, see for instance \eqref{eq:EFTK2}. 

\paragraph{Thermal EFT for squashed Lens spaces}
Using the thermal EFT expansion \eqref{eq:squashEFT} for squashed Lens spaces in combination with the leading coefficients \eqref{eq:cnmCS}, we get a prediction for the partition function on the squashed $S^3/\mathbb{Z}_p$ in the large-squashing regime:
\begin{equation}
    \log \CZ_s = \frac{2\zeta(3)}{\beta^2}+ \frac{\zeta(1)\beta^0}{12}+\Big(\frac{7}{11520}+\frac{\zeta(1)p^2}{256\pi^2}\Big)\beta^2 +O(\beta^4)\,.
\end{equation}
This matches with (3.25) and (4.2) of \cite{DeFrancia:2000xm}, once we interpret the $\zeta(1)\beta^0$ term as in \eqref{eq:zeta1_reg} and similarly regulate the $\zeta(1)\beta^2$ term:
\begin{equation}\label{eq:zeta1b}
    \zeta(1)\beta^2 \to \left(\frac{4}{3}+2\log 2 -\log \beta +\gamma_{E}\right)\beta^2\,.
\end{equation}
For a numerical large squashing expansion, see \cite{Bobev:2017asb}. 

\subsection{Massless fermion}\label{sec:free-fermion}
As a second explicit example, we consider the free massless fermion. Since we follow the same approach as for the conformally coupled scalar, we will mostly focus on the final results. 
\subsubsection{Turning off fugacity} Once more, we will determine the thermal EFT coefficients $c_{n,0}$ in two different ways: from the exact spectrum on $S^1\times S^2$, and via the heat-kernel expansion on $S^1\times \mathbb{H}^2$.
\paragraph{Thermal EFT coefficients from \texorpdfstring{$S^1\times S^2$}{}}
Compared to the bosonic case \eqref{eq:logZsgen} there are two changes: the sum over $k$ comes with a $(-1)^{k+1}$, yielding $\zeta(1+s)(1-2^{-s})$, while the spectral $\zeta$ is now $\sum_l d_l \omega^{-s}_l= 2\zeta(s-1)$, leaving us with
\begin{equation}\begin{split}
    \log \CZ_f(\beta\to 0)= \sum \text{Res}\left( 2\beta^{-s}\Gamma(s)\zeta(s-1)\zeta(1+s)(1-2^{-s}) \right)\,.
    \end{split}
\end{equation}
Comparing this to \eqref{eq:hkexp} at $R=2$, we read off the following thermal EFT coefficients:
\begin{equation}\label{eq:cnfer}
   c_{n,0} = \left(\frac{-1}{8}\right)^n \left(4^{n-1}-1\right) (2 n-1) \pi ^{-2 n-1} \zeta (3-2 n) \zeta (2 n)\,.
\end{equation}
In particular $c_{0,0}= -3\zeta(3)/8\pi$. By taking the limit $n\to 1$, we get $c_{1,0}=\log(2)/48\pi$. 

\paragraph{Thermal EFT coefficients from \texorpdfstring{$S^1\times \mathbb{H}^2$}{}}
Alternatively, the Plancherel measure for a fermion on $\mathbb{H}^2$ determines the following identity contribution to the heat-kernel
\begin{equation}
    K_I(t) = \frac{A}{2\pi}\int^\infty_0 \rmd r \,\rme^{-t r^2} \, r \coth\pi r\, ,
\end{equation}
which lets us extract heat-kernel coefficients by comparing against \eqref{eq:hkexp}:
\begin{equation}\label{eq:knfer}
    k_n = \frac{(-1)^{n+1} \zeta(1-2n)}{ 2^{n+1}\pi \Gamma(n)}\,.
\end{equation}
The fermionic analog of \eqref{eq:ktoc} tells us that
\begin{equation}
  k_n \left(1-4^{1-n}\right)\Gamma\left(\tfrac32-n\right) \zeta(3-2n) = \sqrt{\pi}c_{n,0} \,.
\end{equation}
Using this in combination with \eqref{eq:knfer} we indeed recover the same coefficients \eqref{eq:cnfer}. 

\subsubsection{Turning on fugacity}
Once we include an angular fugacity $\Omega$ on $S^2$, we get access to the coefficient $c_{0,1}$ multiplying $F^2$. In combination with the previously obtained $c_{n,0}$, this then allows us to make a prediction for the thermal partition function of a free massless fermion on $S^3/\mathbb{Z}_p$ at large squashing.
\paragraph{Thermal EFT coefficients from \texorpdfstring{$S^1\times_\Omega S^2$}{}}
For the free massless fermion, see for instance table 7 and 9 of \cite{Bobev:2023ggk}, we find the following leading EFT coefficients:
\begin{equation}\label{eq:cnmFF}
    \begin{tabular}{|C|C|C|}
c_{n,m} & 0 & 1  \\\hline
0 & -\frac{3\zeta(3)}{8\pi}  & -\frac{\log(2)}{128\pi}  \\
1 & \frac{\log(2)}{48\pi} &    \\
2 & -\frac{1}{7680\pi}  &       \\
\end{tabular}
\end{equation}
For a Dirac fermion, these get multiplied by a factor $2$. The three leading coefficients for a Dirac fermion match those in (4.5) of \cite{Benjamin:2023qsc}.\footnote{In (4.5) of \cite{Benjamin:2023qsc} it appears $c_{0,1}$ has a factor $96$ instead of $64$. This seems to be a typo: tracing back the steps starting from their (3.16) and (4.4), one does find $64$.}

\paragraph{Thermal EFT for squashed Lens spaces}
Plugging the coefficients \eqref{eq:cnmFF} back into the thermal EFT \eqref{eq:squashEFT} for the squashed $S^3/\mathbb{Z}_p$, we get the following large-squashing prediction:
\begin{equation}
    \log \CZ_f = \frac{3\zeta(3)}{2\beta^2}-\frac{\log 2}{6} + \left(\frac{1}{480}+ \frac{p^2\log 2}{256\pi^2}\right)\beta^2+O(\beta^4)\,.
\end{equation}
This matches (4.4) of \cite{DeFrancia:2000xm} (after including the overall factor 2 for a Dirac fermion).\footnote{Strictly speaking, having antiperiodic fermions around the Hopf fiber requires even $p$ \cite{DeFrancia:2000xm}. On the simply connected squashed $S^3$ the spin structure is unique, so the sign around the shrinking fiber is not an independent thermal choice.  More generally $S^3/\mathbb{Z}_p$ has one spin structure for odd $p$ and two for even $p$.  Only for even $p$ can one independently choose the antiperiodic spin structure appropriate to an ordinary thermal trace.}

\subsubsection{Relating fermion and scalar on \texorpdfstring{$S^1\times_\Omega S^2$}{}}
From the explicit series expansions for scalar and fermionic partition functions 
\begin{align}\label{eq:logZscal}
    \log \CZ_s(\beta,\Omega)&= \sum_{k=1}^\infty \frac{1}{k}\frac{\cosh  \beta k/2}{\cosh  \beta k - \cos \beta \Omega k}\, ,\\
    \log \CZ_f(\beta,\Omega)&= \sum_{k=1}^\infty \frac{(-1)^{k+1}}{k}\frac{\cos \beta\Omega k/2}{\cosh\beta k-\cos  \beta \Omega k}\, ,\label{eq:logZfer}
\end{align}
one finds the curious relation
\begin{equation}\label{eq:scalvsfer}
    \log \CZ_f(\beta,\Omega) = \log \CZ_s(2\beta\Omega\rmi, 1/\Omega)-\log\CZ_s(\beta\Omega\rmi,1/\Omega)\,.
\end{equation}
For a free periodic fermion, i.e. with $(-1)^F$ insertion, the relation is even more direct: 
\begin{equation}\label{eq:ferZper}
    \log \CZ_{f,\text{per}}(\beta,\Omega) = \log \CZ_{s}(\beta \Omega \rmi, 1/\Omega)\,.
\end{equation}
The relation \eqref{eq:scalvsfer} explains some of the observations made in \cite{Bobev:2023ggk}.\footnote{They expanded $\log\CZ(\beta,\Omega)=\beta^{-2}\sum d_n\tilde{c}_{n,m}\beta^{2n}\Omega^{2m}$, and noticed that $\tilde{c}^f_{n,m} = (-1)^{n+1}\tilde{c}^s_{n,n-m}$ as well as $d^f_n= (4^{n-1}-1)d^s_n$. Both these facts follow directly from the relation \eqref{eq:scalvsfer}.}

These identities have a simple representation-theoretic origin. The conformal scalar and Majorana fermion furnish the singleton representations
\begin{equation}
\mathrm{Rac}=D(\tfrac12,0),
\qquad
\mathrm{Di}=D(1,\tfrac12)
\end{equation}
of $\widetilde{\SO}(2,3)$ \cite{Dirac:1963ta,Dolan:2005wy}. Under its compact subgroup $\SO(2)\times\SO(3)$, they decompose as
\begin{align}\label{eq}
\mathrm{Rac}
&=\bigoplus_{\ell=0}^{\infty}
\Bigl[\tfrac12+\ell\Bigr]\boxtimes[\ell],,
&
\mathrm{Di}
&=\bigoplus_{\ell=0}^{\infty}
[1+\ell]\boxtimes\Bigl[\ell+\tfrac12\Bigr],.
\end{align}
Their characters $\chi_R = \Tr_R q^H y^J$ are the corresponding one-particle traces,
\begin{equation}
\chi_{\mathrm{Rac}}(q,y) =\frac{q^{1/2}(1+q)}
{(1-qy)(1-qy^{-1})}\,,\quad
\chi_{\mathrm{Di}}(q,y) =
\frac{q(y^{1/2}+y^{-1/2})}
{(1-qy)(1-qy^{-1})}.
\label{eq:RacDi-chars}
\end{equation}
With the identification $q=e^{-\beta}, y=e^{\rmi\beta\Omega}$, these are precisely the one-particle factors in \eqref{eq:logZscal} and \eqref{eq:logZfer}. For instance, the periodic fermion trace is
\begin{equation}
\log \CZ_{f,\mathrm{per}} = -\sum_{k=1}^{\infty}\frac1k \chi_{\mathrm{Di}}(q^k,y^k)\,.
\end{equation}
Moreover, the transformation $(\beta,\Omega) \to (\rmi\beta\Omega,\Omega^{-1})$ is equivalent to $(q,y)\to (y^{-1},q)$. The relation \eqref{eq:ferZper} is then implied by 
\begin{equation}
    \chi_{\mathrm{Rac}}(y^{-1},q) = -\chi_{\mathrm{Di}}(q,y)\,.
\label{eq:RacDi-Weyl}
\end{equation}
This fugacity transformation has a clear group-theoretic interpretation. Writing $q=e^{-u}$ and $y=e^v$, it acts as $(u,v)\to (v,-u)$: an order-four Coxeter element of the $B_2$ Weyl group. Thus the relation is a Coxeter-type analytic continuation between the Rac and Di characters, rather than a physical equivalence between the scalar and fermion theories.

\subsection{Critical \texorpdfstring{$\mathrm{O}(N)$}{} model}\label{sec:criticalON}
The large-$N$ critical $\rO(N)$ vector model provides an interacting yet analytically tractable test case for the thermal EFT. For small integer $N$, this model describes some of the best-studied universality classes, with $N=1,2,3$ corresponding respectively to the Ising model, the XY model, and the Heisenberg ferromagnet, while $N\to0$ describes self-avoiding polymers. Universal finite-temperature quantities, such as the free-energy density $f$, can be studied using the $\epsilon$ expansion, Monte Carlo simulations, and independently via a large-$N$ treatment~\cite{Sachdev:1993pr, Chubukov:1993aau}. Remarkably, the latter often gives quantitatively useful results even for relatively small $N$.

At large $N$, the model has also been a productive setting to study the relation between finite-temperature field theory and gravity \cite{Klebanov:2002ja,Hartnoll:2005yc}.
To leading order in $1/N$, a homogeneous Hubbard--Stratonovich saddle reduces the problem to evaluating a Gaussian determinant for the $N$ fundamental scalars, with a self-consistently determined mass $m_\pi$. Beyond leading order, fluctuations of the Hubbard--Stratonovich field generate $1/N$ corrections \cite{Diatlyk:2023msc}. The model has recently been studied in the context of the thermal EFT \cite{David:2024pir,Mauro:2026zus}. Our goal is to independently verify their leading Wilson coefficients, and to derive the next ones.

To do so, we will now extend the large-squashing analysis of Hartnoll and Kumar \cite{Hartnoll:2005yc} to the untwisted, namely $\mathbb Z_p$-invariant, sector of squashed Lens spaces $S^3/\mathbb{Z}_p$, retaining the local large-squashing expansion through $O(\alpha^{-2})$. We work at fixed $p$ and measure lengths in units of the round-sphere radius $a$. It will be useful to introduce the following notation:
\begin{equation}
 A\equiv p^2\alpha\,, \qquad \eta\equiv\frac{a^2m_\pi^2}{A}\,, \qquad
Q_j\equiv j^2+\eta\,,
\end{equation}
with $m_\pi^2$ the saddle-induced mass of the fundamental scalars, and $j\in\mathbb{Z}$ labeling momentum $k=pj$ along the Hopf fiber. With this notation in place, the spectral $\zeta$-function becomes:
\begin{equation}
\zeta_p(s) = (a\mu)^{2s} \sum_{j\in\mathbb Z}\sum_{n=0}^{\infty} \frac{p|j|+1+2n} {\left((p|j|+1+2n)^2+A Q_j-1\right)^s}\, ,
\label{eq:lens-zeta-full}
\end{equation}
where $\mu$ is an arbitrary scale. For fixed $j$, let
\begin{equation}
b_j\equiv p|j|+1\,, \qquad f_j(x)\equiv \frac{x}{\left(x^2+A Q_j-1\right)^s}\,.
\end{equation}
We want a large-$A$ expansion of the spectral $\zeta$. The spacing-two Abel--Plana formula says:\footnote{The relevant branch points lie at $n_\pm=-\frac{b_j}{2} \pm\frac{i}{2}\sqrt{A Q_j-1}$ and hence remain in the left half-plane.  The Abel--Plana representation therefore contains no additional contour contribution.}
\begin{equation}
\sum_{n=0}^{\infty}f_j(b_j+2n) = \frac12\int_{b_j}^{\infty}\hspace{-0.1cm}\rmd x \,f_j(x) +\frac12f_j(b_j) + \frac{\rmi}{2}\int_0^\infty \hspace{-0.1cm}\rmd t\,\frac{f_j(b_j+\rmi t)-f_j(b_j-it)}{\rme^{\pi t}-1} \,.
\label{eq:lens-AP-exact}
\end{equation}
The exact remainder in \eqref{eq:lens-AP-exact} admits the Euler--Maclaurin expansion
\begin{equation}
\frac{\rmi}{2}\int_0^\infty \hspace{-0.1cm}\rmd t\,\frac{f_j(b_j+\rmi t)-f_j(b_j-it)}{\rme^{\pi t}-1}= -\frac16f_j'(b_j)  +\frac1{90}f_j'''(b_j)  -\frac1{945}f_j^{(5)}(b_j)  +\dots,
\label{eq:lens-AP-remainder}
\end{equation}
where the dotted terms are of the next order in the large-$\alpha$ expansion, since $b_j^2/A Q_j=O(\alpha^{-1})$, uniformly in $j$.  This controls the error and justifies summing the Euler--Maclaurin expansion termwise. Keeping terms through $O(\alpha^{-2-s})$, one finds
\begin{align}
\sum_{n=0}^{\infty}f_j(b_j+2n)  ={}&  \frac{A^{1-s}}{4(s-1)}Q_j^{1-s}  +  A^{-s} \left( \frac13-\frac{p^2j^2}{4}  \right)Q_j^{-s}  \nonumber\\
&+ sA^{-1-s}  \left( \frac{p^4j^4}{8}  -\frac{p^2j^2}{2} +\frac4{15}  \right)Q_j^{-s-1} \label{eq:lens-fixed-j-full}\\
&+  s(s+1)A^{-2-s} \left( -\frac{p^6j^6}{24} +\frac{p^4j^4}{3} -\frac{2p^2j^2}{3} +\frac{64}{315} \right)Q_j^{-s-2}+\dots\,. \nonumber
\end{align}
In particular, all non-analytic dependence on $|j|$ cancels in the coefficients displayed above.

Now we can solve the gap equation for $m_\pi^2$, or equivalently $\eta$. At strong coupling, it takes the form $\zeta_p(1)=0$ \cite{Hartnoll:2005yc}. Defining the Epstein--Hurwitz function
\begin{equation}
    E(s;\eta) \equiv \sum_{j\in\mathbb Z} (j^2 + \eta)^{-s} = \sum_{j\in\mathbb Z} Q_j^{-s}\,,
\end{equation}
we find from the leading three terms in \eqref{eq:lens-fixed-j-full} that
\begin{align}
    \frac{\zeta_p (s)}{(a\mu)^{2s}} = &\frac{A^{1-s}}{4(s-1)}E(s-1;\eta) + A^{-s}\left(\left(\frac13+\frac{p^2\eta}{4}\right) E(s;\eta) - \frac{p^2}{4}E(s-1;\eta) \right)\nonumber \\
    &+ sA^{-1-s}\bigg (\left(\frac{4}{15}+\frac{p^2 \eta}{2}+\frac{p^4 \eta^2}{8}\right)E(s+1;\eta) -\left(\frac{p^2}{2} + \frac{p^4 \eta}{4}\right)E(s;\eta)\\&\qquad \qquad\qquad +\frac{p^4}{8}E(s-1;\eta) \bigg)  + \dots\,.\nonumber
    \end{align}
Evaluating this at $s=1$, using that $E(0;\eta)=0$ in $\zeta$-function regularization,
\begin{align}
    \zeta_p (1) = \frac14 E'(0;\eta) + A^{-1}\left(\left(\frac13+\frac{p^2\eta}{4}\right) E(1;\eta)\right)+\dots\,.
    \end{align}
Now we plug in the following identities for $E$
\begin{equation}
  E'(0,\eta) = -2\log(2\sinh(\pi\sqrt{\eta}))\, , \quad  E(1,\eta) = \frac{\pi}{\sqrt{\eta}}\coth(\pi\sqrt{\eta})\,.
\end{equation}
Using these, the mass gap equation to first subleading order in $A^{-1}$ is solved by
\begin{equation}
\eta=\eta_0+\frac{\eta_1}{A}+ \dots\,, \qquad \eta_0=\frac{\log^2 \varphi}{\pi^2}, \qquad \eta_1=p^2\eta_0+\frac43\,,
\end{equation}
where $ \varphi\equiv\frac{1+\sqrt5}{2}$ is the golden ratio.
Equivalently, remembering $\beta^2 = 16\pi^2/p^2(1+\alpha)$
\begin{equation}
a^2m_\pi^2  =  \frac{16\log^2 \varphi}{\beta^2} +\frac43 + O(\beta^{2}).
\label{eq:lens-mass-full}
\end{equation}
Imposing this saddle, we find the following large-squashing expansion for the partition function $\log \CZ =N\zeta_p'(0)/2$ of the critical $\rO(N)$ model on $S^3/\mathbb{Z}_p$ to leading order in $N$:
\begin{equation}\begin{split}
\frac{1}{N}\log \CZ  ={}&  \frac{8\zeta(3)}{5\beta^2}  +  \frac{\beta^2}{16 } \left( \frac{\sqrt5\, p^2 \log\varphi}{12\pi^2}  +\frac{\sqrt5}{45 \log\varphi}  \right)  \nonumber\\
&+  \frac{\beta^4}{32 } \left( \frac{p^2}{45\pi^2}  \left(  1-\frac{\sqrt5}{4\log\varphi}  \right)  +  \frac{16}{2835}  \left(  \frac{1}{\log^2\varphi}  +\frac{\sqrt5}{4\log^3\varphi}  \right)  \right)  +O(\beta^{6})\,.
\label{eq:lens-logZ-full}
\end{split}\end{equation}
Note the absence of a $\beta^0$ term. Comparing to the thermal EFT expansion on \eqref{eq:squashEFT}, we find the following leading coefficients:
\begin{equation}
f\equiv -c_{0,0}=\frac{2N\zeta(3)}{5\pi}\,, \qquad c_1\equiv c_{1,0}=0\,, 
\label{eq:lens-coefficients-lower}
\end{equation}
Note that the above free energy density has a relative factor of $4/5$ compared to the conformally coupled scalar \eqref{eq:scalfreec}. This is the familiar ratio between the strong and weak coupling results \cite{Hartnoll:2005yc}, see also \cite{Romatschke:2019ybu} for a discussion of its universality in large-$N$ $d=3$ bosonic CFTs. Next, the $p^2$ and $p^{0}$ terms at order $\beta^2$ separate $c_{0,1}$ and $c_{2,0}$:
\begin{equation}
 c_2\equiv c_{0,1}=-\frac{N\sqrt5\log\varphi}{96\pi}\,,\qquad c_{2,0}=-\frac{N}{2304\pi\sqrt5\log\varphi}\,.
\label{eq:lens-c2}
\end{equation}
At order $\beta^4$, the new geometric invariants in play are $R F^2 $ and $R^3$.  The coefficients of $p^{2}$ and $p^{0}$ determine $c_{1,1}$ and $c_{3,0}$, respectively:
\begin{equation}
c_{1,1} = \frac{N}{5760\pi} \left( \frac{\sqrt5}{\log\varphi}-4\right)\,,\qquad c_{3,0}  =  -\frac{N(4\log\varphi+\sqrt5)}{725760\pi \log^3\varphi}\,.
\label{eq:lens-c11}
\end{equation}
The lower-order coefficients $f$, $c_1$, and $c_2$ agree with the $S^1\times_\Omega S^2$  computation of \cite{Mauro:2026zus}, while $c_{2,0}$ agrees with the $\Omega=0$ expansion of \cite{David:2024pir}. Our squashed-Lens calculation thus extends this to one further local-EFT order, yielding the $\beta^4$ data $c_{1,1}$ and $c_{3,0}$. 

As a final comment, compare this to the critical large-$N$ Gross--Neveu model, where the parity-symmetric thermal saddle has vanishing auxiliary mass, $m_\star=0$, as found on $S^1\times S^2$ by \cite{Miele:1996rp}. Consequently, to leading order at large-$N$, the thermal partition function on this saddle, and hence the leading thermal EFT coefficients, coincide with those of $N$ free massless Dirac fermions. 
Interactions first modify $\log \CZ$ at $O(N^0)$, through the Hubbard--Stratonovich fluctuation determinant. For the leading correction to the free energy density, see \cite{Diatlyk:2023msc}.

\subsection{Holographic CFT}\label{sec:holo}
As our final example we consider holographic CFTs with an Einstein dual, for which the leading few thermal EFT coefficients were derived in \cite{Benjamin:2023qsc}. In \cite{Bobev:2023ggk} they were mapped to microscopic data in the case of three-dimensional $\mathcal{N}=2$ SCFTs arising from M2-branes. Our goal here is to determine all non-derivative coefficients $c_{n,m}$ in \eqref{eq:genEFT} as well as some of the leading derivative ones $d_i$ in \eqref{eq:Lder}.  

Before studying the thermal EFT expansion, we briefly review the gravitational solutions of interest, namely the AdS-Kerr and Taub-Bolt backgrounds in 4d. These are subcases of the more general AdS-Kerr-Bolt metric studied in appendix~\ref{app:AdSkerrbolt}. It should be noted that for holographic CFTs, the thermal effective action has an extended range of validity, since at large $N$, non-perturbative corrections in $\beta$ are further suppressed by $1/N$ \cite{Horowitz:2017ifu}.

Unless explicitly stated otherwise, we will set $L_\AdS= G=1$ in this section.

\subsubsection{AdS-Kerr}\label{sec:kerr}
The AdS-Kerr solution has metric
\begin{equation}\label{eq:kerr}
    \rmd s^2 = \frac{\cQ}{\rho^2}(\rmd t_{\scriptscriptstyle{E}} - \frac{\alpha}{\Xi}\sin^2\vartheta\rmd \varphi)^2 + \rho^2(\frac{\rmd r^2}{\cQ}+\frac{\rmd \vartheta^2}{\cP}) + \frac{\cP \sin^2\vartheta}{\rho^2} (\alpha \rmd t_{\scriptscriptstyle{E}} + \frac{r^2-\alpha^2}{\Xi}\rmd \varphi)^2\, , 
\end{equation}
where
\begin{equation}
  \cQ = (1+r^2)(r^2-\alpha^2)- 2m r\, , \quad \rho^2 = r^2- \alpha^2\cos^2\vartheta\, ,\quad
  \cP = 1+ \alpha^2\cos^2\vartheta \, , \quad \Xi = 1+\alpha^2\,.
\end{equation}
In terms of the outer horizon radius $r_+$, the regularized on-shell action is given by \cite{Hawking:1998kw, Gibbons:2004ai}:
\begin{equation}\label{eq:kerrOS}
    I = \frac{\beta}{2\,\Xi} (m + \alpha^2 r_+- r^3_+)\, .
\end{equation}
Regularity at the horizon imposes $(t_{\scriptscriptstyle{E}},\varphi)\sim (t_{\scriptscriptstyle{E}}+\beta,\, \varphi - \beta \Omega')$ with 
\begin{equation}\label{eq:betaOmp}
  \beta = \frac{2\pi (r^2_+-\alpha^2)}{2r^3_++r_+(1-\alpha^2)-m}\, , \quad  \Omega' = \frac{\alpha \Xi}{r^2_+-\alpha^2} \,.
\end{equation}
The boundary metric, with $t_{\scriptscriptstyle{E}} = \beta \tau$, is given by
\begin{equation}\label{eq:kerrbdy}
    \rmd s^2_{\text{bdy}} = \beta^2(\rmd \tau - \frac{\alpha \sin^2\vartheta}{\beta\Xi}\rmd\varphi)^2 + \frac{\rmd \vartheta^2}{\cP}+ \frac{\cP \sin^2\vartheta}{\Xi^2} \rmd \varphi^2\,.
\end{equation}
This metric is already in KK-form and would lead to a Weyl rescaled metric and gauge field of the form \eqref{eq:weyl2} and \eqref{eq:kerr_alt}, with $\Omega$ replaced by $\alpha$. The correct expressions to use in the thermal EFT calculations are those involving $\Omega$, which is the horizon velocity with respect to a non-corotating observer at infinity \cite{Hawking:1998kw, Gibbons:2004ai}:
\begin{equation}\label{eq:Omegadef}
    \Omega = \alpha +\Omega' = \frac{\alpha (1+r^2_+)}{r^2_+ - \alpha^2}\,.
\end{equation}  
Indeed, in \eqref{eq:kerrbdy} one still has to pull out the $\Omega'$ rotation by redefining $\varphi = \phi - \beta \Omega' \tau$. In $\theta$-coordinates \eqref{eq:kerr1}, related to $\vartheta$ by a transformation of the form \eqref{eq:thetachange}, the angular velocities simply add up, leading to the replacement of $\alpha$ by $\Omega$.

\subsubsection{AdS-Taub-Bolt}\label{sec:taubbolt}
The boundary metrics of the spherical, flat and hyperbolic AdS-Taub-Bolt solutions are such that only $R^{n}F^{2m}$ terms appear in the thermal EFT; since $R$ and $F^2$ are constant, there are no derivative terms. The on-shell action of these solutions was calculated in \cite{Emparan:1999pm}:
\begin{equation}\label{eq:taubon}
    I = \frac{\beta A}{8\pi} (m+3n^2 r_+ - r^3_+)\, , 
\end{equation}
where $A$ is the area of the relevant $\cM_2$ (e.g. $\Sigma_g$ or $\mathbb{T}^2$ or the Hopf base space) and $r_+$ is the outer horizon radius of the Taub-Bolt solution. This radius is determined by $V(r_+)=0$ (horizon) and $V'(r_+)= 4\pi/\beta$ (regularity), with the metric function $V(r)$ given by
\begin{equation}\label{eq:Vdef}
    V = \frac{k(r^2+n^2)- 2mr +r^4-6n^2r^2-3n^4}{r^2-n^2}\,,
\end{equation}
where $k=-1,0,1$ is the normalized Gaussian curvature of $\cM_2$.

\paragraph{Spherical case}
The spherical AdS-Taub-Bolt solution has metric
\begin{equation}\label{eq:spherKB}
    \rmd s^2 = 4n^2V(r) (\rmd\psi + \cos\theta\,\rmd\phi)^2+ \frac{\rmd r^2}{V(r)} + (r^2-n^2)(\rmd \theta^2+\sin^2\theta\,\rmd\phi^2)\, .
\end{equation}
Its boundary is the squashed $S^3/\mathbb{Z}_p$ \eqref{eq:squashmetbdy}, whose thermal EFT is determined as in \eqref{eq:squashRF}. To calculate the on-shell action \eqref{eq:taubon}, the relevant area is $A=4\pi$. In this case, regularity of the boundary manifold imposes the additional constraint $\beta= 8\pi n/p$.

\paragraph{Flat case}
The flat AdS-Taub-Bolt solution has metric
\begin{equation}\label{eq:tormet}
    \rmd s^2 = V(r) (\rmd\tau + n(x\rmd y - y \rmd x))^2+ \frac{\rmd r^2}{V(r)} + (r^2-n^2)(\rmd x^2+\rmd y^2)\,.
\end{equation}
Its boundary is $S^1\times \mathbb{T}^2$, with associated thermal EFT quantities $R=0$ and $F^2 = 8n^2 \beta^2$.

\paragraph{Hyperbolic case}
The hyperbolic AdS-Taub-Bolt solution has metric
\begin{equation}\label{eq:hypmet}
    \rmd s^2 = V(r) (\rmd\tau + 2n(\cosh\theta -1)\,\rmd\phi)^2+ \frac{\rmd r^2}{V(r)} + (r^2-n^2)(\rmd \theta^2+\sinh^2\theta\,\rmd \phi^2)\,,
\end{equation}
Its boundary is $S^1\times \mathbb{H}^2$, with associated thermal EFT quantities $R=-2\beta^2$ and $F^2 = 8n^2 \beta^2$.

\subsubsection{Turning off fugacity} The thermal EFT coefficients $c_{n,0}$ follow from the on-shell action of $\AdS$-Schwarzschild, which is related to hyperbolic AdS-Taub-Bolt by flipping the sign of $R$ in the thermal EFT.
\paragraph{Thermal EFT coefficients from AdS-Schwarzschild}
The on-shell action is given by
\begin{equation}\label{eq:onshSS}
    I = \frac{\beta}{2}\left(m -r^3_+ \right)\,.
\end{equation}
It can be evaluated in closed form as a function of $\beta$, by first solving $V(r_+)=0$ for $m(r_+)$ and then solving $V'(r_+)=4\pi/\beta$ for $r_+(\beta)$, where $V(r) = 1-2m/r +r^2$:
\begin{equation}
    m = \frac{1}{2} r_+ \left(r^2_+ +1\right)\, ,\quad r_+ = \frac{\sqrt{4 \pi ^2-3 \beta ^2}+2 \pi }{3 \beta }\,.
\end{equation}
Plugging this back into the action gives
\begin{equation}\label{eq:onshSSb}
    I= -\frac{1}{27\beta^2}\left(8\pi^3 - 9\pi\beta^2 + \left(4\pi^2 -3\beta^2\right)^\frac32\right)\,.
\end{equation}
To compare with the thermal EFT expansion, note that the Weyl rescaled metric is simply $\rmd \hat{s}^2 = \beta^{-2}(\rmd\theta^2+\sin^2\theta\rmd\phi^2)$, and hence $R=2\beta^2$. We can then simply read off all $c_{n,0}$ from
\begin{equation}\label{eq:ZeftSS}
    \log\CZ_{\text{EFT}} = - \frac{4\pi}{\beta^{2}}\sum_n c_{n,0}R^n= \frac{1}{27\beta^2}\bigg(8\pi^3 - \tfrac{9\pi}{2}R + \big(4\pi^2 -\tfrac32 R\big)^\frac32\bigg)\,.
\end{equation}
In particular, we find
\begin{equation}\label{eq:cn0holo}
    c_{n,0}= -\frac{3^{n-2}\, \Gamma \left(n-\tfrac{3}{2}\right)}{2^{3 n+1} \pi ^{2 n-3/2} \Gamma (n+1)}\,(1+\delta_{n0}+\delta_{n1})\,.
\end{equation}
The first two terms $c_{0,0}= -4\pi^2/27$ and $c_{1,0}=1/12$ were given before in \cite{Benjamin:2023qsc}, while the others correspond to equation (71) in \cite{Allameh:2024qqp}.
 
\paragraph{Thermal EFT coefficients from hyperbolic AdS-Taub-Bolt} Consider now the hyperbolic Taub-Bolt metric \eqref{eq:hypmet} at $n=0$. In this case the boundary metric is $S^1\times \mathbb{H}_2/\Gamma$. Moreover, the on-shell action reduces to the Schwarzschild form \eqref{eq:onshSS}. The only difference is the function $V(r)=-1 -2m/r + r^2$ in the metric, leading to
\begin{equation}
    m = \frac{1}{2} r_+ \left(r^2_+ -1\right)\, ,\quad r_+ = \frac{\sqrt{4 \pi ^2+3 \beta ^2}+2 \pi }{3 \beta }\,.
\end{equation}
and the following on-shell action as a function of temperature:
\begin{equation}
        I= -\frac{1}{27\beta^2}\left(8\pi^3 + 9\pi\beta^2 + \left(4\pi^2 +3\beta^2\right)^\frac32\right)\,,
\end{equation}
differing from \eqref{eq:onshSSb} by $\beta^2\to - \beta^2$. In the thermal EFT, the Ricci scalar of the Weyl rescaled metric experiences a similar sign flip: $R= -2\beta^2$. Written as a function of $R$, the on-shell action thus reduces to the same form \eqref{eq:ZeftSS}, resulting in $c_{n,0}$ as before in \eqref{eq:cn0holo}.

\subsubsection{Turning on fugacity} By turning on NUT charge $n$ or rotation $\Omega$, we can access more Wilson coefficients. The hyperbolic AdS-Taub-Bolt background yields all non-derivative coefficients $c_{n,m}$, while AdS-Kerr fixes the leading derivative coefficient $d_1$, as well as subleading combinations.

\paragraph{Thermal EFT coefficients from flat AdS-Taub-Bolt}
The metric is given by \eqref{eq:tormet}. The horizon and regularity conditions impose
\begin{equation}
    m = \frac{-3 n^4-6 n^2 r^2_+ +r^4_+}{2 r_+}\, , \quad r_+ = \frac{\sqrt{4 \pi^2+9 n^2\beta^2}+2 \pi }{3 \beta }\,.
\end{equation}
Plugging this back into the on-shell action \eqref{eq:taubon}, we find
\begin{equation}
    I = -\frac{A}{108\pi\beta^2}\left(8 \pi ^3+\left(4\pi^2+9n^2 \beta ^2 \right)^\frac{3}{2}\right)\,.
\end{equation}
In the thermal EFT we have $R=0$ and $F^2= 8n^2\beta^2$. Hence, one obtains the $c_{0,m}$ from
\begin{equation}\label{eq:torusL}
    \log \cZ_{\text{EFT}} = -\frac{A}{\beta^2}\sum_m c_{0,m}F^{2m}= \frac{A}{108\pi\beta^2}\left(8\pi^3 + \left(4\pi^2 + \tfrac{9}{8}F^2\right)^\frac32\right)\,. 
\end{equation}
Explicitly, we find 
\begin{equation}\label{eq:c0nholo}
    c_{0,m}= (-1)^{m+1} \pi^{\frac32}\frac{9^{m-1}\Gamma(m-\frac32)}{2^{5m+1}\pi^{2m}\Gamma(m+1)}(1+\delta_{m0})\,.
\end{equation}
The first two terms $c_{0,0}= -4\pi^2/27$ and $c_{0,1}= -1/32$ were given in \cite{Benjamin:2023qsc}.

\paragraph{Thermal EFT coefficients from hyperbolic AdS-Taub-Bolt}
The metric is given by \eqref{eq:hypmet}. The horizon and regularity conditions impose
\begin{equation}
    m = \frac{-3 n^4-6 n^2 r^2_+ +r^4_+-r^2_+-n^2}{2 r_+}\, , \quad r_+ = \frac{\sqrt{4 \pi ^2+3\beta^2+9n^2 \beta ^2 }+2 \pi }{3 \beta }\,.
\end{equation}
Plugging this back into the on-shell action \eqref{eq:taubon}, we find
\begin{equation}
    I = -\frac{A}{108 \pi  \beta ^2}\left(8\pi^3 +9\pi\beta^2 +\left(4\pi^2+3 \beta ^2 +9n^2\beta^2\right)^\frac{3}{2}\right )\,.
\end{equation}
The thermal EFT has $R= -2\beta^2$ and $F^2 = 8n^2\beta^2$, so we again find a closed form expression:
\begin{equation}\label{eq:hypL}
\log \cZ_{\text{EFT}}= - \frac{A}{\beta^2}\sum_{n,m} c_{n,m}R^n F^{2m} = \frac{A}{108\pi\beta^2}\bigg(8\pi^3 - \tfrac{9\pi}{2}R + \big(4\pi^2 -\tfrac32 R+ \tfrac{9}{8}F^2\big)^\frac32\bigg)\,.
\end{equation}
This reduces to \eqref{eq:torusL} at $R=0$ and to \eqref{eq:ZeftSS} at $F=0$, so that the $c_{n,0}$ and $c_{0,m}$ are consistent with those obtained above. However, we now get access to every other $c_{n,m}$ as well, simply by expanding $\log \cZ_{\text{EFT}}$ to the desired order. 
One finds:
\begin{equation}
c_{n,m}=\frac{(-1)^{m+1}3^{n+2m-2}\Gamma\left(n+m-\tfrac32\right)}
{2^{3n+5m+1}\pi^{2n+2m-\frac32}\Gamma(n+1)\Gamma(m+1)}
\left(1+\delta_{m0}\left(\delta_{n0}+\delta_{n1}\right)\right).
\end{equation}
Setting $m=0$ or $n=0$ respectively reproduces \eqref{eq:cn0holo} and \eqref{eq:c0nholo} above. The lowest few coefficients are provided below: 

\begin{equation}\label{eq:cnmlist}
    \begin{tabular}{|C|C|C|C|C|}
c_{n,m} & 0 & 1 & 2  & 3\\\hline
0 & \frac{-4\pi^2}{27}  & \frac{-1}{32} & \frac{-9}{4096\pi^2} & \frac{27}{262144\pi^4} \\
1 & \frac{1}{12} & \frac{3}{512\pi^2}  & \frac{-27}{65536\pi^4} & \frac{243}{4194304\pi^6}\\
2 & \frac{-1}{256\pi^2}  & \frac{9}{16384\pi^4} & \frac{-243}{2097152\pi^6}  &  \\
3  & \frac{-1}{4096\pi^4} & \frac{27}{262144\pi^6}   &    &\\
4 & \frac{-9}{262144\pi^6}&&
\end{tabular}
\end{equation}

\paragraph{Thermal EFT for spherical AdS-Taub-Bolt}
Consider a holographic CFT on a squashed Lens space $S^3/\mathbb{Z}_p$ with metric \eqref{eq:squashmetbdy}. Its thermal EFT expansion has $R = 2\beta^2$ and $F^2 = p^2\beta^4/8\pi^2$ as in \eqref{eq:squashRF}. Moreover, we already know all coefficients $c_{n,m}$ from the hyperbolic case, as well as $A=4\pi$. In other words, we can simply use  \eqref{eq:hypL} to find the partition function for a holographic CFT on $S^3/\mathbb{Z}_p$:
\begin{equation}
    \log \cZ = \frac{1}{27 \beta^2}\bigg(8\pi^3 - 9\pi\beta^2 + \big(4\pi^2 -3\beta^2+ \tfrac{9 p^2\beta^4}{64\pi^2}\big)^\frac32\bigg)\,.
\end{equation}
Using \eqref{eq:squashmet}, this matches the $p=1$ large-squashing expansion (34) in \cite{Bobev:2016sap}. More generally, for any $p$, one checks directly that $\log \cZ$ equals minus the on-shell action \eqref{eq:taubon} by plugging in the following solutions to  $V(r_+)=0$ (horizon) and $V'(r_+) = 4\pi/\beta =p/2n$ (regularity): 
\begin{equation}\begin{split}
   m&= \frac{ - 3 n^4- 6n^2 r^2_+ + r^4_++ r^2_++n^2 }{2 r_+}\,,\\ r_+ &= \frac{\sqrt{p^2+144 n^4-48 n^2}+p}{12 n}=\frac{\sqrt{4\pi^2-3 \beta^2+9 p^2\beta^4/64\pi^2}+2\pi}{3 \beta}\,.
   \end{split}
\end{equation}

\paragraph{Thermal EFT for AdS-Kerr}
Using the results of section~\ref{sec:kerr}, we can find the metric parameters $m, \alpha$ in terms of the horizon radius $r_+$ and angular velocity $ \Omega$: 
\begin{equation}
    m = \frac{\left(1+r^2_+ \right) \left(r^2_+-\alpha ^2\right)}{2 r_+}\, , \quad \alpha = \frac{-1-r^2_++\sqrt{(1+r^2_+)^2+4 r^2_+ \Omega ^2}}{2 \Omega }\,.
\end{equation}
Moreover, from \eqref{eq:betaOmp}, we have the following large-temperature expansions at fixed $\Omega$:
\begin{equation}
    r_+ = \frac{4 \pi }{3 \beta }-\frac{\beta  \left(1+2 \Omega ^2\right)}{4 \pi }-\frac{\beta ^3 \left(3-6 \Omega^2-6\Omega ^4\right)}{64 \pi ^3}-\frac{9\beta ^5  \left(1+3 \Omega ^2+3 \Omega ^4+ 2 \Omega ^6\right)}{512 \pi ^5}+ O(\beta^7)\,.
\end{equation}
Plugging all of the above back into the on-shell action \eqref{eq:kerrOS} yields
\begin{equation}\begin{split}\label{eq:kerrIexp}
   I = \frac{4\pi}{\beta^2(1+\Omega^2)}\bigg (&\frac{-4\pi^2}{27}+\frac{\beta^2\left(2+\Omega ^2\right)}{12}  -\frac{\beta^4\left(1+\Omega ^2+\Omega ^4\right)}{64 \pi ^2} \\
   &- \frac{\beta^6\left(2+3\Omega^2-3 \Omega ^4-2 \Omega ^6\right)}{1024 \pi ^4} +O\left(\beta^8\right)\bigg) \, .
   \end{split}
\end{equation}
On the other hand, the thermal EFT formalism predicts the leading expansion \eqref{eq:EFTexpK}, with contributions of order $\beta^2$ in \eqref{eq:EFTK2} and $\beta^4$ in \eqref{eq:EFTK4}. Substituting the $c_{n,m}$ listed in \eqref{eq:cnmlist}:
\begin{align}
    \log \cZ_{\text{EFT}} = &\frac{-4\pi}{\beta^2\left(1+\Omega^2\right)}\bigg(\frac{-4\pi^2}{27}+\frac{\beta^2\left(2+\Omega ^2\right)}{12} -\frac{\beta^4}{64\pi^2}\left(1+\tfrac{4}{5} \Omega ^4-\tfrac{1024\pi^2}{15} d_1 \Omega ^2 \left(\Omega ^2+5\right)\right)\nonumber
    \\\nonumber
    &\qquad\;-\frac{\beta^6}{1024\pi^4}\Big(2+\tfrac{168}{35} \Omega ^4+\tfrac{32}{35} \Omega ^6-\tfrac{16384\pi^4}{105}\Omega^2 \big(\left(9 d_2+d_3+6d_6\right) \left(24 \Omega ^2+56\right) \Omega ^2\\
    &\qquad\qquad\quad\qquad+\left(2 d_4+d_5\right) \left(11 \Omega ^4+14 \Omega ^2+35\right)\big)\Big)+O\left(\beta^8\right)\bigg) \,,
    \end{align}
Comparing the top line to \eqref{eq:kerrIexp}, we can read off the $(\nabla F)^2$ coefficient $d_1= -3/1024\pi^2$. This is a non-trivial consistency check of the formalism: after using the AdS-Taub-Bolt $c_{n,m}$, $\log \cZ_\text{EFT}+I$ has an order $\beta^2$ contribution proportional to $\Omega^2(5+\Omega^2)$, which is precisely the sort of term provided by $d_1 (\nabla F)^2$. Slightly different AdS-Taub-Bolt coefficients would have required a slightly different polynomial, which would have made a match impossible. 

At higher order, we cannot fully fix the derivative coefficients --- as is already clear from \eqref{eq:EFTK4} --- but we can constrain certain combinations. All combined, we find
\begin{equation}\label{eq:diKerr}
   d_1 = \frac{-3}{1024\pi^2} \, , \quad 9d_2+d_3+6d_6=\frac{135}{131072\pi^4}\, , \quad 2d_4+d_5 =\frac{-9}{16384\pi^4}\,.
\end{equation}
As a further check of these EFT coefficients, we consider in appendix~\ref{app:AdSkerrbolt} the more general Plebanski-Demianski solution with rotation and non-vanishing NUT charge.

\subsubsection{Higher-derivative corrections}\label{sec:higherderiv}
The $d=4$ (super)gravity action is by itself an EFT. Including four-derivative corrections to the 4d Einstein-Maxwell gravity theory changes the regularized on-shell action to \cite{Bobev:2021oku}:
\begin{equation}\label{eq:I4d}
    I_{4\partial} = (1+64\pi(\lambda_2-\lambda_1)) I_{2\partial}+ 32\pi^2 \lambda_1 \chi(\mathcal{M}_2)\,.
\end{equation}
Here, $\lambda_{1}$ and $\lambda_2$ are the higher-derivative couplings as defined in \cite{Bobev:2021oku}. The effect on the thermal EFT coefficients is then as follows:
\begin{equation}
    c_{n,m} \to c_{n,m}(1+64\pi(\lambda_2-\lambda_1)) + 8\pi \lambda_1 \delta_{n1}\delta_{m0}\,.
\end{equation}
The derivative terms $d_i$ are rescaled similarly. This is consistent with what \cite{Bobev:2023ggk} found for the leading coefficients in Kerr. It is clear why this works: it simply rescales $I_{2\partial}$ by $1+64\pi(\lambda_2-\lambda_1)$ and adds a topological $8\pi\lambda_1\int_{\mathcal{M}_2} \sqrt{\hat g} R = 32\pi^2\lambda_1\chi(\mathcal{M}_2)$, thus arriving precisely at \eqref{eq:I4d}.

\section{Conclusion and outlook}
\label{sec:discussion}
We studied CFT partition functions on curved three-manifolds in two complementary regimes where universal methods are available. For small deformations of the round $S^3$, conformal perturbation theory expresses the response in terms of integrated stress-tensor correlators. At quadratic order, the harmonic-space formula \eqref{eq:FS3spin2} applies to arbitrary metric deformations, and shows that the round sphere is a local maximum of the free energy. It reproduces the known Hopf-squashing result and extends directly to double squashing. We also analyzed sources for scalar and vector operators, and more generally for conserved spin-$s$ currents, finding an alternating sign pattern in their quadratic response that is particularly relevant when supersymmetry links metric and background gauge-field deformations.

At cubic order, we clarified the distinction between metric derivatives of $\log \CZ$ and standard stress-tensor correlators, which differ from the Osborn--Petkou correlator by contact terms \cite{Osborn:1993cr,Parisini:2023nbd}. Keeping this distinction explicit makes the conformal perturbation theory result consistent with holographic and free field evidence that the cubic Hopf-squashing coefficient is universally controlled by $c_T t_4$ \cite{Bueno:2018yzo}. The same method should extend in a straightforward fashion to $d=5$, matching the holographic and free field results of \cite{Bueno:2020odt}. We also derived the linear Hopf-squashing correction \eqref{eq:too-answer-q} to the scalar two-point function by integrating the regulated $\langle T\CO\CO\rangle$ distribution, reproducing the ambient-space result of~\cite{Parisini:2023nbd}.

The second regime is the large-squashing, or small-fiber, limit of circle bundles $S^1\times_{\mathrm f}\mathcal{M}_2$, where the partition function is governed by a two-dimensional thermal effective action constructed from the Weyl-rescaled base metric and Kaluza--Klein field strength \cite{Banerjee:2012iz,Benjamin:2023qsc}. We determined its Wilson coefficients for free conformally coupled scalars and massless fermions, reproducing the large-squashing asymptotics on Lens spaces \cite{DeFrancia:2000xm}. For the large-$N$ critical $\rO(N)$ model, we extended the squashed-sphere analysis of \cite{Hartnoll:2005yc} to Lens spaces and to further subleading orders, determining the local thermal-EFT data through order $\beta^4$. This reproduces the lower-order coefficients of \cite{Mauro:2026zus,David:2024pir} and fixes $c_{1,1}$ and $c_{3,0}$ as in \eqref{eq:lens-c11}. For holographic CFTs, AdS-Taub-Bolt resums the full non-derivative sector $c_{n,m}R^n F^{2m}$, see \eqref{eq:hypL}, while AdS-Kerr fixes the leading derivative term $(\nabla F)^2$ and constrains subleading combinations as in \eqref{eq:diKerr}. Their agreement provides a nontrivial check that the thermal EFT captures the universal large-squashing data.\\

Several directions merit further study:

\begin{itemize}

    \item The integrated $\langle T\CO\CO\rangle$ calculation in section~\ref{sec:SquashS3} illustrates the importance of treating correlators as distributions compatible with the Ward identities \cite{Osborn:1993cr}. An analogous treatment of $\langle TTT\rangle$ should yield an intrinsic CFT derivation of the cubic squashing coefficient. Embedding-space techniques \cite{Costa:2011mg,Costa:2011dw} and generating-function methods for higher-spin correlators \cite{Giombi:2010vg,Giombi:2011rz} may provide efficient tools for this calculation.

   \item It would be useful to understand when the free energy on squashed spheres obeys a global maximality property. The universal negative quadratic response around the round sphere does not exclude a larger free energy at finite squashing. Known counterexamples evade either the bona fide thermal or causal regime. For example, the free Dirac fermion on the simply connected Hopf-squashed $S^3$ uses the spin structure inherited from $S^3$, rather than an antiperiodic thermal spin structure around the small fiber \cite{Bobev:2017asb}. In $d=5$ Gauss--Bonnet holography, global maximality can be violated only for stress-tensor three-point data outside the Hofman--Maldacena conformal-collider bounds \cite{Hofman:2008ar,Bobev:2017asb}. This motivates asking whether global maximality holds for small-circle backgrounds admitting a unitary thermal trace and for CFT data satisfying the usual collider and causality constraints.

    \item Positivity constraints on thermal EFT Wilson coefficients deserve a more systematic study. All examples here and in previous work are, as far as we are aware, compatible with $c_1\geq 0$ and $c_2\leq 0$ \cite{Benjamin:2023qsc,Allameh:2024qqp}.\footnote{This is in the conventional notation, corresponding respectively to $c_{1,0}$ and $c_{0,1}$ in the main text.} This includes free fermions, the $d=3$ Ising model~\cite{Benjamin:2023qsc}, the large-$N$ critical $\rO(N)$ model \cite{Mauro:2026zus}, Einstein-gravity holographic CFTs, and the holographic $\mathcal{N}=2$ SCFTs studied in \cite{Bobev:2023ggk}. It would be valuable either to find a counterexample or to identify a general principle underlying these signs. The qualifier ``thermal'' is essential: non-thermal twists need not satisfy the same constraints~\cite{Allameh:2024qqp}. In $d=4$, a proper counterexample to $c_2 \leq 0$ does exist: free Maxwell theory has $c_2=1/144 >0$.\footnote{We thank Nathan Benjamin and Edgar Shaghoulian for bringing this to our attention.} The $c_1\geq 0$ conjecture of \cite{Allameh:2024qqp} remains open. More generally, it is important to understand how local correlation function data of the CFT on flat space are encoded in the thermal EFT Wilsonian coefficients. 

    \item Although thermal backgrounds break conformal symmetry, their correlators remain constrained by Ward identities and obey nontrivial sum rules \cite{Marchetto:2023xap}. Related constraints arise in EFTs with spontaneously broken boosts \cite{Maldacena:2002vr,Hui:2022dnm}, where positivity bounds follow from the corresponding sum rules \cite{Creminelli:2022onn}. In the hydrostatic generating functional, $c_2$ controls the small-momentum transverse response to the KK gauge field $A_i$, 
  \begin{equation} 
  \frac{\delta^2 W}{\delta A_y(k)\delta A_y(-k)} = -4c_2 \beta^{3-d}k^2+O(k^4)\,, \qquad k=k_x\,. 
  \end{equation} 
  Microscopically, this response contains the separated correlator $\langle T_{\tau y}T_{\tau y}\rangle$. Away from zero frequency, momentum conservation formally relates its transverse momentum dependence to the shear-channel spectral density, suggesting a possible origin of $c_2\leq 0$ in genuine thermal systems. The full metric functional derivative, however, includes seagull and other contact terms required by the diffeomorphism Ward identities. Whether the complete response admits a positive sum rule remains open.

    \item Similar EFT analyses should be possible in higher dimensions.  In even dimensions,  Weyl-anomaly terms enrich and perhaps complicate the discussion. In $d=5$, on the other hand, a different practical obstacle is the absence of known non-supersymmetric unitary interacting CFTs. Supersymmetric examples therefore provide the most accessible testing ground.

    \item The thermal EFT can be enlarged to include background gauge fields for global symmetries. This would connect directly to charge-resolved asymptotic data, as in the critical $\rO(N)$ model at nonzero chemical potential \cite{Diatlyk:2023msc}, and to recent studies of CFTs in external magnetic fields \cite{Boyack:2023uml,Herzog:2025ddq}. It should also permit an analysis of three-dimensional $\mathcal{N}=2$ SCFTs, which necessarily possess a $\UU(1)$ R-symmetry current. Related EFT methods have already proved useful for supersymmetric indices in four dimensions \cite{Cassani:2021fyv,DiPietro:2014bca}.
    
    \item It will be interesting to consider CFTs placed on other curved backgrounds, both Lorentzian and Euclidean, and try to identify appropriate limits controlled either by an EFT or conformal perturbation theory. See \cite{Anand:2025mfh,Komargodski:2026ain} for recent work in this spirit which considers large angular momentum limits in CFTs and the relation to placing them in an appropriate pp-wave background.  

    \item Localization results express supersymmetric Euclidean on-shell actions in terms of NUT and Bolt fixed-point data \cite{BenettiGenolini:2019jdz}. How does the boundary data of the thermal EFT --- the Weyl-rescaled metric and KK field strength --- map onto this fixed-point description if it extends also to non-supersymmetric setups?

    \item In gravity, more general large boundary deformations, such as dipolar potentials \cite{Costa:2015gol} and differential rotation \cite{Markeviciute:2017jcp}, provide a broader setting to test how much of the tidal response is fixed by the dual CFT kinematics. Although the corresponding gravitational solutions generally require numerical methods, their large-deformation instabilities have a clear boundary interpretation: the ensemble becomes singular when the source reaches a limiting velocity or chemical potential. This is already apparent in the rotating thermal EFT on $S^1_\beta\times_\Omega S^2$, whose Lorentzian continuation becomes singular as the thermal Killing vector becomes null at $|\Omega|=1$. Analogous singularities arise in superfluid EFT as the flow approaches the speed of sound and Cherenkov radiation sets in \cite{Gouteraux:2022qix,Kourkoulou:2022doz}.

    \item Our holographic AdS-Kerr-Bolt examples determine particular combinations of the derivative Wilson coefficients, see \eqref{eq:diKerr}. To disentangle them, it would be interesting to consider more general boundary metrics, construct the corresponding asymptotically AdS fillings numerically, and extract their renormalized on-shell actions.
    

    \item Finally, one may wonder whether any of the curved-manifold setups studied here admit useful realizations or analogues in condensed-matter systems near quantum criticality.
\end{itemize}

\section*{Acknowledgments}

We are grateful to Davide Cassani, Matijn Fran\c cois, Sean Hartnoll, Junho Hong, Zohar Komargodski, Enrico Marchetto, Dalimil Maz\'a\v c, Sridip Pal, Valentin Reys, Jorge Santos, David Tong, Inne Van de Plas, Silviu Pufu, and Xuao Zhang, for valuable discussions and to Dalimil Maz\'a\v c for early collaboration on this project. NPB was supported by the FWO projects G003523N, G094523N, and G0E2723N, and the KU Leuven C1 project C16/25/01. KP was supported by the US DOE DE-SC011941, the STFC consolidated grant ST/T000694/1 and Simons Investigator award \#620869. KP is also grateful to the ITF at KU Leuven for the warm hospitality during the completion of this work. 

\appendix
\section{Decomposing the two-point function}\label{app:spherharm}
Here, we present details regarding the spherical harmonics on $S^d$ that were used in the main text. The two-point function $\langle \CO^{i_1\dots i_s}(x)\CO^{j_1\dots j_s}(y)\rangle$ on the sphere is fixed by conformal invariance. Its eigentensors are the spherical harmonics $Y^n(x)$, with eigenvalues $\lambda_n$ obtained by explicitly computing the overlaps $\int Y^n \cdot\langle\CO\CO\rangle \cdot Y^n$. This is manageable for the zonal harmonics, and was done in \cite{Giombi:2013yva} for general spin on $S^3$, and in \cite{Gubser:2002vv, Giombi:2015haa} for spin-0 and spin-1 on $S^d$. We collect and extend these results. For a more mathematical treatment, see \cite{Dobrev:1977qv}.

\subsection{Harmonics on \texorpdfstring{$S^3$}{}}
Spin-$s$ symmetric traceless tensors on $S^3$  decompose under $\SO(4) = \SU(2)_L\times \SU(2)_R$ as
\begin{equation}
    \bigoplus_{n=s}^\infty \bigoplus_{t=-s}^{s}\, (\mathbf{n+1+t},\, \mathbf{n+1-t})\,.
\end{equation}
The basis tensors are the spherical harmonics $Y^{t,nlm}_{\mu_1\dots \mu_s}$, with $|t|, |m|\leq l \leq n $ and \cite{Higuchi:1986wu}:
\begin{equation}\begin{split}\label{eq:S3harmonics}
    \nabla^2 Y^{t,nlm}_{\mu_1\dots \mu_s} &= -\left(n(n+2)+t^2-s(s+1)\right) Y^{t,nlm}_{\mu_1\dots \mu_s}\, ,\\
    \nabla^\nu Y^{t,nlm}_{\nu\mu_2\dots \mu_s}&= -\left(\frac{((n+1)^2-s^2)(s^2-t^2)}{s(2s-1)}\right)^{\tfrac12} Y^{t,nlm}_{\mu_2\dots \mu_s}\,.
    \end{split}
\end{equation}
The spin-$s$ two-point kernel $\langle \CO^{i_1\dots i_s}(x)\CO^{j_1\dots j_s}(y)\rangle$ on $S^3$ has eigenvalues \cite{Giombi:2013yva}:\footnote{Compared to \cite{Giombi:2013yva}, we shifted $n$ by $1$, so that our spherical harmonics start at $n=s$ rather than $s+1$. Moreover, in the formulas below we consider $t\geq 0$ to avoid clutter, since the parity-even two-point kernel has degenerate eigenvalues in the $t$ and $-t$ sectors. }
\begin{equation}\label{eq:S32pt}
    \lambda^{\Delta,\, s,\,t}_{n} =  \frac{\Gamma(2-\Delta)}{\Gamma(\Delta-1)} \frac{\Gamma(t-1+\Delta)}{\Gamma(t+2-\Delta)}\, \lambda^{\Delta,\, s,\,0}_{n}\, , 
\end{equation}
where
\begin{equation}\label{eq:S32pt0}
    \lambda^{\Delta,\, s,\,0}_{n} = c_s(\Delta)\, \frac{\Gamma(n+\Delta)}{\Gamma(n+3-\Delta)} \, ,  \quad   c_s(\Delta) = 4\pi \sin(\pi \Delta) \Gamma(2-2\Delta) \frac{\Gamma(\Delta)\Gamma(s+2-\Delta)}{\Gamma(\Delta + s)\Gamma(2-\Delta)}\,.
\end{equation}
Here, $s$ is the spin of the operator and $t$ is the depth. In other words, $t=0$ is scalar/longitudinal while $s=|t|$ is transverse traceless. The mode number is indicated only by $n$, since the eigenvalue does not depend on the $\SO(3)$ quantum numbers. The formula has been demonstrated for $s=0,1,2,3$ and conjectured for general $s$ in \cite{Giombi:2013yva}. They did not write $c_s(\Delta)$ explicitly for general $s$, as it eventually dropped out in their calculation of interest. In the above we have therefore extrapolated it from the $s\leq 3$ results. Granting this, the final result simplifies to
\begin{equation}
    \lambda^{\Delta,\, s,\,t}_{n} = -2\pi\sin(\pi \Delta)\Gamma(3-2\Delta)\,\frac{\Gamma(n+\Delta)}{\Gamma(n+3-\Delta)} \frac{\Gamma(\Delta+t-1)}{\Gamma(\Delta+s)} \frac{\Gamma(s+2-\Delta)}{\Gamma(t+2-\Delta)}\,.
\end{equation}
For conserved currents $\Delta=s+1$ only the $t=s$ eigenvalue is non-vanishing, corresponding to transverse traceless spherical harmonics. In that case, the above reduces to
\begin{equation}\label{eq:S3cons}
   \lambda^s_n  =  \frac{(-1)^s\,\pi^2}{(2s)!} \frac{\Gamma(n+s+1)}{\Gamma(n-s+2)} \,.
\end{equation}
All eigenvalues quoted above assume unit normalization of the two-point function; in the main text they should be multiplied by the corresponding coefficient, such as $c_J$ for spin-1 currents, and $c_T$ for the stress tensor.
\subsection{Harmonics on \texorpdfstring{$S^d$}{}}
For spin-$0$, starting from (A.5) in \cite{Drummond:1977uy}, we have
\begin{equation}\label{eq:Sd2pt0}
    \lambda^{\Delta}_n= 2^{d-2\Delta}\pi^{\frac{d}{2}}\frac{\Gamma(\frac{d}{2}-\Delta)}{\Gamma(\Delta)}\frac{\Gamma(n+\Delta)}{\Gamma(n+d-\Delta)}\,.
\end{equation}
Using Euler reflection and Legendre duplication, this reduces correctly to \eqref{eq:S32pt0} when $d=3$. For spin-$1$, the transverse result is given by (3.12) in \cite{Giombi:2015haa}:
\begin{equation}
    \lambda^{\Delta, 1}_n = 2^{d-2\Delta}\pi^{\frac{d}{2}}(\Delta-1)\frac{\Gamma(\frac{d}{2}-\Delta)}{\Gamma(\Delta+1)}\frac{\Gamma(n+\Delta)}{\Gamma(n+d-\Delta)}\, ,
\end{equation}
which for a conserved current becomes
\begin{equation}\label{eq:Sdspin1}
    \lambda^{1}_n = -\pi^{\frac{d}{2}}\frac{\Gamma(2-\frac{d}{2})\Gamma(n+d-1)}{2^{d-3}\Gamma(d)\Gamma(n+1)}\, .
\end{equation}
For $d=3$ this reduces to \eqref{eq:S3cons} evaluated at $s=1$. Observe that for any $d$, the spin-$1$ eigenvalue $\lambda^{\Delta,1}_n$ is related to the scalar result by a factor $(\Delta-1)/\Delta$. More generally, the conformal two-point kernel for a symmetric traceless spin-$s$ primary is the spinning conformal intertwiner (or Knapp--Stein operator). Its transverse-traceless component has the same $n$-dependence as in the scalar case, while its spin-dependent normalization is fixed by the TT tensor structure \cite{Dobrev:1977qv,Schaub:2024rnl}. We therefore obtain the following TT eigenvalue:
\begin{equation}\label{eq:lambdagen}
  \lambda^{\Delta,s}_n  = \frac{2^{d-2\Delta}\pi^{\frac{d}{2}}}{\Delta+s-1}\frac{\Gamma(\frac{d}{2}-\Delta)}{\Gamma(\Delta-1)}\frac{\Gamma(n+\Delta)}{\Gamma(n+d-\Delta)} \,.
\end{equation}
In particular, in odd $d$, for conserved spin-2 with $\Delta=s+d-2=d$, this becomes: 
\begin{equation}\label{eq:Sd2ptspin2}
  \lambda^{2}_n  = (-1)^{\tfrac{d+1}{2}}\frac{(n)_d\,\cS_{d+1} }{2^{d+1}(d+1)\Gamma(d-1)}\,,
\end{equation}
where $(n)_d$ is the Pochhammer symbol $\Gamma(n+d)/\Gamma(n)$ and $\cS_{d+1}= 2\pi^{\tfrac{d}{2}+1}/\Gamma(\tfrac{d}{2}+1)$.

\section{Ward recurrence relation}\label{app:C}
The spherical harmonic decomposition \eqref{eq:lambdagen} followed from results of \cite{Drummond:1977uy, Giombi:2013yva, Giombi:2015haa} which were obtained by direct integration of the conformal two-point function against zonal harmonics. In this appendix, we aim to understand the resulting structure from a physical point of view, in analogy with the conformal two-point function in momentum space, for which it suffices to impose conformal Ward identities \cite{Bzowski:2019kwd}. The spherical harmonic decomposition of the two-point function on $S^d$ is likewise governed by a recurrence relation, equivalent to the $K$-type recurrence of the conformal intertwiner \cite{Dobrev:1977qv}. Below, we give a direct derivation starting from the conformal Ward identities. Since the $n$-dependence in \eqref{eq:lambdagen} is independent of the spin, we focus on the scalar case to illustrate the idea. 

Given a decomposition of the form
\begin{equation}
    \langle \CO(x) \CO(y) \rangle = \sum_{nlm} f_n \,Y^{nlm}(x)Y^{nlm}(y)\,,
\end{equation}
we want to determine the $f_n$. In stereographic coordinates, conformal invariance imposes
\begin{equation}\label{eq:confinv}
  \Omega^\Delta\,( x\cdot \partial_x)\, \Omega^{-\Delta} \,\langle \CO(x) \CO(y) \rangle = - 2 \Delta \,\langle \CO(x) \CO(y) \rangle\,, \quad \Omega(x) = \tfrac12(1+x^2)\,.
\end{equation}
We are free to pick $y=0$, so that only the zonal harmonics contribute. These are the following normalized Gegenbauer polynomials: 
\begin{equation}
    C^{\frac{d-1}{2}}_n(\sigma)/\sqrt{\cN_{n,d}}\,,\qquad\sigma=\cos\theta=\frac{1-x^2}{1+x^2}\,.
\end{equation}
In $d=2$, these become the Legendre $\mathsf{P}_n$. The normalization is given by
\begin{equation}
   \cN_{n,d} =  \frac{\Gamma(n+d-1)}{(2n+d-1)\Gamma(n+1)} = \frac{C^{\frac{d-1}{2}}_n(1)}{(2n+d-1)}\,.
\end{equation}
For now, only the $n$-dependence matters. The overall normalization is not fixed by \eqref{eq:confinv}, but can be found by considering the coincident limit, or by evaluating the constant mode integral. Rewriting and simplifying \eqref{eq:confinv} then yields
\begin{equation}\begin{split}
    &-\Delta \,\sigma\,\sum_n f_n (2n+d-1)C^{\frac{d-1}{2}}_n(\sigma) + \sum_n  (2n+d-1)f_n(1-\sigma^2)\partial_\sigma C^{\frac{d-1}{2}}_n(\sigma)\\
    &=   \Delta \sum_n (2n+d-1)f_n C^{\frac{d-1}{2}}_n(\sigma)\,.
\end{split}
\end{equation}
At this point, we can use the Gegenbauer recursion relations: 
\begin{align}
    (1-\sigma^2)\partial_\sigma C^{\frac{d-1}{2}}_n(\sigma) &= (n+d-2)C^{\frac{d-1}{2}}_{n-1}(\sigma)- n \sigma C^{\frac{d-1}{2}}_n(\sigma)\, ,\\
    (2n+d-1)\sigma\, C^{\frac{d-1}{2}}_n(\sigma) &=(n+1)C^{\frac{d-1}{2}}_{n+1}(\sigma) + (n+d-2)C^{\frac{d-1}{2}}_{n-1}(\sigma)\,\label{eq:recursion}.
\end{align}
Doing so leads to
\begin{equation}
\begin{split}
    &-\sum_n(n+\Delta) (n+1) f_n C^{\frac{d-1}{2}}_{n+1}(\sigma) + \sum_n  (n+d-1-\Delta)(n+d-2)f_n C^{\frac{d-1}{2}}_{n-1}(\sigma)\\
    &=   \Delta \sum_n (2n+d-1)f_n C^{\frac{d-1}{2}}_n(\sigma)\,.
\end{split}
\end{equation}
If we now compare the coefficients of each Gegenbauer $C^{\frac{d-1}{2}}_n(\sigma)$, we get the condition
\begin{equation}
 (n+d-\Delta)(n+d-1)f_{n+1} -n(n+\Delta-1)f_{n-1}  = \Delta(2n+d-1) f_n \,
\end{equation}
which has the following regular solution for $n\geq 0$:
\begin{equation}
    f_n = C\cdot \frac{\Gamma(n+\Delta)}{\Gamma(n+d-\Delta)}\,,
\end{equation}
as can be checked by noting that
\begin{equation}
    (n+\Delta)(n+d-1)-n(n+d-1-\Delta)=\Delta(2n+d-1)\,.
\end{equation}
This explains the structure of \eqref{eq:lambdagen}, at least for scalars. Equivalently, the recurrence is the compact-picture $K=\SO(d+1)$-type recurrence for the scalar conformal intertwiner \cite{Dobrev:1977qv}.

\section{Comments on the scalar three-point function}\label{app:3pt}
This appendix reviews the integrated scalar three-point function on $S^d$.
The corresponding conformal integral was evaluated long ago in
\cite{Drummond:1977dn}; we reproduce the derivation in conventions adapted to
the constant scalar deformations considered in the main text. We then give a harmonic
representation of the scalar three-point function on $S^3$, whose constant-source limit provides a check of the integrated result.

\subsection{Integrating against the constant mode}
Here we explain how to arrive at \eqref{eq:3ptint}. We want to calculate the following integral:
\begin{equation}
  \mathcal{I} =  \int_{S^d}\int_{S^d}\int_{S^d} \langle \CO(x)\CO(y)\CO(z)\rangle\, .
\end{equation}
Taking $z=0$, and writing $\Delta= d-\epsilon$, this becomes
\begin{equation}
\begin{split}
    \mathcal{I} &= 2^{3\epsilon-d}\,\frac{2\pi^{\frac{d+1}{2}}}{\Gamma(\frac{d+1}{2})}\,\int  \frac{\rmd^dx\,\rmd^dy}{(1+x^2)^\epsilon(1+y^2)^\epsilon|x-y|^{d-\epsilon}|x|^{d-\epsilon}|y|^{d-\epsilon}}\\
    &=2^{3\epsilon-d}\,\frac{2\pi^{\frac{d+1}{2}}}{\Gamma(\frac{d+1}{2})}\,\int  \frac{\rmd^du\,\rmd^dv}{(1+u^2)^\epsilon(1+v^2)^\epsilon(u^2+v^2-2u\cdot v)^{\frac{d-\epsilon}{2}}}\,.
    \end{split}
\end{equation}
In the second equality, we substituted $x=u^{-1}$ and $y=v^{-1}$. Next, we introduce Schwinger parameters $\tau_{1,2,3}$ for the respective denominators:
\begin{equation}
    \begin{split}
       \mathcal I= \frac{2^{3\epsilon+1-d}\pi^{\frac{d+1}{2}}}{\Gamma(\frac{d+1}{2})\Gamma(\epsilon)^2\Gamma(\frac{d-\epsilon}{2})}\,\int& \rmd^du\,\rmd^dv\, \rmd \tau_1\rmd\tau_2\rmd\tau_3 \,(\tau_1\tau_2)^{\epsilon-1}\tau_3^{\frac{d-\epsilon}{2}-1}\,\rme^{-\tau_1-\tau_2}\\&\, \cdot \exp{-\begin{pmatrix}
          u & v 
       \end{pmatrix}\cdot \begin{pmatrix}
         \tau_1+\tau_3  & -\tau_3\\ -\tau_3 &\tau_2+ \tau_3
       \end{pmatrix}\cdot \begin{pmatrix}
           u\\ v
       \end{pmatrix}}\,.
    \end{split}
\end{equation}
At this point, we perform the Gaussian integral:
\begin{equation}
       \mathcal I= \frac{2^{3\epsilon+1-d}\pi^{\frac{3d+1}{2}}}{\Gamma(\frac{d+1}{2})\Gamma(\epsilon)^2\Gamma(\frac{d-\epsilon}{2})}\,\int \rmd \tau_1\rmd\tau_2\rmd\tau_3 \,\frac{(\tau_1\tau_2)^{\epsilon-1}\tau_3^{\frac{d-\epsilon}{2}-1}\,\rme^{-\tau_1-\tau_2}}{(\tau_1\tau_2+\tau_3(\tau_1+\tau_2))^\frac{d}{2}}\,.
\end{equation}
Now we go to Feynman-like parameters: $\tau_1= \frac{\rho}{1+\sigma}, \tau_2 = \frac{\rho \sigma}{1+\sigma}, \tau_3 = \frac{\rho \sigma \tau}{(1+\sigma)^2}$ to find 
\begin{equation}
\begin{split}
      \mathcal{I} &= \frac{2^{3\epsilon+1-d}\pi^{\frac{3d+1}{2}}}{\Gamma(\frac{d+1}{2})\Gamma(\epsilon)^2\Gamma(\frac{d-\epsilon}{2})}\,\int \rmd \rho\rmd\sigma\rmd\tau \,\frac{\sigma^{\frac{\epsilon}{2}-1}\rho^{\frac{3\epsilon-d}{2}-1}\tau^{\frac{d-\epsilon}{2}-1}\,\rme^{-\rho}}{(1+\sigma)^{\epsilon}(1+\tau)^\frac{d}{2}}\,\\
       &=\frac{2^{3\epsilon+1-d}\pi^{\frac{3d+1}{2}}\Gamma(\frac{3\epsilon-d}{2})\Gamma(\frac{\epsilon}{2})^3}{\Gamma(\frac{d+1}{2})\Gamma(\frac{d}{2})\Gamma(\epsilon)^3}= \frac{8\pi^{\tfrac{3}{2}(d+1)}\Gamma(\frac{3\epsilon-d}{2})}{\Gamma(d)\Gamma(\frac{1+\epsilon}{2})^3}\,.
       \end{split}
\end{equation}
In the last step we used the Legendre duplication formula. Including the appropriate normalization $F^{(3)}(0)= - C_{\scriptscriptstyle{\CO\CO\CO}} \mathcal{I}$, we have arrived at \eqref{eq:3ptint}.

For completeness, we now allow the scalar primaries to have different
dimensions $\Delta_i$.  The resulting integral agrees with the general
conformal integral of \cite{Drummond:1977dn}.
\begin{equation}
  \mathcal I=  \int_{S^d}\int_{S^d}\int_{S^d} \langle \CO_{\Delta_1}(x)\CO_{\Delta_2}(y)\CO_{\Delta_3}(z)\rangle\, .
\end{equation}
Placing again one insertion at $z=0$, and writing $\Sigma = \Delta_1+\Delta_2+\Delta_3$, we get:
\begin{equation}
\begin{split}
    \mathcal{I} &= \frac{2^{1+2d-\Sigma}\pi^{\frac{d+1}{2}}}{\Gamma(\frac{d+1}{2})}\int \frac{\rmd^d x \rmd^d y}{(1+x^2)^{d-\Delta_1}(1+y^2)^{d-\Delta_2}|x|^{\Sigma-2\Delta_2}|y|^{\Sigma-2\Delta_1}|x-y|^{\Sigma-2\Delta_3}}\\
     &= \frac{2^{1+2d-\Sigma}\pi^{\frac{d+1}{2}}}{\Gamma(\frac{d+1}{2})}\int \frac{\rmd^d u \rmd^d v}{(1+u^2)^{d-\Delta_1}(1+v^2)^{d-\Delta_2}|u-v|^{\Sigma-2\Delta_3}}\,.
    \end{split}
\end{equation}
Next, we introduce the same Schwinger parameters as before and do the same substitutions to solve the integrals. The final result is
\begin{equation}
    \int_{S^d}\int_{S^d}\int_{S^d} \langle \CO_{\Delta_1}(x)\CO_{\Delta_2}(y)\CO_{\Delta_3}(z)\rangle = \frac{2^{1+2d - \Sigma}\pi^{\frac{3d+1}{2}}\Gamma\left(d-\frac{\Sigma}{2}\right)\prod_i \Gamma\left(\frac{d-\Sigma}{2}+\Delta_i\right)}{\Gamma(\frac{d+1}{2})\Gamma(\frac{d}{2})\prod_i \Gamma(d-\Delta_i)}\,.
\end{equation}
Note that the RHS has poles at particular scaling dimensions, e.g. when $\Sigma \in 2d + 2\mathbb{N}$. In such fine-tuned situations, which can arise in SCFTs, some care is warranted. However, in 3d $\mathcal{N}=2$ SCFTs, these problematic correlators seem to vanish, so that the poles are avoided.

\subsection{Decomposing into harmonics on \texorpdfstring{$S^3$}{}}

Below, we decompose the three-point function $\langle \CO_\Delta\CO_\Delta\CO_\Delta \rangle$ on $S^3$ into spherical harmonics. To begin, note that it is the product of two-point functions with half the scaling dimension:
\begin{equation}
  \langle
  \cO_\Delta(x)\cO_\Delta(y)\cO_\Delta(z)
  \rangle
  =
  C_{\cO\cO\cO}\,
  K_\delta(x,y)K_\delta(y,z)K_\delta(z,x),
  \qquad
  \delta=\frac{\Delta}{2}.
  \label{eq:OOO-factorization}
\end{equation}
The harmonic decomposition of the two-point function $K_\delta$ is given in \eqref{eq:S32pt0}: 
\begin{equation}
  K_\delta(x,y)  =\langle \CO_\delta(x)\CO_\delta(y)\rangle = \sum_P \lambda^\delta_P\,Y_P(x)\overline{Y_P(y)}.
  \label{eq:OOO-two-point-decomp}
\end{equation}
To avoid clutter, we condensed the notation to $P=(j_P,m_P,\tilde m_P)$, where $j_P=n_P/2$, and $\lambda^\delta_P=\lambda^\delta_{n_P}$. The bar denotes conjugation; in what follows we use   $\overline{Y_{j m\tilde m}}   =   (-1)^{m-\tilde m}Y_{j,-m,-\tilde m}$. The goal will be to project \eqref{eq:OOO-factorization} onto spherical harmonics and compute
\begin{align}
    \mathcal{I}_{ABC} &= \int\hspace{-0.25cm}\int\hspace{-0.25cm}\int_{S^3} Y_A(x) Y_B(y) Y_C(z)\,\langle\cO_\Delta(x)\cO_\Delta(y)\cO_\Delta(z) \rangle\\
    &= C_{\scriptscriptstyle{\CO\CO\CO}}\sum_{P,Q,R} \lambda^\delta_P \lambda^\delta_Q \lambda^\delta_R \int\hspace{-0.25cm}\int\hspace{-0.25cm}\int_{S^3} Y_A(x) Y_B(y) Y_C(z) \,Y_P(x)\overline{Y_P(y)} \,Y_Q(y)\overline{Y_Q(z)}\,Y_R(z)\overline{Y_R(x)}\nonumber
\,.
\end{align}
Now we insert the triple overlap of scalar harmonics \cite{Cutkosky:1983jd}:
\begin{equation}
  \cG_{ABC}
  \equiv
  \int_{S^3}Y_A Y_B Y_C
  =
  \gamma_{ABC}
  \begin{pmatrix}
    j_A & j_B & j_C \\ m_A & m_B & m_C
  \end{pmatrix}
  \begin{pmatrix}
    j_A & j_B & j_C \\ \tilde m_A & \tilde m_B & \tilde m_C
  \end{pmatrix},
  \label{eq:OOO-gaunt-factor}
\end{equation}
with
\begin{equation}
  \gamma_{ABC}
  =
  \frac{(-1)^{j_A+j_B+j_C}}{\sqrt{2}\pi}
  \sqrt{d_A d_B d_C}.
\end{equation}
The projection onto three external harmonics therefore gives
\begin{equation}
  \mathcal I_{ABC}
  =
  C_{\cO\cO\cO}
  \sum_{P,Q,R}
  \lambda^\delta_P\lambda^\delta_Q\lambda^\delta_R\,
  \cG_{A P\bar R}\,\cG_{B Q\bar P}\,\cG_{C R\bar Q},
  \label{eq:OOO-gaunt}
\end{equation}

The sums over the internal magnetic quantum numbers can now be done by applying the Racah recoupling identity to each of the two $\SU(2)$ factors, resulting in:
\begin{equation}
\begin{split}
&\sum_{\substack{m_P,m_Q,m_R\\
                 \tilde m_P,\tilde m_Q,\tilde m_R}}
(-1)^{m_R-\tilde m_R+m_P-\tilde m_P+m_Q-\tilde m_Q}
\\
&\quad\times
\begin{pmatrix}
  j_A & j_P & j_R \\ m_A & m_P & -m_R
\end{pmatrix}
\begin{pmatrix}
  j_B & j_Q & j_P \\ m_B & m_Q & -m_P
\end{pmatrix}
\begin{pmatrix}
  j_C & j_R & j_Q \\ m_C & m_R & -m_Q
\end{pmatrix}
\\
&\quad\times
\begin{pmatrix}
  j_A & j_P & j_R \\ \tilde m_A & \tilde m_P & -\tilde m_R
\end{pmatrix}
\begin{pmatrix}
  j_B & j_Q & j_P \\ \tilde m_B & \tilde m_Q & -\tilde m_P
\end{pmatrix}
\begin{pmatrix}
  j_C & j_R & j_Q \\ \tilde m_C & \tilde m_R & -\tilde m_Q
\end{pmatrix}
\\
&=
\begin{pmatrix}
  j_A & j_B & j_C \\ m_A & m_B & m_C
\end{pmatrix}
\begin{pmatrix}
  j_A & j_B & j_C \\ \tilde m_A & \tilde m_B & \tilde m_C
\end{pmatrix}
\begin{Bmatrix}
  j_A & j_B & j_C \\ j_Q & j_R & j_P
\end{Bmatrix}^{\!2}.
\end{split}
\label{eq:OOO-racah}
\end{equation}
Now we can plug this back into \eqref{eq:OOO-gaunt} to find
\begin{equation}
  \mathcal I_{ABC}
  =
  C_{\scriptscriptstyle{\cO\cO\cO}}\,\cG_{ABC}
  \sum_{j_P,j_Q,j_R}
  \lambda^\delta_P\lambda^\delta_Q\lambda^\delta_R\,
  \frac{\gamma_{A P R}\,\gamma_{B Q P}\,\gamma_{C R Q}}
       {\gamma_{A B C}}\,
  \begin{Bmatrix}
    j_A & j_B & j_C \\ j_Q & j_R & j_P
  \end{Bmatrix}^{\!2}.
  \label{eq:OOO-sixj}
\end{equation}
This form makes the selection rules transparent: the answer vanishes unless the external harmonics can couple to a singlet in both $\SU(2)$ factors,
\begin{equation}
    m_A+m_B+m_C=0,
  \qquad
  \tilde m_A+\tilde m_B+\tilde m_C=0,
  \qquad
  |j_A-j_B|\leq j_C\leq j_A+j_B,
\end{equation}
The remaining sums are over the internal spins $j_P,j_Q,j_R$, restricted by the admissibility conditions of the $6j$-symbol. In general we cannot simplify this further, although it should be noted that for fixed external harmonics, the internal sum can always be split into two finite sums and one infinite sum. 

Finally, consider $j_A=j_B=j_C=0$, for which the external harmonics are
constant.  Then the relevant $6j$-symbol is
\begin{equation}
    \begin{Bmatrix}
    0 & 0 & 0 \\ j_Q & j_R & j_P
    \end{Bmatrix}
    =
    \frac{(-1)^{2j_P}\delta_{j_Pj_Q}\delta_{j_Qj_R}}
    {\sqrt{2j_P+1}}\, .
\end{equation}
The phase drops out of \eqref{eq:OOO-sixj}, which collapses to a single sum. In earlier notation $n=2j_P$:
\begin{equation}\label{eq:3sum}
    \mathcal I_{000}=
  \sum_{n=0}^{\infty}(n+1)^2\bigl(\lambda^\delta_n\bigr)^3.
\end{equation}
The sum converges for $\delta < 1$.  With $a=1+\delta$ and $b=4-\delta$,
one may write the result as
\begin{align}
\mathcal I_{000}
&=
64\pi^3\sin^3\!\left(\frac{\pi\Delta}{2}\right)
\frac{\Gamma(2-\Delta)^3\Gamma(a)^3}{\Gamma(b)^3}
\Bigg[
2\,{}_4\mathsf F_3(2,a,a,a;b,b,b;1)
\nonumber\\
&\hspace{18mm}
 +{}_5\mathsf F_4(2,2,a,a,a;1,b,b,b;1)
\nonumber\\
&\hspace{18mm}
 +\left(\frac{b-1}{a-1}\right)^3
 {}_4\mathsf F_3(1,a-1,a-1,a-1;
 b-1,b-1,b-1;1)
\Bigg],
\label{eq:OOO-constant-hypergeometric}
\end{align}
where we remember that $\Delta=2\delta$ is the scaling dimension of $\CO$. Somewhat miraculously, this agrees with the more familiar closed form \eqref{eq:3ptint}, as can be verified numerically.

\section{Details on the regulated \texorpdfstring{$\langle T\CO\CO\rangle$}{} distribution}\label{app:TOOreg}

This appendix verifies that the regularization prescription \eqref{eq:too-kernel} defines a distribution that agrees with the standard conformal three-point function away from coincident points and satisfies the trace and diffeomorphism Ward identities. 

For convenience, we reproduce the regulated correlator:
\begin{align}\label{eq:too-kernelApp}
q^{2\Delta}\langle T_{ij}(w)\CO(0)\CO(z)\rangle ={}& \frac{3\Delta q}{8\pi}D_{ij}\!\left(\frac1{rs}\right)-\frac{\Delta q}{2\pi} \left(
\frac1sD_{ij}\!\left(\frac1r\right) +\frac1rD_{ij}\!\left(\frac1s\right)\right)
\nonumber\\
&-\frac{\Delta}{3}\delta_{ij} \left(\delta^{(3)}(w)+\delta^{(3)}(w-z)\right)\,,
\end{align}
where we introduced the following notation:
\begin{equation}
r=|w|\,, \qquad s=|w-z|\,, \qquad q=|z|\,, \qquad D_{ij} = \partial_i\partial_j- \frac13\delta_{ij}\partial^2\,.
\end{equation}
All derivatives act with respect to $w$.

\paragraph{Separated points}
It is useful to spell out how \eqref{eq:too-kernelApp} reduces to the usual separated three-point function.  Away from $w=0,z$ one has
\begin{equation}
    D_{ij}\!\left(\frac1r\right)     =\frac1r\left(    \frac{3w_i w_j}{r^4}-\frac{\delta_{ij}}{r^2}
    \right),     \qquad D_{ij}\!\left(\frac1s\right)=\frac1s\left(\frac{3(w-z)_i(w-z)_j}{s^4}
    -\frac{\delta_{ij}}{s^2}     \right).
    \label{eq:too-single-D}
\end{equation}
For the product, the traceless derivative is
\begin{align}
    D_{ij}\!\left(\frac1{rs}\right) =\frac1{rs}\Bigg[ &\frac{3w_iw_j}{r^4} +\frac{3(w-z)_i(w-z)_j}{s^4}
    +\frac{w_i(w-z)_j+w_j(w-z)_i}{r^2s^2}  \nonumber\\
    &-\delta_{ij}\left( \frac1{r^2}+\frac1{s^2}  +\frac23\frac{w\cdot(w-z)}{r^2s^2}    \right)
    \Bigg].
    \label{eq:too-product-D}
\end{align}
This simplifies upon combining \eqref{eq:too-single-D} and \eqref{eq:too-product-D},
\begin{equation}
    3D_{ij}\!\left(\frac1{rs}\right) -4\left (\frac1sD_{ij}\!\left(\frac1r\right)    +\frac1rD_{ij}\!\left(\frac1s\right) \right )=\frac1{rs}\left[-3X_iX_j+\delta_{ij}X^2 \right]\,,
    \label{eq:too-combine}
\end{equation}
where we have defined
\begin{equation}
  X_i=\frac{w_i}{r^2}-\frac{(w-z)_i}{s^2} \,, \qquad X^2=\frac{q^2}{r^2s^2}\,.
    \label{eq:too-X2}
\end{equation}
Then, the non-contact part at separated points gives the expected result \cite{Osborn:1993cr}:
\begin{equation}
    q^{2\Delta}\,\langle T_{ij}(w) \CO(0)\CO(z) \rangle  = -\frac{3\Delta q^3}{8\pi r^3s^3} \left(\frac{X_iX_j}{X^2}-\frac13\delta_{ij}\right).
\end{equation}
The overall normalization is fixed by the Ward identities, as we will review next.

\paragraph{Coincident limits and Ward identities}
Taking the trace of \eqref{eq:too-kernelApp} gives
\begin{equation}
\langle T(w) \CO(0) \CO(z)\rangle = - \frac{\Delta}{\;\; q^{2\Delta}}\left(\delta^{(3)}(w) + \delta^{(3)}(w-z)\right )\, ,
\label{eq:TOO-primary-trace}
\end{equation}
which is the Weyl Ward identity for scalar primaries of dimension $\Delta$. This already tells us we have chosen the correct contact terms. Next, we should check the divergence of \eqref{eq:too-kernelApp}, which is where the distributional regularization comes into play:
\begin{align}\label{eq:divergence}
    \partial^i\langle T_{ij}(w) \CO(0)\CO(z) \rangle ={}&\frac{3\Delta q}{8\pi q^{2\Delta}}\partial^i D_{ij}\!\left(\frac1{rs}\right) -\frac{\Delta q}{2\pi q^{2\Delta}} \partial^i \left(   \frac1sD_{ij}\!\left(\frac1r\right)     +\frac1rD_{ij}\!\left(\frac1s\right)
    \right)\nonumber\\
    &-\frac{\Delta}{3q^{2\Delta}}\partial_j  \left(\delta^{(3)}(w)+\delta^{(3)}(w-z)\right) \,.
\end{align}
Useful relations are
\begin{equation}\begin{split}
    \partial^i D_{ij} &= \frac23\,\partial_j\partial^2\, , \qquad \partial^2\,\frac1r = -4\pi\delta^{(3)}(w)\,,
\qquad \partial^2\,\frac1s = -4\pi\delta^{(3)}(w-z)\,,\\
&\partial^2\,\frac1{rs} = 2\,\frac{w\cdot(w-z)}{r^3s^3} -\frac{4\pi}{q} \left(\delta^{(3)}(w)+\delta^{(3)}(w-z)\right)\,.
\end{split}
\end{equation}
We also have the distributional rule 
\begin{equation}
    f(w)\,\partial_j\delta^{(3)}(w-a)  =
f(a)\,\partial_j\delta^{(3)}(w-a)- \partial_j f|_{w=a}\,\delta^{(3)}(w-a)\,.
\end{equation}
Using these identities one obtains
\begin{align}
  \partial^i D_{ij}\!\left(\frac1{rs}\right) &= \frac{4}{3}\partial_j\,\frac{w\cdot(w-z)}{r^3s^3} -\frac{8\pi}{3q}\partial_j \left(\delta^{(3)}(w)+\delta^{(3)}(w-z)\right) \\
\frac1s \partial^i D_{ij}\!\left(\frac1r\right)     +\frac1r \partial^i D_{ij}\!\left(\frac1s\right)&= \partial_j\frac{w\cdot(w-z)}{r^3s^3}- \frac{8\pi}{3q}\partial_j\left(\delta^{(3)}(w)+\delta^{(3)}(w-z)\right)\\
&\qquad +\frac{4\pi z_j}{ q^3}\left(\delta^{(3)}(w)-\delta^{(3)}(w-z)\right)\,.\nonumber
\end{align}
Plugging these back in \eqref{eq:divergence} one is left with
\begin{equation}
    \partial^i \langle T_{ij}(w)\CO(0)\CO(z)\rangle
= \left( -\delta^{(3)}(w) + \delta^{(3)}(w-z) \right)\frac{2\Delta z_j}{\;q^{2\Delta+2}}\,,
\end{equation}
which agrees with the correct Ward identity:
\begin{equation}
    \partial^i\langle T_{ij}(w)\,\CO(0)\CO(z)\rangle
=\left (\delta^{(3)}(w)-\delta^{(3)}(w-z)\right )\,\partial_j \langle \mathcal O(0)\,\mathcal O(z)\rangle\,.
\end{equation}

\section{CFTs on \texorpdfstring{$\mathbb{R} \times S^2$}{}}\label{app:RxS2}
In this appendix, we apply the harmonic-space method to static deformations of a three-dimensional CFT on $\mathbb{R}\times S^2$. We first consider scalar sources, and then  deformations of the spatial sphere, for which we also determine the first-order shift of the scalar energy gap.

Concretely, we will consider $\mathbb{R}\times S^2$, with metric
\begin{equation}
    \rmd s^2 = \rmd \tau^2 + \rmd \Omega^2_2 = \Omega^{-2}(r)(\rmd r^2 + r^2 \rmd \Omega^2_2)\, , \qquad \Omega(r)=r=\rme^\tau\,.
\end{equation}

For a given CFT, we can calculate the free energy density. Since $\mathbb{R}\times S^2$ is $S_\beta^1\times S^2$ with $\beta \to \infty$, this computes the vacuum energy:
\begin{equation}
    E_{\text{vac}} = - \lim_{\beta\to \infty}\partial_\beta \log \CZ\, .
\end{equation}
The vacuum energy changes as follows under $\tau$-independent deformations $\cL + j \cO$:
\begin{equation}\label{eq:dEvac2}
    E''_{\text{vac}}(0)= -\int \rmd \tau \int_{S^2} \int_{S^2}j(\Omega_1)\cdot\langle \CO(\tau,\Omega_1)\CO(0,\Omega_2)\rangle\cdot j(\Omega_2)\,.
\end{equation}
Our approach will be to decompose $j$ in $S^2$ spherical harmonics $j= \sum a_{lm}Y_{lm}$ and to do the same for the two-point kernel. This will provide a purely CFT derivation of the holographic result of \cite{Fischetti:2017sut}, where global properties were also studied. For related discussions, see \cite{Fischetti:2018shp, Fischetti:2020knf}, as well as \cite{Cheamsawat:2020awh}, who found a striking quantitative agreement between the fermionic and
holographic results for finite deformations.

\subsection{Scalar deformation}
The conformal two-point function on $\mathbb{R}\times S^2$ takes the form
\begin{equation}\label{eq:2ptRS2}
    \langle \CO(\tau_1, \Omega_1) \CO(\tau_2, \Omega_2)\rangle = \frac{r_1^\Delta\, r_2^\Delta}{|\mathbf{r}_1-\mathbf{r}_2|^{2\Delta}} = \frac{1}{2^\Delta|\cosh(\tau_1-\tau_2) -  \Omega_1\cdot\Omega_2|^{\Delta}}\,.
\end{equation}
To decompose this kernel in spherical harmonics, we should integrate against Legendre polynomials. Using 7.228 and 7.132.3 of \cite{Gradshteyn:1702455} we find the following eigenvalues:\footnote{To check numerically, one should note that $Q_\nu^\mu(x)$ with $x>1$ in \cite{Gradshteyn:1702455} corresponds to $\mathtt{LegendreQ[\nu,\mu,3,x]}$ in Mathematica. Instead of 7.132.3 one could also use 7.151.2, which is in principle equivalent,  but appears to have a typo: the factor $2^{\alpha-\mu}$ should be $2^{2-\mu}$.}
\begin{equation}\label{eq:lambdaRS2}
   \lambda^\Delta_l = \int^\infty_{-\infty} \rmd \tau \int^1_{-1}\rmd x\; \frac{\mathsf{P}_l(x)}{2^\Delta(\cosh\tau -x)^\Delta} = \frac{\Gamma(\frac{1}{2})\Gamma(\frac{3}{2}-\Delta)}{2\Gamma(\Delta)}\frac{\Gamma(\frac{l+\Delta}2)^2}{\Gamma(\frac{3+l-\Delta}{2})^2}\, ,
\end{equation}
with $\Omega_1\cdot \Omega_2 = \cos\theta= x$, leading to the following decomposition:
\begin{equation}
   \int \rmd \tau\, \langle \CO(\tau,\Omega_1) \CO(0,\Omega_2)\,\rangle  = 2\pi \sum_{lm}\lambda^\Delta_l\, Y_{lm}(\Omega_1)\bar{Y}_{lm}(\Omega_2) \,.
\end{equation}
This trivializes \eqref{eq:dEvac2}; we find
\begin{equation}
    E''_{\text{vac}}(0) = -2\pi\sum_{lm} \, \lambda^\Delta_l |a_{lm}|^2\,.
\end{equation}
Note that $\lambda^\Delta_l>0$ for $\Delta=1$ or $\Delta= 3-\epsilon$, which thus leads to a decrease in $E_{\text{vac}}$.
Incidentally, integrating the 2-point function in $\tau$ first, yields an Appell $\text{F}_1$ function:
\begin{equation}
    \int \rmd \tau\, \langle \CO(\tau, \Omega_1) \CO(0, \Omega_2)\rangle  = \frac{2}{\Delta}\text{F}_1(\Delta,\Delta,\Delta,\Delta+1;\rme^{\rmi \theta}, \rme^{-\rmi\theta})\,
\end{equation}
Integrating this against Legendre polynomials would seem like a tougher problem, but now we know the result \eqref{eq:lambdaRS2} from the opposite order of integration.

\subsection{Metric deformation}
We follow the same strategy for metric deformations of the form
\begin{equation}\label{eq:S2met}
    \rmd s^2 = \rmd \tau^2 + (1+\chi(\Omega))\rmd \Omega_2^2, \quad \chi(\Omega)= \sum_{lm} \chi_{lm}Y_{lm}(\Omega)\,.
\end{equation}
Under a general $\tau$-independent squashing of $S^2$:
\begin{equation}
    E''_{\text{vac}}(0) = -\frac14 \int \rmd \tau \int_{S^2}\int_{S^2} h^{\mu\nu}(\Omega_1)\cdot\langle T_{\mu\nu}(\tau,\Omega_1)T_{\rho\sigma}(0,\Omega_2)\rangle \cdot h^{\rho \sigma}(\Omega_2)\,.
\end{equation}
For the particular deformation in \eqref{eq:S2met}, we can use $T^\mu_\mu = 0$ to reduce this to
\begin{equation}
    E''_{\text{vac}}(0) = -\frac14 \int \rmd \tau \int_{S^2}\int_{S^2} \chi(\Omega_1)\cdot\langle T_{\tau\tau}(\tau,\Omega_1)T_{\tau\tau}(0,\Omega_2)\rangle \cdot \chi(\Omega_2)\,.
\end{equation}
From the general expression of the stress-tensor two-point function, we find
\begin{equation}
    \langle T_{\tau\tau}(\tau,\Omega_1)T_{\tau\tau}(0,\Omega_2)\rangle = c_T\cdot\frac{(1-\Omega_1\cdot\Omega_2\cosh\tau)^2-\tfrac13 (\cosh\tau-\Omega_1\cdot\Omega_2)^2}{8(\cosh\tau-\Omega_1\cdot\Omega_2)^5}\,.
\end{equation}
Hence, if we can calculate, with $z=\cosh\tau$, the following eigenvalue
  \begin{equation}\label{eq:lambda_intS2}
   \lambda_l = 2\int^\infty_{1} \rmd z\, (z^2-1)^{-\frac12}\int^1_{-1}\rmd x \,\mathsf{P}_l(x)\,\frac{(1-xz)^2-\frac13(z-x)^2}{8(z-x)^5}\, ,
\end{equation}
with $\mathsf{P}_l(x)$ a Legendre polynomial, then the change in vacuum energy will be given as
\begin{equation}\label{eq:Evacc}
     E''_{\text{vac}}(0) = -\frac{\pi c_T}{2}  \sum_{lm} \, \lambda_l |\chi_{lm}|^2\,.
\end{equation}
To evaluate \eqref{eq:lambda_intS2} we write it as a sum of terms $I_l(\alpha,\Delta)$ that can be evaluated using 7.228 and 7.132.3 in \cite{Gradshteyn:1702455}:
\begin{equation}\label{eq:defI}\begin{split}
   I_l(\alpha, \Delta) &\equiv  \int^\infty_{1} \rmd z\, (z^2-1)^{\alpha-\frac12}\int^1_{-1}\rmd x \,\frac{\mathsf{P}_l(x)}{(z-x)^\Delta}\\& = \frac{2^{\Delta -2} \Gamma \left(\alpha +\frac{1}{2}\right) \Gamma \left(\alpha -\Delta +\frac{3}{2}\right) \Gamma \left(\frac{l+\Delta }{2}\right) \Gamma \left(\frac{l+\Delta }{2}-\alpha\right)}{\Gamma (\Delta ) \Gamma \left(\frac{1}{2} (l-\Delta +3)\right) \Gamma \left(\frac{1}{2} (l -\Delta +3)+\alpha\right)}.
   \end{split}
\end{equation}
To this end, we expand $(1-zx)^2 = (z-x)^2 + (1-z^2)(1-x^2)$ and use
\begin{equation}
    \frac{1-x^2}{(z-x)^5}= \frac{1-z^2}{(z-x)^5}+ \frac{2x}{(z-x)^4}+\frac{1}{(z-x)^3}\,.
\end{equation}
For the linear piece in $x$ we use the recurrence relation 
$ (2l+1)x\, \mathsf{P}_l(x) =(l+1)\mathsf{P}_{l+1}(x) + l\mathsf{P}_{l-1}(x)$. Every resulting term of $\lambda_l$ in \eqref{eq:lambda_intS2} is now of the form $I_l(\alpha, \Delta)$:
\begin{equation}
    \lambda_l = \frac{1}{6} I_l(0,3) + \frac{1}{4} I_l(2,5) - \frac{1}{4} I_l(1,3) -\frac{l}{4l+2} I_{l-1}(1,4) - \frac{l+1}{4l+2} I_{l+1}(1,4)\, .
\end{equation}
Performing the sum leads to some remarkable simplifications:
\begin{equation}
    \lambda_l = \frac{\pi}{48}(l-1)(l+1)(l+2)\frac{\Gamma(\frac{l+1}{2})^2}{l\,\Gamma(\frac{l}{2})^2}\,.
\end{equation}
This result agrees with (3.39) of \cite{Fischetti:2017sut}, who obtained it from a holographic calculation.\footnote{It is also consistent with the pure-AdS polar Love number of \cite{Franzin:2024cah}, see in particular (3.9) and (5.7).} Note that $\lambda_l >0$ for $l>1$ and vanishes for $l=0$ (rescaling $S^2$) or $l=1$ (diffeomorphism generated by a CKV). This establishes that $E''_{\text{vac}}(0)$ as determined by \eqref{eq:Evacc} is negative for unitary CFTs and thus the squashing considered in \eqref{eq:S2met} decreases the vacuum energy. Note also that $\lambda_l \sim l^3$ for large $l$, which is the expected UV scaling in $d=3$.

\paragraph{Influence of metric deformation on scalar gap}
We can also ask whether the $\tau$-independent metric deformation \eqref{eq:S2met} changes the lowest energy appearing in a scalar two-point function. Concretely, we look at the late-time exponent $\omega_{\min}$ defined by
\begin{equation}
  \langle \CO(T\to\infty,\Omega)\CO(0)\rangle_\chi
  \sim \rme^{-\omega_{\min}T}\,.
\end{equation}
The first-order correction in $\chi$ to $\omega_{\min}$ is
\begin{equation}\begin{split}
  \delta_\chi \omega_{\min} &= -\lim_{T\to \infty} \frac{1}{T} \delta_\chi \log \langle \CO(T, \Omega)\CO(0) \rangle  \\
  &=  -\lim_{T\to \infty} \frac{1}{2T}   \int \rmd\tau\,\rmd\Omega'\,   \chi(\Omega')\, \langle T_{\tau\tau}(\tau,\Omega')\CO(T,\Omega)\CO(0)\rangle\, \\
  &=  -\frac{1}{2}  \int \,\rmd\Omega'\,   \chi(\Omega')\, \langle \CO|T_{\tau\tau}(\Omega')|\CO\rangle\,  = -\frac{\Delta}{2}\,
  \frac{1}{4\pi}\int_{S^2}\rmd\Omega\,\chi(\Omega).
\end{split}
\end{equation}
The second line uses that contact terms can be dropped in the $T\to \infty$ limit --- the change in $\omega_{\min}$ must come from inserting $T_{\tau\tau}$ in the long middle region between the operators $\CO$. In the third line we used that in the large-$T$ limit the three-point function becomes a matrix element in a scalar primary state. This state is rotationally invariant, so the one-point function is independent of $\Omega$.  The integral over the sphere is then fixed by the Hamiltonian,
\begin{equation}
  \int_{S^2}\rmd\Omega\,T_{\tau\tau}=H,
  \qquad
  H|\CO\rangle=\Delta|\CO\rangle\,.
\end{equation}
Hence, to first order only the constant mode affects $\omega_{\min}$. It simply changes the radius of the sphere to $R=(1+\chi)^{1/2}$, and changes $\omega_{\min}=\Delta/R$ accordingly. The conclusion is not new; a direct evaluation of the integrated three-point function can be found in section 5.3 of \cite{Hickling:2016mzp}.

\section{AdS-Kerr-Bolt}\label{app:AdSkerrbolt}
In this appendix, we consider the analytically known Plebanski-Demianski solutions as a further testing ground of the thermal EFT formalism \cite{Plebanski:1976gy}. We are specifically interested in stationary solutions with both rotation and NUT charge. We will derive their on-shell action and compare to the thermal EFT prediction. 

Starting from the metric as written in (17) in \cite{Griffiths:2005qp}, we put the acceleration to zero and consider vanishing electric and magnetic charges $e=g=0$. Finally, we go to Euclidean signature by taking $t = -\rmi t_{\scriptscriptstyle{E}}$ as well as $ a = \rmi \alpha$ and $\ell = \rmi n$. In the end, we are left with:
\begin{equation}\label{eq:KBsol}
  \rmd s^2= \frac{\mathcal{Q}}{\rho^2} \left(\rmd t_{\scriptscriptstyle{E}} - \mathcal{R}\rmd \varphi\right)^2+\frac{\rho^2}{\mathcal{Q}}\rmd r^2+ \frac{\rho^2}{\mathcal{P}}\rmd \vartheta^2 + \sin^2\vartheta \frac{\mathcal{P}}{\rho^2}\left(\alpha \rmd t_{\scriptscriptstyle{E}} + \left(r^2-(\alpha + n)^2\right)\rmd \varphi\right)^2\, ,
\end{equation}
where $\mathcal P$, $\mathcal R$ and $\rho$ are given by
\begin{equation}\label{eq:PBbuild}
    \mathcal{P}= 1+ 4\alpha n \cos\vartheta+\alpha^2\cos^2\vartheta\, , \quad \mathcal{R}= \alpha \sin^2\vartheta +4n \sin^2\tfrac{\vartheta}{2}\, , \quad \rho^2= r^2 - (n+\alpha\cos\vartheta)^2\, , 
\end{equation}
while $\mathcal Q$ takes the form
\begin{equation}
    \mathcal{Q} = r^4 + r^2(1-\alpha^2-6n^2)-2mr +(n^2-\alpha^2)(1-3n^2)\,.
\end{equation}
At $n=0$, this reduces to AdS-Kerr \eqref{eq:kerr} up to a rescaling of $\varphi$ by $\Xi = 1+\alpha^2$. At $\alpha=0$, it becomes the spherical AdS-Taub-NUT/Bolt solution \eqref{eq:spherKB}, after writing $t_{\scriptscriptstyle E}=2n(\psi+\varphi)$. 

Let us now zoom in on the horizon at $r=r_+$. By going to corotating coordinates 
\begin{equation}\label{eq:KBrot}
    \rmd \varphi = \rmd \phi - \Omega'\,\rmd t_{\scriptscriptstyle{E}}  \,, \qquad \Omega' = \frac{\alpha }{r^2_+ - (\alpha + n)^2}\, , 
\end{equation}
and focusing on the contracting Euclidean time circle, we find:
\begin{equation}\begin{split}
    \rmd s^2 \to\;\, &\frac{\mathcal{Q}'_+ (r-r_+)}{\rho^2_+}\left(1+\frac{\alpha(\alpha\sin^2\vartheta + 4n \sin^2\tfrac{\vartheta}{2})}{r^2_+ - (n+\alpha)^2}\right)^2 \rmd t^2_{\scriptscriptstyle{E}} +  \frac{\rho^2_+}{\mathcal{Q}'_+(r-r_+)}\,\rmd r^2\\
    & = \rho^2_+ \left (\frac{\mathcal{Q}'_+ (r-r_+)}{(r^2_+ - (n+\alpha)^2)^2} \rmd t^2_{\scriptscriptstyle{E}} +  \frac{\rmd r^2}{\mathcal{Q}'_+(r-r_+)}\right)\\
    &=\rho^2_+ \left (\frac{\mathcal{Q}'^2_+ x^2}{4(r^2_+ - (n+\alpha)^2)^2} \rmd t^2_{\scriptscriptstyle{E}} +  \rmd x^2 \right)\,.
    \end{split}
\end{equation}
In the last step we defined $x^2 = 4(r-r_+)/\mathcal{Q}'_+$.  The inverse temperature is therefore
\begin{equation}\label{eq:KBtemp}
    \beta = \Delta t_{\scriptscriptstyle{E}} = \frac{4\pi(r^2_+ - (n+\alpha)^2)}{\mathcal{Q}'_+}\,.
\end{equation} 
Despite appearances in intermediate steps, the $\vartheta$-dependence ultimately dropped out. 

The reader may have noticed that the metric \eqref{eq:KBsol} has conical singularities at the north and south poles of the $S^2$, corresponding to $\vartheta=0$ and $\pi$. We postpone a related discussion of global regularity to appendix~\ref{sec:spindle} --- for the purpose of verifying the thermal EFT expansion~\eqref{eq:genEFT} for holographic CFTs, a local analysis is sufficient.

\subsection{Calculating the on-shell action}
We follow the method of \cite{Emparan:1999pm} to calculate the on-shell action of the AdS-Kerr-NUT/Bolt solution \eqref{eq:KBsol}. Setting $G=1$, we find the following contributions from the bulk action
\begin{equation}
    I_{\text{bulk}} = -\frac{1}{16\pi }\int \rmd^4 x \sqrt{g}(R+6)= \frac{\beta \Delta\phi}{4\pi }\left(r^3-r^3_+ - (r-r_+)(3n^2 + \alpha^2)\right)\,,
\end{equation}
the GHY boundary action, with $K = h^{ij}\nabla_i n_j$, is
\begin{equation}
    I_{\text{surf}} =-\frac{1}{8\pi }\int \rmd^3x \sqrt{h} K = -\frac{\beta \Delta \phi}{4\pi }\left(3r^3+r\left(2-11n^2-\tfrac53 \alpha^2\right)-3m\right)\,, 
\end{equation}
and the counterterm action (the $R^2_h$ terms do not contribute in $4d$ since $\sqrt{h}R^2 \,\propto\, 1/r$)
\begin{equation}
    I_{\text{ct}} = \frac{1}{8\pi }\int \rmd^3 x\sqrt{h}\left(2+ \frac{R_h}{2}\right)  = \frac{\beta \Delta \phi}{4\pi }\left(2r^3 + r\left(2-8n^2-\tfrac23 \alpha^2\right)-2m \right)\,.
\end{equation}
Adding them up leaves us with a finite on-shell action
\begin{equation}\label{eq:KBonsh}
    I =\frac{\beta \Delta \phi}{4\pi }\left(m +r_+\left(3n^2+\alpha^2\right) -r^3_+ \right) \,.
\end{equation}
It reduces to the Taub-Bolt/NUT action \eqref{eq:taubon} at $\alpha=0$ and the Kerr action \eqref{eq:kerrOS} at $n=0$.

\paragraph{The on-shell action at large temperature} We will expand the on-shell action at small $\beta$ and fixed $\alpha, n$. To do so, we first use the horizon condition $\cQ_+ = 0$ to solve for $m$. Then we invert the temperature relation \eqref{eq:KBtemp} to solve for $r_+$ in a small $\beta$ expansion:
\begin{align}
    r_+ = &\frac{4 \pi }{3 \beta }-\frac{\beta}{4\pi}  \left(2 \alpha ^2-3 n^2+6 \alpha  n+1\right)\\&-\frac{3\beta ^3}{64\pi^3}  \left(10( \alpha ^4+ \alpha ^2)+9 n^3(n-4 \alpha  )+6 \left(8 \alpha ^2-1\right) n^2+18 \left(3 \alpha ^3+\alpha \right) n+1\right)+\dots\,.\nonumber
    \end{align}
For our purposes, the expansion above must be done up to order $\beta^5$. Then, plugging all of this back into \eqref{eq:KBonsh}, we obtain the on-shell action at small $\beta$:
\begin{equation}\begin{split}\label{eq:KBexp}
   I/\Delta\phi  = &-\frac{8 \pi ^2}{27 \beta ^2}+\frac{1}{6} \left(3 \alpha ^2-3 n^2+6 \alpha  n+2\right)\\
   &+\frac{\beta ^2}{32\pi^2} \left(4 \alpha ^4-9 n^4+6 \left(6 \alpha ^2+1\right) n^2+24 \alpha ^3 n-1\right)\\&  +\frac{\beta ^4}{256\pi^4} \Big(52 \alpha ^6+24 \alpha ^4 \left(45 n^2+2\right)+216 \alpha ^3 n \left(3 n^2+1\right)\\
   &\qquad\qquad\;+108 \alpha ^2 n^2 \left(2-3 n^2\right)+\left(3 n^2-1\right)^3+432 \alpha ^5 n\Big)\,.
\end{split}\end{equation}
A few comments are in order. First, in the spherical Taub-Bolt we have $\alpha=0$ and a further condition $\beta= 8\pi n$ is imposed by regularity of the boundary three-manifold. Each term involving $n$ in the above expression then ends up being more subleading in the small-$\beta$ expansion. Second, compared to the Kerr expansion \eqref{eq:kerrIexp}, we are now expanding at fixed $\alpha$ rather than at fixed $\Omega=\alpha+\Omega'$. Using $\Delta\phi= 2\pi/(1+\alpha^2)$ in combination with the relation $\alpha(\Omega,\beta)$ obtained by inverting \eqref{eq:Omegadef}, the above expansion at $n=0$ is mapped to \eqref{eq:kerrIexp}.

\subsection{Thermal EFT expansion}
Now we compare \eqref{eq:KBexp} to the thermal EFT expansion \eqref{eq:genEFT}. To do so requires the Weyl-invariant geometric data on the two-dimensional base. Using \eqref{eq:KBrot}, the KK metric \eqref{eq:KKmet} at the boundary takes the following form:
\begin{equation}
    \rmd s^2 = A\rmd \theta^2 + B\rmd \tau^2 -2 C\rmd \tau \rmd \phi + D \rmd \phi^2 = B\left(\rmd \tau - \tfrac{C}{ B}\rmd \phi\right)^2+A\rmd\theta^2+\left(D-\tfrac{C^2}{B}\right)\rmd\phi^2\,,
\end{equation}
where $A,B,C,D$ are determined in terms of the metric building blocks \eqref{eq:PBbuild}:
\begin{equation}\begin{split}
A &= \cP^{-1}\, ,\quad B = \beta^2\Omega'^2 \cP \sin ^2\theta  +\beta^2\left(\Omega'  \mathcal{R}+1\right)^2\,,\\&C=\beta\Omega'\cP  \sin ^2(\theta ) +\beta\mathcal{R} \left(\Omega'  \mathcal{R}+1\right)
\,,\quad   D= \cP\sin ^2\theta  +\mathcal{R}^2\,.
\end{split}
\end{equation}
Since we are expanding at fixed $\alpha,n$, each interaction term in the EFT now contributes not just at a single order in $\beta$ but at all subleading orders too. This is due to the implicit $\beta$-dependence in $\Omega'$. At $n=0$, this effect is removed by switching to a fixed $\Omega= \alpha +\Omega'$ expansion. For general $\alpha, n$, the volume term can be evaluated exactly:
\begin{align}
   \frac{1}{\Delta\phi} \int \sqrt{\hat g} &= \frac{1}{\beta^2}\frac{4 n \Omega +2}{(4 n \Omega +1) (\Omega  (2 \alpha +\alpha  \Omega  (\alpha +4 n)+4 n+\Omega )+1)}\\&\approx \frac{2}{\beta ^2}-\frac{9 \alpha  (\alpha +3 n)}{4 \pi ^2}-\frac{27 \alpha  \beta ^2 \left(\alpha  \left(5 \alpha ^2+7\right)-18 n^3+6 \alpha  n^2+6 \left(5 \alpha ^2+2\right) n\right)}{128 \pi ^4}.\nonumber
    \end{align}
The other contributions can be found as a small-$\beta$ expansion. For instance,
\begin{equation}\begin{split}
   \frac{1}{\Delta\phi} \int \sqrt{\hat g}F^2 &= \frac{16}{3} \left(\alpha ^2+3 n^2\right)-\frac{6 \alpha  \beta ^2 \left(-\alpha +3 n^3-\alpha  n^2+\alpha ^2 n\right)}{\pi ^2} +\dots\,,\\
   \frac{1}{\Delta\phi}\int \sqrt{\hat g}R&=4 \left(\alpha ^2+1\right)-\frac{9 \alpha  \beta ^2 n \left(\alpha ^2-4 \alpha  n+1\right)}{2 \pi ^2}+\dots\,.
   \end{split}
\end{equation}
As an example of derivative terms, we have
\begin{equation}\begin{split}
    \frac{1}{\Delta\phi}\int \sqrt{\hat g}(\nabla F)^2=&\frac{32 \beta ^2}{15}\alpha ^2  (\alpha ^2+5)+\frac{12 \alpha ^2 \beta ^4 }{5 \pi ^2}\big(\alpha ^4+2 \alpha ^2\\&+2 \left(13 \alpha ^2-5\right) n^2+\left(\alpha ^2+5\right) \alpha  n+5\big)+\dots\, ,
    \end{split}
\end{equation}
as well as
\begin{align}
    \frac{1}{\Delta\phi}\int \sqrt{\hat g}(\nabla R)^2 &= \frac{768\beta^4}{35} \alpha ^2 \left(\alpha ^2 \left(3 \alpha ^2+7\right)+7 \left(9 \alpha ^2+5\right) n^2\right)+\dots\,,\\
    \frac{1}{\Delta\phi}\int \sqrt{\hat g}R(\nabla F)^2 &=\frac{64\beta^4}{105} \alpha ^2 \left(11 \alpha ^4+14 \alpha ^2 \left(24 n^2+1\right)+35\right)+\dots\,.
\end{align}
The other terms can be evaluated similarly. Comparing the resulting EFT expansion to the on-shell action \eqref{eq:KBexp} imposes constraints on the EFT coefficients. Crucially, they are precisely those obtained before from hyperbolic Taub-Bolt (for the $c_{n,m}$, see \eqref{eq:cnmlist}) and Kerr (for the $d_i$, see \eqref{eq:diKerr}), which provides a further consistency check. 

One might have hoped that the Kerr-Bolt metric would allow us to disentangle $(\nabla R)^2$ from $F^2(\nabla F)^2$  and $\nabla R \nabla F^2$ contributions, as well as $R(\nabla F)^2$ from $(\nabla\nabla F)^2$. However, just like in Kerr, their contributions end up being proportional. 


\subsection{Smooth Seifert boundaries over a spindle}\label{sec:spindle}
We will now discuss global regularity properties of the metric \eqref{eq:KBsol}.
Let us begin by zooming in near the poles of the $S^2$. Defining
\begin{equation}
\mathcal P_\pm=1\pm4\alpha n+\alpha^2\,, \qquad
\rho_\pm^2=r^2-(n\pm\alpha)^2\,,
\end{equation}
the transverse metric near $\vartheta=0$, with shrinking generator $\partial_\varphi$, is
\begin{equation}
\rmd s_\perp^2\simeq \frac{\rho_+^2}{\mathcal P_+}
\left(\rmd\vartheta^2+\mathcal P_+^2\vartheta^2\rmd\varphi^2\right)\,.
\end{equation}
Near $\vartheta=\pi$ it is instead $\partial_\varphi+4n\partial_{t_{\scriptscriptstyle E}}$ which shrinks. Introducing $t_-=t_{\scriptscriptstyle E}-4n\varphi$, and setting $\epsilon=\pi-\vartheta$, the transverse metric at fixed $t_-$ becomes
\begin{equation}
\rmd s_\perp^2\simeq \frac{\rho_-^2}{\mathcal P_-} 
\left(\rmd\epsilon^2+\mathcal P_-^2\epsilon^2\rmd\varphi^2\right)\,.
\end{equation}
Global regularity requires a compatible lattice of $(\Delta t_{\scriptscriptstyle E},\Delta \varphi)$ identifications in the two commuting Killing directions. From the analysis above, we conclude that the locally smooth axial cycles have the following period vectors:
\begin{equation}
N=\left(0\,,\;\frac{2\pi}{\mathcal P_+}\right)\,,
\qquad
S=\left(\frac{8\pi n}{\mathcal P_-}\,,\;
\frac{2\pi}{\mathcal P_-}\right)\,.
\end{equation}
Similarly, let $H$ denote the thermal circle, with period vector
\begin{equation}
H=(\beta\,,\;-\beta\Omega')\,.
\end{equation}
A smooth three-manifold with a locally free thermal-circle action is obtained when the lattice of identifications is generated by $H$ and a second vector $A$, such that
\begin{equation}
N=q_N A-p_N H\,,
\qquad
S=q_S A-p_S H\,,
\end{equation}
for integers $q_N,q_S>0$ and $p_N,p_S$ satisfying $\gcd(p_N,q_N)=\gcd(p_S,q_S)=1$. 

The most familiar examples have $q_N = q_S = 1$. Let us start with Kerr: at $n=0$, we can take $A$ such that $N=S=A=(0,\Delta\varphi)$, with $\Delta \varphi = 2\pi/(1+\alpha^2)$, so that $p_N=p_S=0$. The resulting Seifert fibration is trivial. For Taub-Bolt, we have instead $\alpha=0$, so that $N=(0,2\pi)$ and $S=(8\pi n, 2\pi)$. We take $A=N$, so that $p_N=0$. Since $H=(\beta, 0)$, regularity requires $\beta= 8\pi n/p$, where $p=-p_S$ labels the Lens space $S^3/\mathbb{Z}_p$.

In general, the integers $q_N$ and $q_S$ are the orders of the exceptional fibers at the two poles. Near an exceptional fiber of order $q$, the total space is locally
\begin{equation}
(D^2\times S^1)/\mathbb Z_q\,,
\qquad
(z,\tau)\sim
\left(\rme^{2\pi \rmi/q}z,\tau+\frac{2\pi p}{q}\right),
\qquad
\gcd(p,q)=1\,.
\end{equation}
Although the quotient of the transverse disc is a cone, the combined action on $D^2\times S^1$ is free, and hence the total space is a smooth three-dimensional manifold --- a Seifert manifold. The quotient by the thermal circle is known as a spindle orbifold \cite{Ferrero:2020laf, Crisafio:2024fyc}:
\begin{equation}
\mathcal M_3/S^1_H = \mathbb{WCP}^1_{[q_N,q_S]}\, .
\end{equation}

To connect back to our thermal EFT discussion, note that in corotating coordinates $\phi=\varphi+\Omega' t_{\scriptscriptstyle E}$, the cycles become $H=(\beta,0)$ and $A=(\delta,\Delta\phi)$. This determines the angular period $\Delta\phi$ on the orbifold base, which is the quantity appearing in the on-shell action \eqref{eq:KBonsh}.

\paragraph{Example} Not every smooth Seifert manifold over a spindle admits a compatible AdS-Kerr-Bolt filling, but for certain discrete parameters, this is possible. As a concrete example, consider the solution \eqref{eq:KBsol} with
\begin{equation}
\alpha=1\,,\qquad
n=\frac16,\qquad
m=\frac{145}{54}\, .
\end{equation}
Then $\mathcal Q_+=0$ determines the horizon radius $r_+ = 11/6$. Local regularity requires
\begin{equation}
\Omega' =\frac{\alpha}{r_+^2-(\alpha+n)^2}
=\frac12\, ,
\qquad
\beta=\frac{4\pi\left(r_+^2-(\alpha+n)^2\right)}
{\mathcal Q'+}
=\frac{3\pi}{7}\, .
\end{equation}
Moreover, the two local axial factors are
\begin{equation}
\mathcal P_+=\frac83\, ,
\qquad
\mathcal P_-=\frac43\, .
\end{equation}
The cycles $H,N,S$ thus have the following period vectors:
\begin{equation}
H=\left(\frac{3\pi}{7}\,,\;-\frac{3\pi}{14}\right)\,,\qquad
N=\left(0\,,\;\frac{3\pi}{4}\right)\,,\qquad
S=\left(\pi\,,\;\frac{3\pi}{2}\right)\,.
\end{equation}
Choosing an appropriate second generator $A$ of global identifications, we get
\begin{equation}
A=\left(\frac{2\pi}{7},\frac{3\pi}{28}\right)\, ,\qquad N=3A-2H,\qquad
S=8A-3H\, .
\end{equation}
Since $\gcd(2,3)=\gcd(3,8)=1$, the three-dimensional conformal boundary is smooth.  The thermal circle generated by $H$ acts locally freely, with exceptional fibers of orders $3$ and $8$, and its quotient is the spindle $\mathbb{WCP}^{1}_{[3,8]}$. Since the $H$-circle collapses at the horizon, the latter can be thought of as the spindle. In the corotating coordinate $\phi=\varphi+\frac12 t{\scriptscriptstyle E}$, the two lattice generators take the simple form
\begin{equation}
H=\left(\frac{3\pi}{7}\,,\;0\right)\,,
\qquad
A=\left(\frac{2\pi}{7}\,,\;\frac{\pi}{4}\right)\,.
\end{equation}
Thus, the angular coordinate on the orbifold base has period $\Delta\phi=\pi/4$. 

\linespread{1.11}
\bibliography{CFTonM3}
\bibliographystyle{JHEP}
\end{document}